\documentclass[12pt]{article}
\usepackage{amsfonts,bm}
\usepackage{graphicx}
\usepackage{amsmath}
\usepackage{setspace}
\usepackage{float}
\usepackage[margin=1.1in]{geometry}
\usepackage{ntheorem}
\usepackage{epsfig}
\usepackage{subfigure}
\theoremstyle{plain}
\theorembodyfont{}
\theoremsymbol{}
\theoremprework{}
\theorempostwork{}
\theoremseparator{.}

\begin{document}


\title{POD/DEIM Nonlinear model order reduction of an ADI implicit shallow water equations model}
\author{R. \c Stef\u anescu, I.M. Navon \\The Florida State University, Department of Scientific Computing,\\ Tallahassee, Florida 32306, USA\\rstefanescu@fsu.edu, inavon@fsu.edu}

\date{}
\maketitle

\begin{abstract}
    In the present paper we consider a 2-D shallow-water equations (SWE) model on a $\beta$-plane solved using an alternating direction fully implicit (ADI) finite-difference scheme 
    on a rectangular domain. The scheme was shown to be unconditionally stable for the linearized equations.

    The discretization yields a number of nonlinear systems of algebraic equations. We then use a proper orthogonal decomposition (POD) to reduce the dimension of the SWE model. Due to the model nonlinearities, the computational complexity of the reduced model still depends on the number of variables of the full shallow - water equations model. By employing the discrete empirical interpolation method (DEIM) we reduce the computational complexity of the reduced order model due to its depending on the nonlinear full dimension model and regain the full model reduction expected from the POD model.

    To emphasize the CPU gain in performance due to use of POD/DEIM, we also propose testing an explicit Euler finite difference scheme (EE) as an alternative to the ADI implicit scheme for solving the swallow water equations model.

    We then proceed to assess the efficiency of POD/DEIM as a function of number of spatial discretization  points, time steps, and POD basis functions. As was expected, our numerical experiments showed that the CPU time performances of POD/DEIM schemes are proportional to the number of mesh points. Once the number of spatial discretization points exceeded $10000$ and for $90$ DEIM interpolation points, the CPU time was decreased by a factor of $10$ in case of POD/DEIM implicit SWE scheme and by a factor of $15$ for the POD/DEIM explicit SWE scheme in comparison with the corresponding POD SWE schemes. Moreover, our numerical tests revealed that if the number of points selected by DEIM algorithm reached 50, the approximation errors due to POD/DEIM and POD reduced systems have the same orders of magnitude supporting the theoretical results existing in the literature.

\end{abstract}

{\bf Keyword:} shallow water equations; proper orthogonal decomposition; reduced-order models (ROMs); finite difference methods; discrete empirical interpolation method (DEIM);

\section{Introduction}
The shallow water equations are the simplest form of the equations of motion that can be used to describe the horizontal (motion) structure of the atmosphere. They describe the evolution of an incompressible and inviscid fluid in response to gravitational and rotational accelerations and their solutions represent East West propagating Rossby waves and inertia - gravity waves. 

To avoid the limitations imposed by the Courant Friedrichs - Lewy (CFL) stability conditions restricting the time steps in explicit finite difference approximations, implicit scheme must be considered. We propose here the alternating direction implicit (ADI) method introduced by Gustafsson in \cite{Gus1971}. Linear and nonlinear versions of ADI scheme may be found in studies proposed by Fairweather and Navon in \cite{FN1980} and Navon and De Villiers in \cite{NVG1986}. Kreiss and Widlund in \cite{KreWid1966} established the convergence of alternating direction implicit methods for elliptic problems.  Such methods reduce multidimensional problem to systems of one dimensional problems (Douglas and Gunn \cite{DoGu1964}, Yanenko \cite{Yan1971} and Marchuck \cite{March1974}).

The nonlinear algebraic systems corresponding to the discrete model were solved using the quasi-Newton method proposed in Gustafsson \cite{Gus1971}. This quasi-Newton method performs an LU decomposition done every $M$-th time step, where $M$ is a fixed integer. Since back substitution is a fast operation the scheme will be efficient as long as the number of iterations is small.

The major issue in large scale complex modelling is that of reducing the computational cost while preserving numerical accuracy. Among the model reduction techniques, the proper orthogonal decomposition (POD) method provides an efficient means of deriving the reduced basis for high-dimensional nonlinear flow systems. The POD method has been widely and successfully applied to signal analysis and pattern recognition as Karhunen–-Lo\`{e}ve, statistics as principal component analysis (PCA), geophysical fluid dynamics and meteorology as empirical orthogonal functions (EOF) etc. The POD method was applied also to SWE model and we mention here the work of Cao et al. \cite{Caoetal2006}, Vermeulen and Heemink \cite{VermHeem2006}, Daescu and Navon \cite{DaeNav2008} and Altaf et al. \cite{Altafetal2009}.

In this paper we reduced the dimension of the SWE model by employing the POD method.   However due to the nonlinearities of the implicit SWE model the computational complexity of the reduced shallow water equations model still depends on the number of variables of the full shallow - water equations model. To mitigate this problem, we apply the discrete empirical interpolation method (DEIM) to address the reduction of the nonlinear components and thus reduce the computational complexity by implementing the POD/DEIM method.

DEIM is a discrete variant of the empirical interpolation method (EIM) proposed by Barrault et al. \cite{BMN2004} for constructing an approximation of a non-affine parameterized function, which was proposed in the context of reduced-basis discretization of nonlinear partial differential equations. The application was suggested and analysed by Chaturantabut and Sorensen in \cite{ChaSor2012}, \cite{ChaSor2010}, \cite{ChaSor2011} and \cite{Cha2008}.

The paper is organized as follows. In Section $2$ we introduce the Gustafsson ADI fully implicit method applied to the shallow water equations model and briefly describe its algorithmic components since they are already available in archived literature. In Section $3$ we describe in some detail the snapshot POD procedure and its implementation to the ADI method for the SWE model. Section $4$ addresses the snapshot POD combined with DEIM methodology and provides the detailed algorithmic description of the DEIM implementation. In Section $5$ we present the numerical experiments related to the POD/DEIM procedure for both explicit and implicit schemes applied to the SWE models.

The POD/DEIM procedure amounts to replace orthogonal projection with an interpolation projection of the nonlinear terms that requires the evaluation of only a few selected components of the nonlinear terms.

We evaluate the efficiency of DEIM as a function of number of spatial discretization points, time steps and basis functions for this quadratically nonlinear problem and additional studies about the conservation of the integral invariants of the SWE, root mean square errors(RMSEs) and correlation coefficients between full model, POD and POD/DEIM systems were performed.

\section{Brief description of the Gustafsson ADI  method.}

In meteorological and oceanographic problems, one is often not interested in small time steps because the discretization error in time is small compared to the discretization error in space. The fully implicit scheme considered in this paper is first order in both time and space and it is stable for large CFL condition numbers (we tested the stability of the scheme for a CFL condition number equal to $7.188$). It was proved by Gustafsson in \cite{Gus1971} that the method is unconditionally stable for the linearized version of the SWE model.

Here we shortly describe the Gustafsson shallow water alternating direction implicit method (Gustafsson \cite{Gus1971}, Fairweather and Navon \cite{FN1980}, Navon and de Villiers \cite{NVG1986}). We are solving the SWE model using the $\beta$-plane approximation on a rectangular domain.
\begin{equation}\label{eq1}
\frac{\partial w}{\partial t}=A(w)\frac{\partial w}{\partial x}+B(w)\frac{\partial w}{\partial y}+C(y)w,
\end{equation}
$$0\leq x\leq L,~0\leq y\leq D,~t\in(0,t_f],$$
where $w=(u,v,\phi)^T$ is a vector function, $u,v$ are the velocity components in the $x$ and $y$ directions, respectively, $h$ is the depth of the fluid, $g$ is the acceleration due to gravity and $\phi = 2\sqrt{gh}$.

The matrices $A,~B$ and $C$ are expressed

$$ A=-\left(\begin{array}{ccc}
           u&0&\phi/2\\
           0&u&0\\
           \phi/2&0&u \end{array}\right),
~B=-\left(\begin{array}{ccc}
           v&0&0\\
           0&v&\phi/2\\
           0&\phi/2&v \end{array}\right),
~ C=\left(\begin{array}{ccc}
           0&f&0\\
           -f&0&0\\
           0&0&0 \end{array}\right),$$

where $f$ is the Coriolis term given by
$$f=\hat f + \beta(y-D/2),~\beta=\frac{\partial f}{\partial y},~y\in[0,D],$$
with $\hat f$ and $\beta$ constants.

We assume periodic solutions in the x-direction
$$w(x,y,t)=w(x+L,y,t),~x=0,~y\in[0,D],~t\in(0,t_f],$$
while in the $y-$direction
$$v(x,0,t)=v(x,D,t)=0,~x\in[0,L],~t\in(0,t_f].$$
Initially $w(x,y,0)=\psi(x,y),~\psi:\mathbb{R}\times\mathbb{R}\rightarrow \mathbb{R},~(x,y)\in[0,L]\times[0,D]$. Note that no boundary conditions are necessary for $u$ and $\phi$ at $y=0,D$.

Now we introduce a mesh of $N_x\cdot N_y$ equidistant points on $[0,L]\times[0,D]$, with $\Delta x=L/(N_x-1),~\Delta y=D/(N_y-1)$. We also discretize the time interval $[0,t_f]$ using $NT$ equally distributed points and $\Delta t=t_f/(NT-1)$. Next we define vectors of unknown variables of dimension $n_{xy}=N_x\cdot N_y$ containing approximate solutions such as
$${\boldsymbol u}(t_n)\approx u(x_i,y_j,t_n),{\boldsymbol v}(t_n)\approx v(x_i,y_j,t_n),{\bm{{\phi}}}(t_n)\approx \phi(x_i,y_j,t_n)\in\mathbb{R}^{n_{xy}},$$
$$i=1,2,..,N_x,~j=1,2,..,N_y,~n=1,2,..NT. $$

Then Gustafsson's nonlinear ADI finite difference shallow water equations scheme (ADI FD SWE) is defined by

I. First step - get solution at $t_{n+\frac{1}{2}}$
\begin{equation}\label{eq2}
\begin{split}
&\hspace{-15mm}{\bf u}(t_{n+\frac{1}{2}})+\frac{\Delta t}{2}F_{11}\biggl({\bf u}(t_{n+\frac{1}{2}}),{\boldsymbol \phi}(t_{n+\frac{1}{2}})\biggr)={\bf u}(t_n)-\frac{\Delta t}{2}F_{12}\biggl({\bf u}(t_n),{\bf v}(t_n)\biggr)+\\
&~~~~~~~~~~~~~~~~~~~~~~~~~~~~~~~~~~~~~~~~~~\frac{\Delta t}{2} {[\underbrace{{\bf f},{\bf f},..,{\bf f}}_{N_x}]}^T*{\bf v}(t_n),\\
&\hspace{-15mm}{\bf v}(t_{n+\frac{1}{2}})+\frac{\Delta t}{2}F_{21}\biggl({\bf u}(t_{n+\frac{1}{2}}),{\bf v}(t_{n+\frac{1}{2}})\biggr)+\frac{\Delta t}{2}{[\underbrace{{\bf f},{\bf f},..,{\bf f}}_{N_x}]}^T*{\bf u}(t_{n+\frac{1}{2}})={\bf v}(t_n)-\\
&~~~~~~~~~~~~~~~~~~~~~~~~~~~~~~~~~~~~~~~~~~\frac{\Delta t}{2} F_{22}\biggl({\bf v}(t_n),{\boldsymbol \phi}(t_n)\biggr),\\
&\hspace{-15mm}{\boldsymbol \phi}(t_{n+\frac{1}{2}})+\frac{\Delta t}{2}F_{31}\biggl({\bf u}(t_{n+\frac{1}{2}}),{\boldsymbol \phi}(t_{n+\frac{1}{2}})\biggr)={\boldsymbol\phi}(t_n)-\frac{\Delta t}{2}F_{32}\biggl({\bf v}(t_n), {\boldsymbol \phi}(t_n)\biggr),
\end{split}
\end{equation}
with "*" denoting the componentwise multiplication, ${\bf f}$ is a $N_y$-dimensional vector storing the Coriolis components $f(y_j),~j=1,2,..N_y$ and the nonlinear functions $F_{11},F_{12},F_{21}, F_{22},F_{31},F_{32}: \mathbb{R}^{n_{xy}} \times$ $\mathbb{R}^{n_{xy}} \rightarrow \mathbb{R}^{n_{xy}}$ are defined as follows
$$F_{11}({\bf u},{\boldsymbol \phi})={\boldsymbol u}*A_x{\boldsymbol u}+\frac{1}{2}{\boldsymbol \phi}*A_x{\boldsymbol\phi},$$
$$F_{12}({\boldsymbol u},{\boldsymbol v})={\boldsymbol v}*A_y{\boldsymbol u}, F_{21}({\boldsymbol u},{ \boldsymbol v})={\boldsymbol u}*A_x{\boldsymbol v},$$
$$F_{22}({\boldsymbol v},{\boldsymbol\phi})={\boldsymbol v}*A_y{\boldsymbol v}+\frac{1}{2}{\boldsymbol \phi}*A_y{\boldsymbol\phi},$$
$$F_{31}({\boldsymbol u},{\boldsymbol \phi})=\frac{1}{2}{\boldsymbol\phi}* A_x {\boldsymbol u}+{\boldsymbol u}*{A_x\boldsymbol \phi},$$
$$F_{32}({\boldsymbol v},{\boldsymbol \phi})=\frac{1}{2}{\boldsymbol\phi}*A_y{\boldsymbol v}+{\boldsymbol v}*A_y{\boldsymbol\phi},$$
where $A_x,A_y\in \mathbb{R}^{n_{xy}\times n_{xy} }$ are constant coefficient matrices for discrete first-order and second-order differential operators which take into account the boundary conditions.

II. Second step - get solution at $t_{n+1}$
\begin{equation}\label{eq3}
\begin{split}
&\hspace{-9mm}{\bf u}(t_{n+1})+\frac{\Delta t}{2}F_{12}\biggl({\bf u}(t_{n+1}),{\bf v}(t_{n+1})\biggr)-\frac{\Delta t}{2} {[\underbrace{f,f,..,f}_{N_x}]}^T*{\bf v}(t_{n+1})={\bf u}(t_{n+\frac{1}{2}})-\\
&~~~~~~~~~~~~~~~~~~~~~~~~~~~~~~~~~~~~~~~~~~\frac{\Delta t}{2}F_{11}\biggl({\bf u}(t_{n+\frac{1}{2}}),{\boldsymbol \phi}(t_{n+\frac{1}{2}})\biggr),\\
&\hspace{-9mm}{\bf v}(t_{n+1})+\frac{\Delta t}{2} F_{22}\biggl({\bf v}(t_n),{\boldsymbol \phi}(t_n)\biggr)={\bf v}(t_{n+\frac{1}{2}})-\frac{\Delta t}{2}F_{21}\biggl({\bf u}(t_{n+\frac{1}{2}}),{\bf v}(t_{n+\frac{1}{2}})\biggr)-\\
&~~~~~~~~~~~~~~~~~~~~~~~~~~~~~~~~~~~~~~~~~\frac{\Delta t}{2}{[\underbrace{f,f,..,f}_{N_x}]}^T*{\bf u}(t_{n+\frac{1}{2}}),\\
&\hspace{-9mm}{\boldsymbol \phi}(t_{n+1})+\frac{\Delta t}{2}F_{32}\biggl({\bf v}(t_{n+1}), {\boldsymbol \phi}(t_{n+1})\biggr)={\boldsymbol\phi}(t_{n+{\frac{1}{2}}})-\frac{\Delta t}{2}F_{31}\biggl({\bf u}(t_{n+\frac{1}{2}}),{\boldsymbol \phi}(t_{n+\frac{1}{2}})\biggr).
\end{split}
\end{equation}

The nonlinear systems of algebraic equations (\ref{eq2}) and (\ref{eq3}) are solved using the quasi-Newton method. Thereby we rewrite (\ref{eq2}) and (\ref{eq3}) in the form
$$ g(\alpha) = 0$$
where $\alpha$ is the vector of unknowns.
Due to the fact that no more than two variables are coupled to each other on the left-hand side of equations (\ref{eq2}) and (\ref{eq3}), we first solve system (\ref{eq2}) for ${\bf u}=[u_1,u_2,...,u_{n_{xy}}]$ and ${\boldsymbol\phi}=[\phi_1,\phi_2,...,\phi_{n_{xy}}]$ i.e. the first and the third equations in (\ref{eq2}) and define
    $$ \alpha = (u_1,\phi_1,u_2,\phi_2,...,u_{n_{xy}},\phi_{n_{xy}})\in\mathbb{R}^{2n_{xy}}.$$
The iterative Newton method is given by
    \begin{equation*}
    \alpha^{(m+1)}=\alpha^{(m)}-J^{-1}(\alpha^{(m)})g(\alpha^{(m)}),
    \end{equation*}
where the superscript denotes the iteration and $J\in \mathbb{R}^{2n_{xy}\times 2n_{xy}}$ is the Jacobian
$$ J=\frac{\partial g}{\partial \alpha}.$$

Owing to the structure of the Gustafsson algorithm for the SWE, the Jacobian matrix is either block cyclic tridiagonal or block tridiagonal.

$J^{-1}g$ is solved by first applying an $LU$ decomposition to $J$. Then it is computed by backsubstitution in two stages. First $z$ is solved from
    $$Lz=g,$$
and then $J^{-1}g$ is obtained from
$$U(J^{-1}g)=z.$$
In the quasi-Newton method, the computationally expensive $LU$ decomposition is performed only once every $M-th$ time-step, where $M$ is a fixed integer.

Because the backsubstitution is a fast operation, the quasi-Newton method is computationally efficient especially when the number of nonlinear iterations at each time step is small. Gustafsson proved in \cite{Gus1971} that even one quasi-Newton iteration is sufficient at each time step.

The quasi-Newton formula is
\begin{equation*}
    \alpha^{(m+1)}=\alpha^{(m)}-\hat J^{-1}(\alpha^{(m)})g(\alpha^{(m)}),\textrm { where }
\end{equation*}
$$\hat J = J(\alpha^{(0)})+O(\Delta t).$$

The method works when $M$, the number of time-steps between successive updating of the $LU$ decomposition of the Jacobian matrix $J$, is a relatively small number. For our numerical experiments we took $M=6$.

The second part of the system (\ref{eq2}), the second equation in (\ref{eq2}) is solved for ${\bf v}=[v_1,v_2,...,v_{n_{xy}}]$ by employing the same quasi-Newton method. Thus $\alpha$ is defined as
    $$\alpha = (v_1,v_2,...,v_{n_{xy}})\in\mathbb{R}^{n_{xy}}.$$

In order to obtain the SWE numerical solution at $t_{n+1}$ we applied the same quasi-Newton technique for system (\ref{eq3}). This time the coupled variables were
    $$ \alpha = (v_1,\phi_1,v_2,\phi_2,...,v_{n_{xy}},\phi_{n_{xy}})\in\mathbb{R}^{2n_{xy}}$$
for the second and third equation in (\ref{eq3}), while $\bf u$ was solved from the remaining equation.

\section{The POD version of SWE model}

Proper orthogonal decomposition provides a technique for deriving low order model of dynamical systems. It can be thought of as a Galerkin approximation in the spatial variable built from functions corresponding to the solution of the physical system at specified time instances. Noack et al. \cite{Noack2010} proposed a system reduction strategy for Galerkin models of fluid flows leading to dynamic models of lower order based on a partition in slow, dominant and fast modes. Let us denote by $Y=[u^1,~u^2,...,~u^{NT}] \in \mathbb{R}^{n_{xy}\times NT}$ an ensemble of $NT-$ time instances of the numerical solution obtained from ADI FD SWE scheme at $t_1$,$t_2$,..,$t_{NT}$ for the horizontal component of the velocity. Due to possible linear dependence, the snapshots themselves are not appropriate as a basis. Instead three methods can be employed, singular value decomposition (SVD) for $Y\in \mathbb{R}^{n_{xy}\times NT}$, eigenvalues decomposition for $YY^T\in \mathbb{R}^{n_{xy}\times n_{xy}}$ or eigenvalue decomposition for $Y^TY\in \mathbb{R}^{NT\times NT}$ (see \cite{DimApr2008} and \cite{DimAprSte2010}) and the leading generalized eigenfunctions are chosen as a basis, referred to as POD basis. Error estimates for proper orthogonal decomposition models for nonlinear dynamical systems may be found in \cite{HinVol2005}.

Here we built the POD decomposition of each variables separately and we present only the construction of the POD basis corresponding to $u$ since we applied a similar procedure to determine the POD bases for $v$ and $\phi$. Taking into account that $NT\ll n_{xy}$, we choose to construct the POD basis $U\in\mathbb{R}^{n_{xy}\times k},~k\in \mathbb{N}^*$ by solving the eigenvalue problem
$$ Y^TY\hat u_i=\lambda_i\hat u_i,~i=1,2,..,NT,$$
and retaining the set of right singular vectors of $Y$ corresponding to the $k$ largest singular values, i.e. $U=\{u_i\}_{i=1}^k$, $u_i=\frac{1}{\sqrt{\lambda_i}}Y\hat u_i$.

Similarly, let $V,\Phi\in \mathbb{R}^{n_{xy}\times k} $ be the POD basis matrices of the vertical component of the velocity and geopotential, respectively. Now we can approximate ${\bf u},{\bf v}$ and ${\boldsymbol \phi}$ as follows
    $$ {\bf u}(t_n)\approx U\mathbf{\tilde u}(t_n),~{\bf v}(t_n)\approx V\mathbf{\tilde v}(t_n),~{\bm{{ \phi}}}(t_n)\approx \Phi{\bm{{\tilde \phi}}}(t_n),$$
    $$\mathbf{ \tilde u}(t_n),\mathbf{ \tilde v}(t_n),{\bm{{\tilde \phi}}}(t_n) \in \mathbb{R}^{k},~n=1,2,..,NT.$$

The POD reduced-order system is constructed by applying the Galerkin projection method to ADI FD SWE discrete model (\ref{eq2}) and (\ref{eq3}) by first replacing ${\bf u},{\bf v},{\bm\phi}$ with their approximations $U{\bm\tilde u}$, $V{\bm\tilde v}$, $\Phi{\bm\tilde \phi}$, respectively, and then premultiplying the corresponding equations by $U^T$, $V^T$ and $\Phi^T$.

The resulting POD reduced system for the first step $t_{n+\frac{1}{2}}$ of the ADI FD SWE scheme is

\begin{equation}\label{eq4}
\begin{split}
&\hspace{-8mm}{\mathbf{\tilde u}}(t_{n+\frac{1}{2}})+\frac{\Delta t}{2}U^T\tilde F_{11}\biggl({\mathbf{\tilde u}}(t_{n+\frac{1}{2}}),{\bm{{\tilde \phi}}}(t_{n+\frac{1}{2}})\biggr)={\mathbf{\tilde u}}(t_n)-\frac{\Delta t}{2}U^T\tilde F_{12}\biggl({\mathbf{\tilde u}}(t_n),{\mathbf{\tilde v}}(t_n)\biggr)\\
&~~~~~~~~~~~~~~~~~~~~~~~~~~~~~~~~~~~~~~~~~~+\frac{\Delta t}{2} U^T\biggl({[\underbrace{{\bf f},{\bf f},..,{\bf f}}_{N_x}]}^T*{V \mathbf{\tilde v}}(t_n)\biggr),\\
&\hspace{-8mm}{\mathbf{\tilde v}}(t_{n+\frac{1}{2}})+\frac{\Delta t}{2}V^T\tilde F_{21}\biggl({\mathbf{\tilde u}}(t_{n+\frac{1}{2}}),{\mathbf{\bm\tilde v}}(t_{n+\frac{1}{2}})\biggr)+\frac{\Delta t}{2}V^T\biggl({[\underbrace{{\bf f},{\bf f},..,{\bf f}}_{N_x}]}^T*{U \mathbf{\tilde u}}(t_{n+\frac{1}{2}})\biggr)\\
&~~~~~~~~~~~~~~~~~~~~~~~~~~~~~~~~~~~~~~~~~~={\mathbf{\tilde v}}(t_n)-\frac{\Delta t}{2}V^T\tilde F_{22}\biggl({\mathbf{\tilde v}}(t_n),{\bm{{\tilde \phi}}}(t_n)\biggr),\\
&\hspace{-8mm}{\bm{{\tilde \phi}}}(t_{n+\frac{1}{2}})+\frac{\Delta t}{2}\Phi^T\tilde F_{31}\biggl({\mathbf{\tilde u}}(t_{n+\frac{1}{2}}),{\bm{{\tilde \phi}}}(t_{n+\frac{1}{2}})\biggr)={\bm{{\tilde \phi}}}(t_n)-\frac{\Delta t}{2}\Phi^T\tilde F_{32}\biggl({\mathbf{\tilde v}}(t_n), {\bm{{\tilde \phi}}}(t_n)\biggr),
\end{split}
\end{equation}
where $\tilde F_{11},\tilde F_{12},\tilde F_{21},\tilde F_{22},\tilde F_{31}, \tilde F_{32}: \mathbb{R}^{k} \times$ $\mathbb{R}^{k} \rightarrow \mathbb{R}^{k}$ are defined by
\begin{equation}\label{eq5}
\begin{split}
&\tilde F_{11}({\mathbf{\tilde u}},{\bm{{\tilde \phi}}})=(U{\mathbf{ \tilde u}})*(\underbrace{A_xU}{\mathbf{\tilde u}})+\frac{1}{2}({\Phi{\bm{{\tilde \phi}}}})*(\underbrace{A_x\Phi}{\bm{{\tilde \phi}}}),\\
&\tilde F_{12}({\mathbf{\tilde u}},{\mathbf{\tilde v}})=(V{\mathbf{\tilde v}})*(\underbrace{A_yU}{\mathbf{\tilde u}}), \tilde F_{21}({\mathbf{\tilde u}},{\mathbf{\tilde v}})=(U{\mathbf{\bm \tilde u}})*(\underbrace{A_xV}{\mathbf{ \tilde v}}),\\
&\tilde F_{22}({\mathbf{\tilde v}},{\bm{{\tilde \phi}}})=(V{\mathbf{ \tilde v}})*(\underbrace{A_yV}{\mathbf{\tilde v}})+\frac{1}{2}(\Phi{\bm{{\tilde \phi}}})*(\underbrace{A_y\Phi}{\bm{{\tilde \phi}}}),\\
&\tilde F_{31}({\mathbf{\tilde u}},{\bm{{\tilde \phi}}})=\frac{1}{2}(\Phi{\bm{{\tilde \phi}}})* (\underbrace{A_x U}{\mathbf{\tilde u}})+(U{\mathbf{\bm \tilde u}})*(\underbrace{A_x\Phi}{\bm{{\tilde \phi}}}),\\
&\tilde F_{32}({\mathbf{\tilde v}},{\bm{{\tilde \phi}}})=\frac{1}{2}(\Phi{\bm{{\tilde \phi}}})*(\underbrace{A_yV}{\mathbf{\tilde v}})+(V{\mathbf{ \tilde v}})*(\underbrace{A_y\Phi}{\bm{{\tilde \phi}}}).
\end{split}
\end{equation}

The second step of the POD reduced system for the ADI FD SWE scheme is depicted below

\begin{equation}\label{eq6}
\begin{split}
&\hspace{-8mm}{\mathbf{\tilde u}}(t_{n+1})+\frac{\Delta t}{2}U^T\tilde F_{12}\biggl({\mathbf{\tilde u}}(t_{n+1}),{\mathbf{\tilde v}}(t_{n+1})\biggr)-\frac{\Delta t}{2} U^T\biggl({[\underbrace{{\bf f},{\bf f},..,{\bf f}}_{N_x}]}^T*{V \mathbf{\tilde v}}(t_{n+1})\biggr)={\mathbf{\tilde u}}(t_{n+\frac{1}{2}})-\\
&~~~~~~~~~~~~~~~~~~~~~~~~~~~~~~~~~~~~~~~~~~\frac{\Delta t}{2}U^T\tilde F_{11}\biggl({\mathbf{\tilde u}}(t_{n+\frac{1}{2}}),{\bm{{\tilde \phi}}}(t_{n+\frac{1}{2}})\biggr)\\
&\hspace{-8mm}{\mathbf{\tilde v}}(t_{n+1})+\frac{\Delta t}{2}V^T\tilde F_{22}\biggl({\mathbf{\tilde v}}(t_{n+1}),{\bm{{\tilde \phi}}}(t_{n+1})\biggr)={\mathbf{\tilde v}}(t_{n+\frac{1}{2}})-\frac{\Delta t}{2}V^T\tilde F_{21}\biggl({\mathbf{\tilde u}}(t_{n+\frac{1}{2}}),{\mathbf{\bm\tilde v}}(t_{n+\frac{1}{2}})\biggr)\\
&~~~~~~~~~~~~~~~~~~~~~~~~~~~~~~~~~~~~~~~~~~-\frac{\Delta t}{2}V^T\biggl({[\underbrace{{\bf f},{\bf f},..,{\bf f}}_{N_x}]}^T*{U \mathbf{\tilde u}}(t_{n+\frac{1}{2}})\biggr),\\
&\hspace{-8mm}{\bm{{\tilde \phi}}}(t_{n+1})+\frac{\Delta t}{2}\Phi^T\tilde F_{32}\biggl({\mathbf{\tilde v}}(t_{n+1}), {\bm{{\tilde \phi}}}(t_{n+1})\biggr)={\bm{{\tilde \phi}}}(t_{n+\frac{1}{2}})-\frac{\Delta t}{2}\Phi^T\tilde F_{31}\biggl({\mathbf{\tilde u}}(t_{n+\frac{1}{2}}),{\bm{{\tilde \phi}}}(t_{n+\frac{1}{2}})\biggr).
\end{split}
\end{equation}

The initial conditions are obtain by multiplying the following three equations with $U^T,~V^T,~\Phi^T$
$${\bf u}(t_1)\approx U\mathbf{\tilde u}(t_1),~
{\bf v}(t_1)\approx V\mathbf{\tilde v}(t_1),~
{\bm{{\phi}}}(t_1)\approx \Phi{\bm{{\tilde \phi}}}(t_1).$$
We get
$$\mathbf{\tilde u}(t_1)\approx U^T{\bf u}(t_1),~
\mathbf{\tilde v}(t_1)\approx V^T{\bf v}(t_1),~
{\bm{{\tilde \phi}}}(t_1)\approx \Phi^T{\bm{{\phi}}}(t_1).$$

Next we define $A_1,A_2\in\mathbb{R}^{n_{xy}\times k}$ such as
  $$\hspace{-5mm}A_1(:,i)={[\underbrace{f,f,..,f}_{N_x}]}^T*V(:,i),~A_2(:,i)={[\underbrace{f,f,..,f}_{N_x}]}^
  T*U(:,i),i=1,..,k$$ and the linear terms in (\ref{eq4}) and (\ref{eq6}), $ U^T\biggl({[\underbrace{f,f,..,f}_{N_x}]}^T*V \mathbf{\tilde v}\biggr)$ and $V^T\biggl({[\underbrace{f,f,..,f}_{N_x}]}^T*U \mathbf{\tilde u}\biggr)$ can be rewritten as $\underbrace{U^TA_1}\mathbf{\tilde v}$ and $\underbrace{V^TA_2}\mathbf{\tilde u}$ respectively.

The coefficient matrices $U^TA_1,~V^TA_2\in \mathbb{R}^{k\times k}$ defined in the linear terms of the POD reduced system as well as the coefficient matrices in the nonlinear functions (i.e. $A_xU,A_yU,$ $A_xV,A_yV,A_x\Phi,A_y\Phi\in \mathbb{R}^{n\times k}$ grouped by the curly braces in (\ref{eq5})) can be precomputed, saved and re-used in all time steps of the interval of integration $[0,t_f]$. However, performing the componentwise multiplications in (\ref{eq5}) and computing the projected nonlinear terms in (\ref{eq4}) and (\ref{eq6})
    \begin{equation}\label{eq7}
    \begin{split}
    &\underbrace{U^T}_{k\times n_{xy} }\underbrace{\tilde F_{11}(\mathbf{\tilde u},{\bm{{\tilde \phi}}})}_{n_{xy}\times 1},~~~~~U^T\tilde F_{12}(\mathbf{\tilde u},\mathbf{\tilde v}),~~~~~V^T\tilde F_{21}(\mathbf{\tilde u},\mathbf{\tilde v}),\\
    &~~~~V^T\tilde F_{22}(\mathbf{\tilde v},{\bm{{\tilde \phi}}}),~~~~~\Phi^T\tilde F_{31}(\mathbf{\tilde u},{\bm{{\tilde \phi}}}),~~~~~\Phi^T\tilde F_{32}(\mathbf{\tilde v},{\bm{{\tilde \phi}}}),
    \end{split}
    \end{equation}
    still have computational complexities depending on the dimension $n_{xy}$ of the original system from both evaluating the nonlinear functions and performing matrix multiplications to project on POD bases. If we denote the complexity for evaluating the nonlinear function $\tilde F_{11}$ by $\alpha(n_{xy})$, then the complexity for computing $U^T\tilde F_{11}(\mathbf{\tilde u},{\bm{{\tilde \phi}}})$ is approximately $O(\alpha(n_{xy})+4n_{xy}k)$.

    By employing the discrete empirical interpolation method we aim to remove this dependency and regain the full model reduction expected from the POD model.The projected nonlinear functions can be approximated by DEIM in a form that enables precomputation so that the computational cost is decreased and independent of the original system. Only a few entries of the nonlinear term corresponding to the specially selected interpolation indices from DEIM algorithm described in the next section must be evaluated at each time step.

    DEIM approximation is applied to each of the nonlinear functions $\tilde F_{11},\tilde F_{12},\tilde F_{21},\tilde F_{22},\tilde F_{31},$ $\tilde F_{32}$ defined in (\ref{eq5}).

\section{The POD/DEIM method and its application to the ADI/SWE model}
\subsection{Discrete Empirical Interpolation Method}

DEIM is a discrete variation of the Empirical Interpolation method (EIM) proposed by Barrault et al. \cite{BMN2004}. The application was suggested and analyzed by Chaturantabut and Sorensen in \cite{ChaSor2012}, \cite{ChaSor2010}, \cite{Cha2008}. In \cite{ChaSor2012}, authors present an error estimate of the POD/DEIM method. This discrete empirical interpolation method provides an efficient way to approximate nonlinear functions. It was also incorporated into the reduced-basis techniques to provide a better reduced-basis treatment (in terms of CPU time) of nonaffine and non-linear parameterized PDEs. DEIM was succesfully applied in conjunction with POD for models governing the voltage dynamics of neurons in \cite{KCS2010}, the integrated circuits with semiconductors with modified nodal analysis and drift diffusion (see \cite{HK2012}) and dynamics of the concentration of lithium ions in lithium ion batteries in \cite{LW2012}. In order to improve the stability of POD/DEIM reduced order schemes in case of a nonlinear transmission line, a micromachined switch and a nonlinear thermal model for a RF amplifier a few modifications to the DEIM based model reduction were proposed by Hochman et al. in \cite{HBW2011}.

Next we describe the DEIM approximation procedure applied to a nonlinear function. Let  $f:D\rightarrow \mathbb{R}^n,~D\subset \mathbb{R}^n$ be a nonlinear function. If $U = [u_1,..,u_{m}]$, $u_i \in \mathbb{R}^n$, $i=1,..,m$ is a linearly independent set, for $ m\leq n$, then for $\tau\in D$, the DEIM approximation of order $m$ for $f(\tau)$ in the space spanned by $\{u_l\}_{l=1}^m$ is given by
\begin{equation}\label{eq8}
f(\tau)\approx U{\bf c}(\tau),~U\in\mathbb{R}^{n\times m},~{\bf c}(\tau) \in \mathbb{R}^m.
\end{equation}
The basis U can be constructed effectively by applying the POD method on the nonlinear snapshots $f(\tau^{t_i}),~\tau^{t_i}\in D~(~\tau\textrm{ may be a function defined from } [0, T] \rightarrow D,\textrm{ and }\tau^{t_i} \textrm{ is the }$ value of  $\tau \textrm{ evaluated at }t_i) ,~i=1,..,n_s,~n_s>0$. Next, interpolation is used to determine the coefficient vector ${\bf c}(\tau)$ by selecting $m$ rows $\rho_1,..,\rho_m,~\rho_i\in \mathbb{N}^{*}$,  of the overdetermined linear system (\ref{eq8}) to form a $m-$by$-m$ linear system
$$P^TU{\bf c}(\tau)=P^Tf(\tau),$$
where $P=[e_{\rho_1},..,e_{\rho_m}]\in \mathbb{R}^{n\times m}, ~e_{\rho_i}=[0,..0,\underbrace{1}_{\rho_i},0,..,0]^T\in\mathbb{R}^n.$
The DEIM approximation of $f\in \mathbb{R}^n$ becomes
$$ f(\tau)\approx U(P^TU)^{-1}P^Tf(\tau).$$
Now the only unknowns that need to be specified are the indices $\rho_1,\rho_2,...,\rho_m$ or the matrix $P$  whose dimensions are $n \times m$. These are determined by the following pseudo - algorithm
\vskip 0.5cm
{\bf{DEIM}}: Algorithm for Interpolation Indices
\vskip 0.5cm
{\bf INPUT}: $\{u_l\}_{l=1}^m\subset\mathbb{R}^n$ (linearly independent):
\vskip 0.2cm
{\bf OUTPUT}: $\vec\rho=[\rho_1,..,\rho_m]\in\mathbb{N}^m$
\vskip 0.2cm
\begin{enumerate}
\item $[|\psi|~~ \rho_1]=max|u_1|,{\psi}\in \mathbb{R}$ and $\rho_1$ is the component position of the largest absolute value of $u_1$, with the smallest index taken in case of a tie.
\item $U=[u_1]\in \mathbb{R}^n,~P=[e_{\rho_1}]\in \mathbb{R}^n,~\vec\rho=[\rho_1]\in \mathbb{N}.$
\vskip 0.1cm

\item For $l=2,..,m$ do
\vskip 0.1cm

\begin{enumerate}

\vskip 0.1cm
\item. Solve $(P^TU){ c}=P^Tu_l \textrm{  for } {c}\in \mathbb{R}^{l-1};~U,P\in\mathbb{R}^{n\times(l-1)}. $
\vskip 0.1cm
\item . $r=u_l-U{c},~r\in \mathbb{R}^n.$
\vskip 0.1cm
\item.  $[|\psi|~~ \rho_l]=max\{|r|\}.$
\vskip 0.1cm
\item. $U \leftarrow [U~~u_l],~P \leftarrow [P~~e_{\rho_l}],~\vec\rho \leftarrow\left[\begin{array}{cc}
           \vec\rho\\
           \rho_l\\
           \end{array}\right].$
\end{enumerate}
\vskip 0.06cm

end For.
\end{enumerate}

The DEIM procedure inductively constructs a set of indices from a linearly independent set. An error analysis in \cite{Cha2008} shows that the POD basis is a suitable choice for this algorithm and the order of the input basis $\{u_l\}_{l=1}^m\subset\mathbb{R}^n$ according to the dominant singular values must be utilized. Initially the algorithm searches for the largest value of the first POD basis $|u_1|$ and the corresponding index represents the first DEIM interpolation index $\rho_1\in\{1,2,..,n\}$. The remaining interpolation indices $\rho_l,~l=2,3..,m$ are selected so that each of them corresponds to the entry of the largest magnitude of $|r|$ defined in step $(A3)-b$. The vector $r$ can be viewed as the residual or the error between the input basis $u_l,~l=2,3..,m$ and its approximation $Uc$ from interpolating the basis $\{u_1,u_2,..,u_{l-1}\}$ at the indices ${\rho_1},{\rho_2},..,{\rho_{l-1}}$. The linear independence of the input basis $\{u_l\}_{l=1}^m$ guarantees that, in each iteration, $r$ is a nonzero vector and the output indices $\{\rho_i\}_{i=1}^m$ are not repeating.

An error bound for the DEIM approximation is provided in Chaturantabut and Sorensen \cite{ChaSor2012} and \cite{Cha2008}. An example of DEIM approximation of a highly nonlinear function defined on a discrete $1D$ spatial domain can be found in \cite{ChaSor2010}, underlying the DEIM efficiency.

\subsection{The DEIM SWE model}

The DEIM approximation presented earlier in this section is used to approximate the nonlinear terms in POD ADI SWE model described in (\ref{eq7}) so that the nonlinear approximations will have a computational complexity proportional to the number of reduced variables obtained with POD. Thus, the application of DEIM in POD framework will allow the construction of faster reduced order models increasing the performances of reduced order hierarchy models such as the one presented in Noack et al. \cite{Noack2003}.

Let $U^{F_{11}}\in \mathbb{R}^{n_{xy}\times m},~m\leq n_{xy}$, be the POD basis matrix of rank $m$ for snapshots of the nonlinear function $\tilde F_{11}$ (obtained from ADI FD SWE scheme). Using the DEIM algorithm we select a set of $m$ DEIM indices corresponding to $U^{F_{11}}$, denoting by $[\rho_1^{F_{11}},..,\rho_m^{F_{11}}]^T\in \mathbb{N}^{m}$, and determine the matrix $P_{F_{11}}\in\mathbb{R}^{n_{xy}\times m}$. The DEIM approximation of $\tilde F_{11}$ assumes the form
        $$ \tilde F_{11}\approx U^{F_{11}}(P_{F_{11}}^TU^{F_{11}})^{-1}\tilde F_{11}^m,$$
so the projected nonlinear term $U^T\tilde F_{11}({\bf \tilde u},{\bm{{\tilde \phi}}})$ in the POD reduced system can be approximated as
$$U^T\tilde F_{11}({\bf \tilde u},{\bm{{\tilde \phi}}})\approx \underbrace{ U^TU^{F_{11}}(P_{F_{11}}^TU^{F_{11}})^{-1}}_{E_1\in \mathbb{R}^{k\times m}}\underbrace{\tilde F_{11}^m({\bf \tilde u},{\bm{{\tilde \phi}}})}_{m\times 1}, $$
where $\tilde F_{11}^m({\bf \tilde u},{\bm{{\tilde \phi}}})=P_{F_{11}}^T\tilde F_{11}({\bf \tilde u},{\bm{{\tilde \phi}}})$ and ${E_1\in \mathbb{R}^{k\times m}}$.

 Since $\tilde F_{11}$ is a pointwise function (introduced in (\ref{eq5})), $\tilde F_{11}^m :\mathbb{R}^k\times\mathbb{R}^k\rightarrow \mathbb{R}^m$ can be defined as
    $$\tilde F_{11}^m({\bf \tilde u},{\bm{{\tilde \phi}}})=(P_{F_{11}}^TU{\bf \tilde u})*(\underbrace{P_{F_{11}}^TA_xU}{\bf \tilde u})+\frac{1}{2}({P_{F_{11}}^T\Phi{\bm{{\tilde \phi}}}})*(\underbrace{P_{F_{11}}^TA_x\Phi}{\bm{{\tilde \phi}}}).$$

If we denote by $U^{F_{12}},U^{F_{21}},U^{F_{22}},U^{F_{31}},U^{F_{32}}\in \mathbb{R}^{n_{xy}\times m}$ the POD bases matrices of rank $m$ for the snapshots of the nonlinear functions $\tilde F_{12},\tilde F_{21}, \tilde F_{22},\tilde F_{31},\tilde F_{32}$, we obtain in a similar manner the DEIM approximations for the rest of the projected nonlinear terms in (\ref{eq7})
 $$U^T\tilde F_{12}({\bf \tilde u},{\bf \tilde v})\approx \underbrace{ U^TU^{F_{12}}(P_{F_{12}}^TU^{F_{12}})^{-1}}_{E_2\in \mathbb{R}^{k\times m}}\underbrace{\tilde F_{12}^m({\bf \tilde u},{\bf\tilde v})}_{m\times 1}, $$
    $$V^T\tilde F_{21}({\bf\tilde u},{\bf\tilde v})\approx \underbrace{ V^TU^{F_{21}}(P_{F_{21}}^TU^{F_{21}})^{-1}}_{E_3\in \mathbb{R}^{k\times m}}\underbrace{\tilde F_{21}^m({\bf \tilde u},{\bf \tilde v})}_{m\times 1}, $$
    $$V^T\tilde F_{22}({\bf\tilde v},{\bm{{\tilde \phi}}})\approx \underbrace{ V^TU^{F_{22}}(P_{F_{22}}^TU^{F_{22}})^{-1}}_{E_4\in \mathbb{R}^{k\times m}}\underbrace{\tilde F_{22}^m({\bf \tilde v},{\bm{{\tilde \phi}}})}_{m\times 1}, $$
    $$\Phi^T\tilde F_{31}({\bf \tilde u},{\bf \tilde \phi})\approx \underbrace{ \Phi^TU^{F_{31}}(P_{F_{31}}^TU^{F_{31}})^{-1}}_{E_5\in \mathbb{R}^{k\times m}}\underbrace{\tilde F_{31}^m({\bf \tilde u},{\bm{{\tilde \phi}}})}_{m\times 1}, $$
    $$\Phi^T\tilde F_{32}({\bf \tilde v},{\bm{{\tilde \phi}}})\approx \underbrace{ \Phi^TU^{F_{32}}(P_{F_{32}}^TU^{F_{32}})^{-1}}_{E_6\in \mathbb{R}^{k\times m}}\underbrace{\tilde F_{32}^m({\bf \tilde v},{\bm{{\tilde \phi}}})}_{m\times 1}, $$
 where ${E_2,E_3,E_4,E_5,E_6\in \mathbb{R}^{k\times m}}$ and
 \begin{equation}\label{eq9}
\begin{split}
&\tilde F_{12}^m({\bf\tilde u},{\bf\tilde v})=(P_{F_{12}}^TV{\bf\tilde v})*(\underbrace{P_{F_{12}}^TA_yU}{\bf \tilde u}),~\tilde F_{21}^m({\bf \tilde u},{\bf \tilde v})=(P_{F_{21}}^TU{\bf\tilde u})*(\underbrace{P_{F_{21}}^TA_xV}{\bf\tilde v}),\\
&\tilde F_{22}^m({\bf\tilde v},{\bm{{\tilde \phi}}})=(P_{F_{22}}^TV{\bf \tilde v})*(\underbrace{P_{F_{22}}^TA_yV}{\bf\tilde v})+\frac{1}{2}(P_{F_{22}}^T\Phi{\bf\tilde \phi})*(\underbrace{P_{F_{22}}^TA_y\Phi}{\bm{{\tilde \phi}}}),\\
&\tilde F_{31}^m({\bf \tilde u},{\bm{{\tilde \phi}}})=(P_{F_{31}}^T\Phi{\bm{{\tilde \phi}}})* (\underbrace{P_{F_{31}}^TA_x U}{\bf \tilde u})+(P_{F_{31}}^TU{\bf \tilde u})*(\underbrace{P_{F_{31}}^TA_x\Phi}{\bm{{\tilde \phi}}}),\\
&\tilde F_{32}^m({\bf\tilde v},{\bm{{\tilde \phi}}})=\frac{1}{2}(P_{F_{32}}^T\Phi{\bm{{\tilde \phi}}})*(\underbrace{P_{F_{32}}^TA_yV}{\bf\tilde v})+(P_{F_{32}}^TV{\bf\tilde v})*(\underbrace{P_{F_{32}}^TA_y\Phi}{\bm{{\tilde \phi}}}).
\end{split}
\end{equation}

Each of the $k\times m$ coefficient matrices grouped by the curly brackets in (\ref{eq9}), as well as $E_i,~i=1,2,..,6$ can be precomputed and reused at all time steps, so that the computational complexity for each of the approximate nonlinear terms is $O(\alpha(m)+4mk)$, thus not depending on the full-order dimension $n_{xy}$. Finally, the POD/DEIM reduced system for the first step of ADI FD SWE model is of the form
\begin{equation}\label{eq10}
\begin{split}
&\hspace{-4mm}{\bf \tilde u}(t_{n+\frac{1}{2}})+\frac{\Delta t}{2}E_1\tilde F_{11}^m\biggl({\bf\tilde u}(t_{n+\frac{1}{2}}),{\bm{{\tilde \phi}}}(t_{n+\frac{1}{2}})\biggr)={\bf\tilde u}(t_n)-\frac{\Delta t}{2}E_2\tilde F_{12}^m\biggl({\bf\tilde u}(t_n),{\bf\tilde v}(t_n)\biggr)+\frac{\Delta t}{2} U^TA_1{\bf\tilde v}(t_n),\\
&\hspace{-4mm}{\bf\tilde v}(t_{n+\frac{1}{2}})+\frac{\Delta t}{2}E_3\tilde F_{21}^m\biggl({\bf\tilde u}(t_{n+\frac{1}{2}}),{\bf\tilde v}(t_{n+\frac{1}{2}})\biggr)+\frac{\Delta t}{2}V^TA_2{ \bf\tilde u}(t_{n+\frac{1}{2}})={\bf\tilde v}(t_n)-\frac{\Delta t}{2}E_4\tilde F_{22}^m\biggl({\bf\tilde v}(t_n),{\bm{{\tilde \phi}}}(t_n)\biggr),\\
&\hspace{-4mm}{\bm{{\tilde \phi}}}(t_{n+\frac{1}{2}})+\frac{\Delta t}{2}E_5\tilde F_{31}^m\biggl({\bf\tilde u}(t_{n+\frac{1}{2}}),{\bm{{\tilde \phi}}}(t_{n+\frac{1}{2}})\biggr)={\bm{{\tilde \phi}}}(t_n)-\frac{\Delta t}{2}E_6\tilde F_{32}^m\biggl({\bf\tilde v}(t_n), {\bm{{\tilde \phi}}}(t_n)\biggr),~n=1,..,NT-1,
\end{split}
\end{equation}

while the second step is introduced below

\begin{equation}\label{eq11}
\begin{split}
&\hspace{-4mm}{\bf \tilde u}(t_{n+1})+\frac{\Delta t}{2}E_2\tilde F_{12}^m\biggl({\bf\tilde u}(t_{n+1}),{\bf\tilde v}(t_{n+1})\biggr)-\frac{\Delta t}{2} U^TA_1{\bf\tilde v}(t_{n+1})={\bf\tilde u}(t_{n+\frac{1}{2}})-\frac{\Delta t}{2}E_1\tilde F_{11}^m\biggl({\bf\tilde u}(t_{n+\frac{1}{2}}),{\bm{{\tilde \phi}}}(t_{n+\frac{1}{2}})\biggr),\\
&\hspace{-4mm}{\bf\tilde v}(t_{n+1})+\frac{\Delta t}{2}E_4\tilde F_{22}^m\biggl({\bf\tilde v}(t_{n+1}),{\bm{{\tilde \phi}}}(t_{n+1})\biggr)={\bf\tilde v}(t_{n+\frac{1}{2}})-\frac{\Delta t}{2}E_3\tilde F_{21}^m\biggl({\bf\tilde u}(t_{n+\frac{1}{2}}),{\bf\tilde v}(t_{n+\frac{1}{2}})\biggr)-\frac{\Delta t}{2}V^TA_2{ \bf\tilde u}(t_{n+\frac{1}{2}}),\\
&\hspace{-4mm}{\bm{{\tilde \phi}}}(t_{n+1})+\frac{\Delta t}{2}E_6\tilde F_{32}^m\biggl({\bf\tilde v}(t_{n+1}), {\bm{{\tilde \phi}}}(t_{n+1})\biggr)={\bm{{\tilde \phi}}}(t_{n+\frac{1}{2}})-\frac{\Delta t}{2}E_5\tilde F_{31}^m\biggl({\bf\tilde u}(t_{n+\frac{1}{2}}),{\bm{{\tilde \phi}}}(t_{n+\frac{1}{2}})\biggr),~n=1,..,NT-1.
\end{split}
\end{equation}

The initial conditions remain the same as in the case of POD reduced system. The nonlinear algebraic systems (\ref{eq10}) and (\ref{eq10}) as well as (\ref{eq4}) and (\ref{eq6}) obtained by employing POD/DEIM and POD methods on ADI FD SWE were first splitted into subsystems according to the left-hand side of the equations where  no more than two variables are coupled to each other and this was done in the same manner as in the case of Gustafsson ADI FD SWE nonlinear systems. The derived systems were solved using the Newton method.

\section{Numerical experiments}

In this section, we present two main experiments for the two - dimensional shallow water equations model to validate the fesability and efficiency of the POD/DEIM method in comparison with POD technique. For all tests we derived the initial conditions from the initial height condition No. 1 of Grammeltvedt \cite{Gram1969} i.e.

\begin{equation*}
\hspace{-10mm}h(x,y)=H_0+H_1+\tanh\biggl(9\frac{D/2-y}{2D}\biggr)+H_2\textrm{sech}^2\biggl(9\frac{D/2-y}{2D}\biggr)\sin\biggl(\frac{2\pi x }{L}\biggr),
\end{equation*}
$$0\leq x\leq L,~0\leq y\leq D.$$

The initial velocity fields were derived from the initial height field using the geostrophic relationship
    $$u = \biggl(\frac{-g}{f}\biggr)\frac{\partial h}{\partial y},~v = \biggl(\frac{g}{f}\biggr)\frac{\partial h}{\partial x}. $$

\begin{figure}[h]
\centering
\includegraphics[trim = 34mm 80mm 32mm 85mm, clip, width=7.4cm]{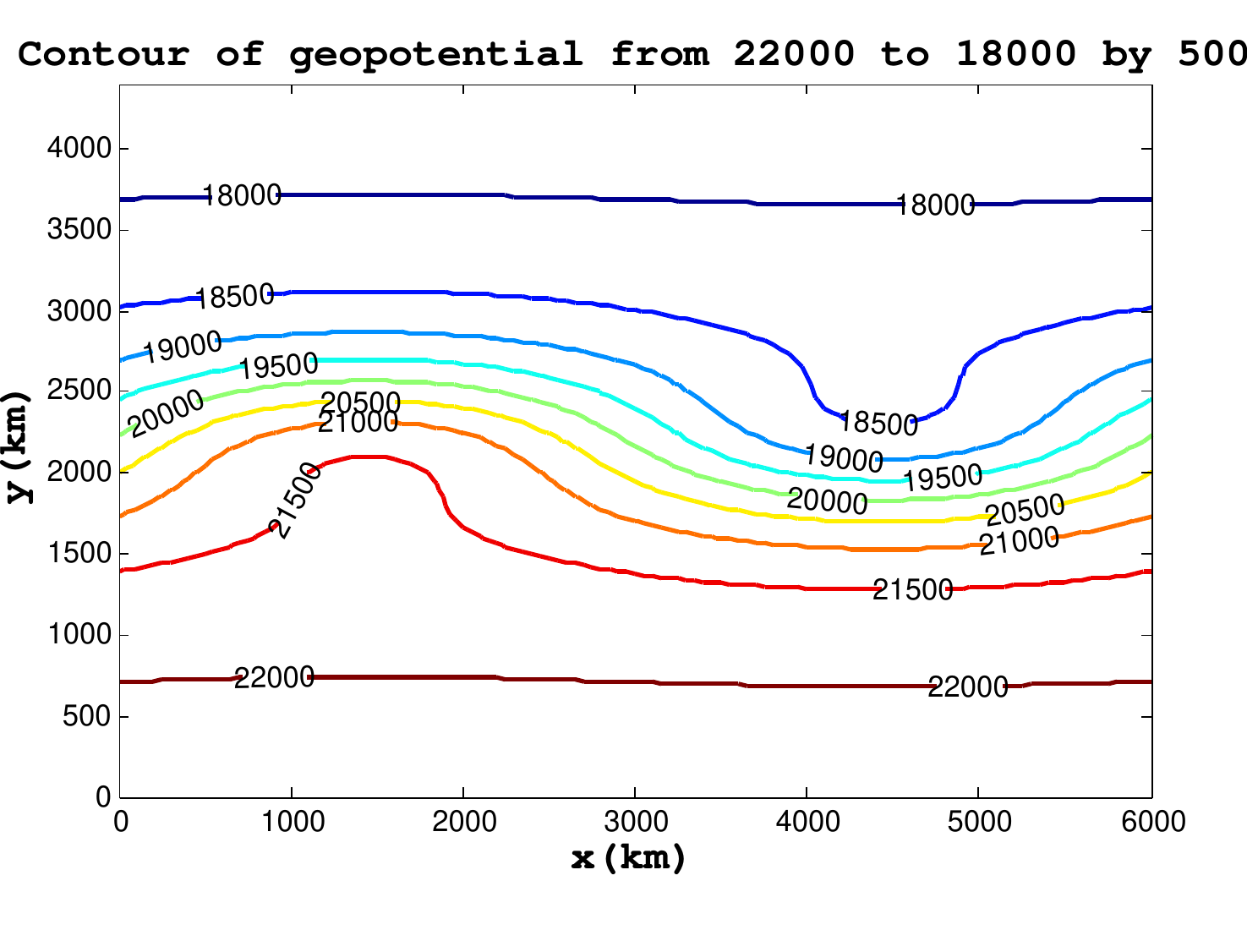}
\includegraphics[trim = 34mm 80mm 32mm 85mm, clip, width=7.4cm]{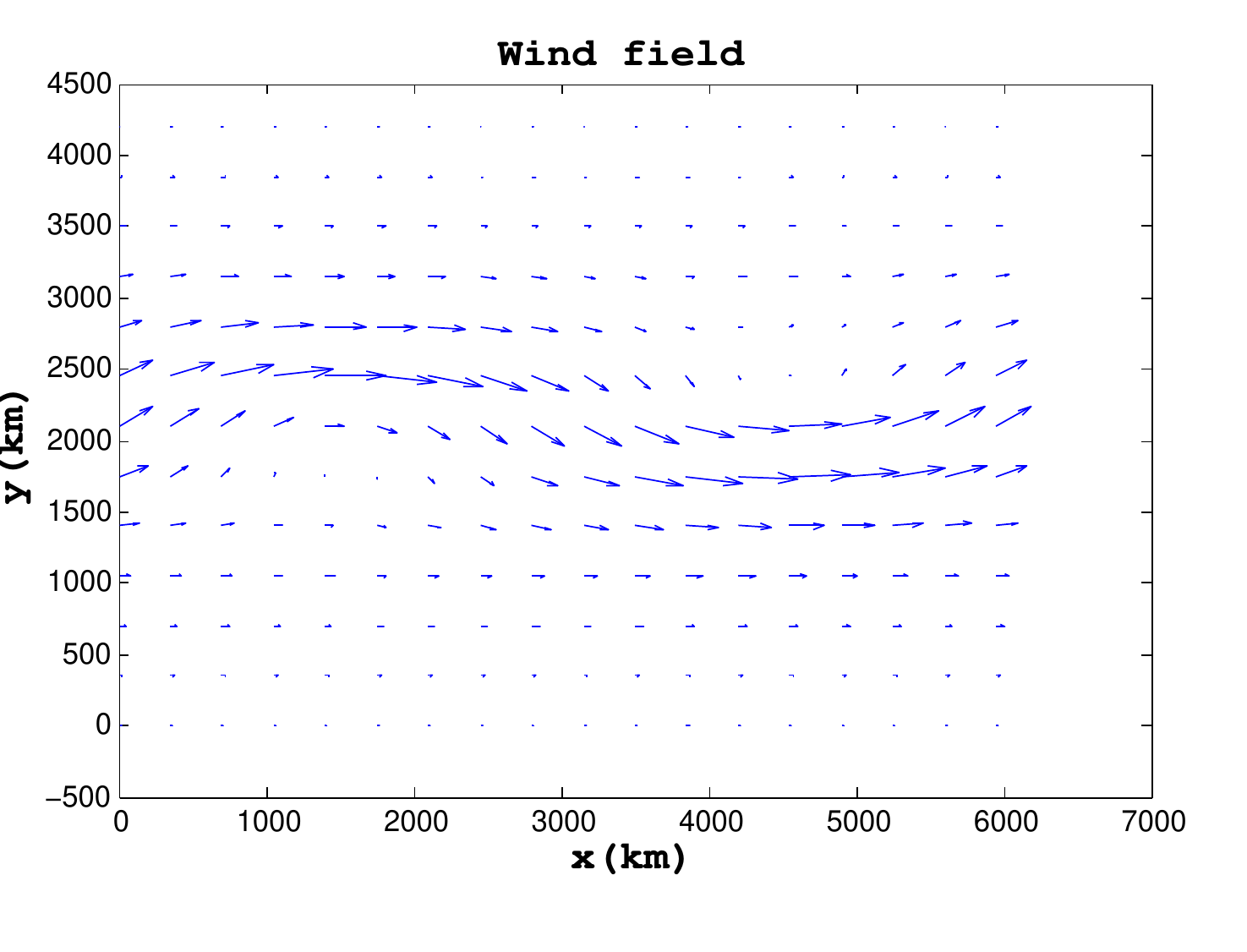}
\caption{\label{fig:DEIM5}Initial condition: Geopotential height field for the Grammeltvedt initial
condition (left). Wind field (the velocity unit is 1km/s) calculated from the geopotential field by
using the geostrophic approximation (right).}
\end{figure}

Figure \ref{fig:DEIM5} depicts the initial geopotential isolines and the geostrophic wind field.

 The constants used were $ L=6000km,~D=4400km,~\hat f=10^{-4}s^{-1}~,\beta=1.5\cdot10^{-11}s^{-1}m^{-1},~g=10 m s^{-2},~H_0=2000m,~H_1=220m,~H_2=133m.$

For the first test, the domain was discretized using a mesh of $301\times 221$ points, with $\Delta x = \Delta y = 20km$. Thus the dimension of the full-order discretized model is $66521$. The integration time window was $24 h$ and we used $91$ time steps ($NT = 91$) with $\Delta t = 960 s$.

ADI FD SWE scheme proposed by Gustafsson in \cite{Gus1971} was first employed in order to obtain the numerical solution of the SWE model. The implicit scheme allowed us to integrate in time using a larger time step deduced from the following Courant-Friedrichs-Levy (CFL) condition
 $$\sqrt{gh}(\frac{\Delta t}{\Delta x}) < 7.188.$$

\begin{figure}[h]
\centering
\includegraphics[trim = 15mm 60mm 12mm 65mm, clip, width=7.4cm]{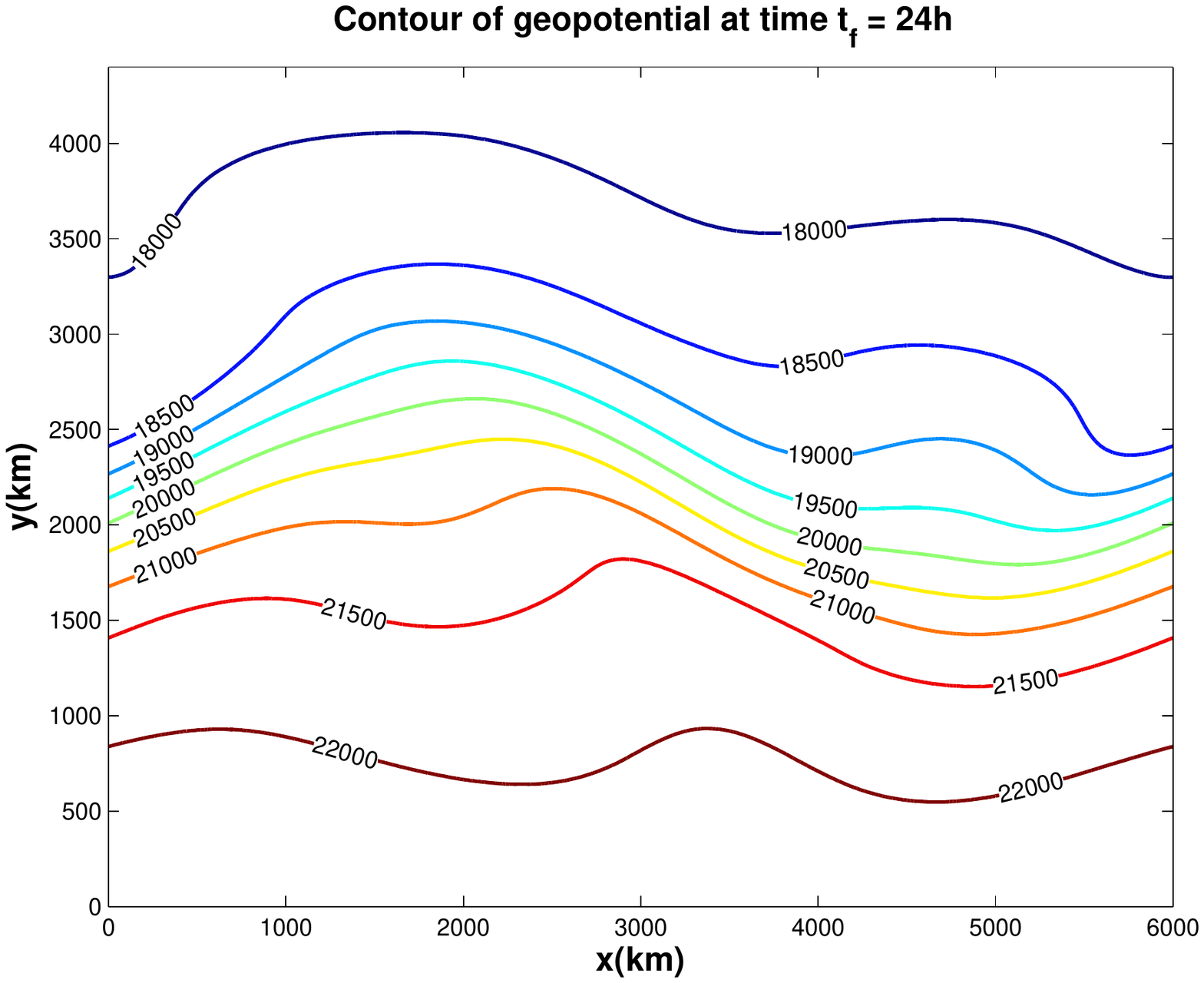}
\includegraphics[trim = 15mm 60mm 12mm 65mm, clip, width=7.4cm]{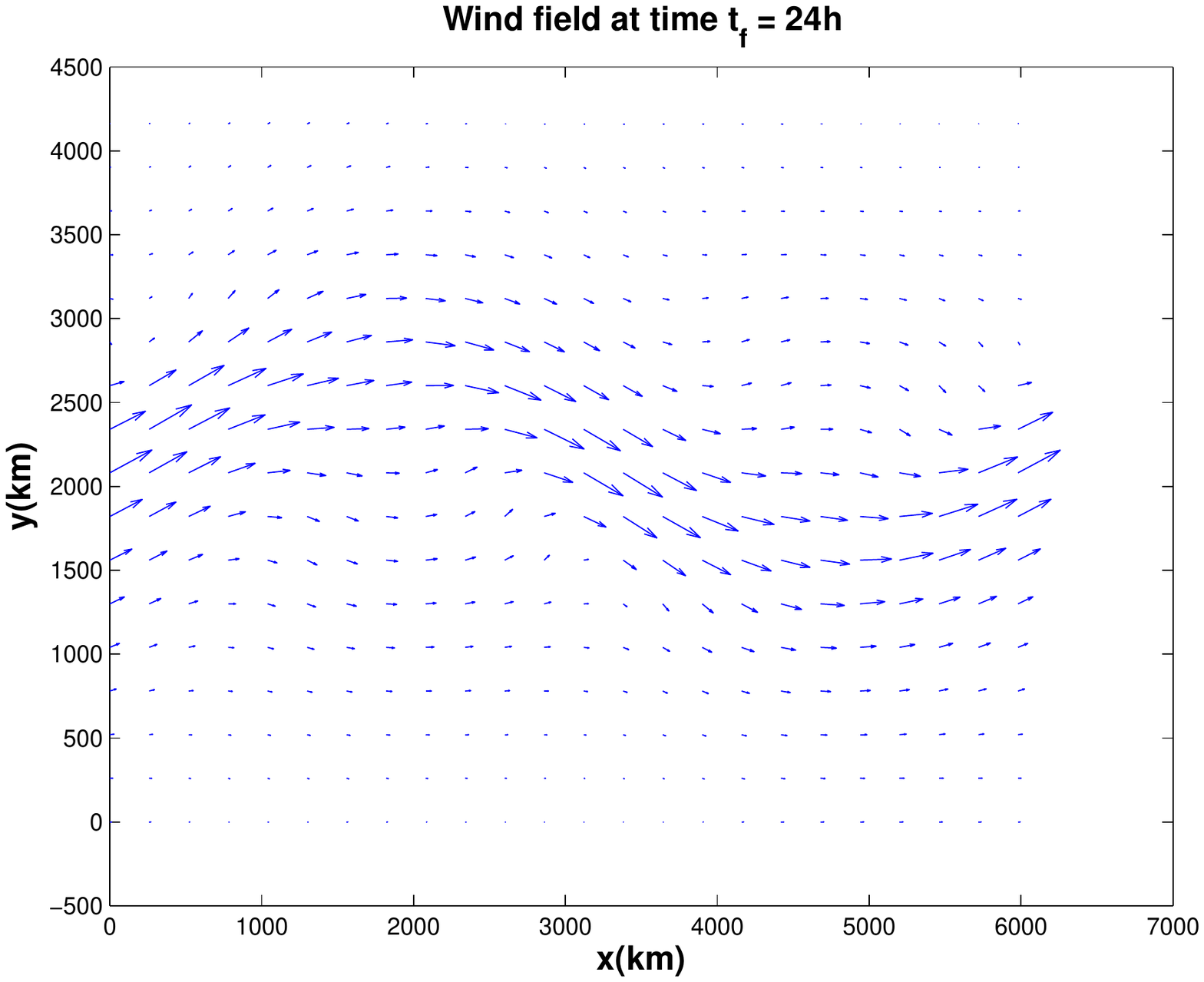}
\caption{\label{fig:DEIM6}The geopotential field (left) and the wind field (the velocity unit is 1km/s) at $t = t_f =24h$ obtained using the ADI FD SWE scheme for $\Delta t = 960 s$. }
\end{figure}

The nonlinear algebraic systems of ADI FD SWE scheme were solved with the Quasi-Newton method and the LU decomposition was performed only once every $6-th$ time step. The SWE solutions at $t=24h$ are illustrated in
Figure \ref{fig:DEIM6}.

\begin{figure}[H]
\centering
\includegraphics[trim = 15mm 60mm 12mm 65mm, clip, width=7.4cm]{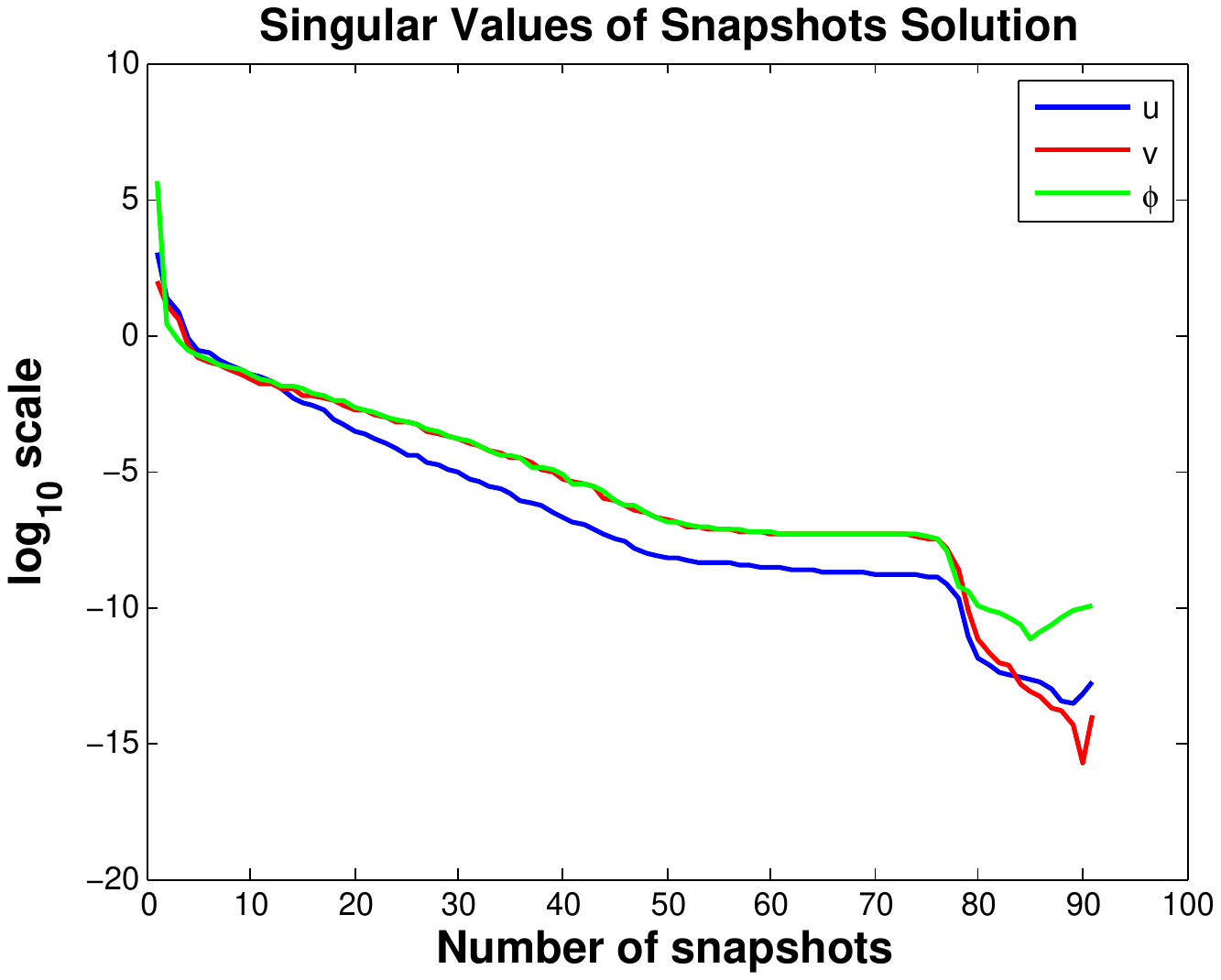}
\includegraphics[trim = 15mm 60mm 12mm 65mm, clip, width=7.4cm]{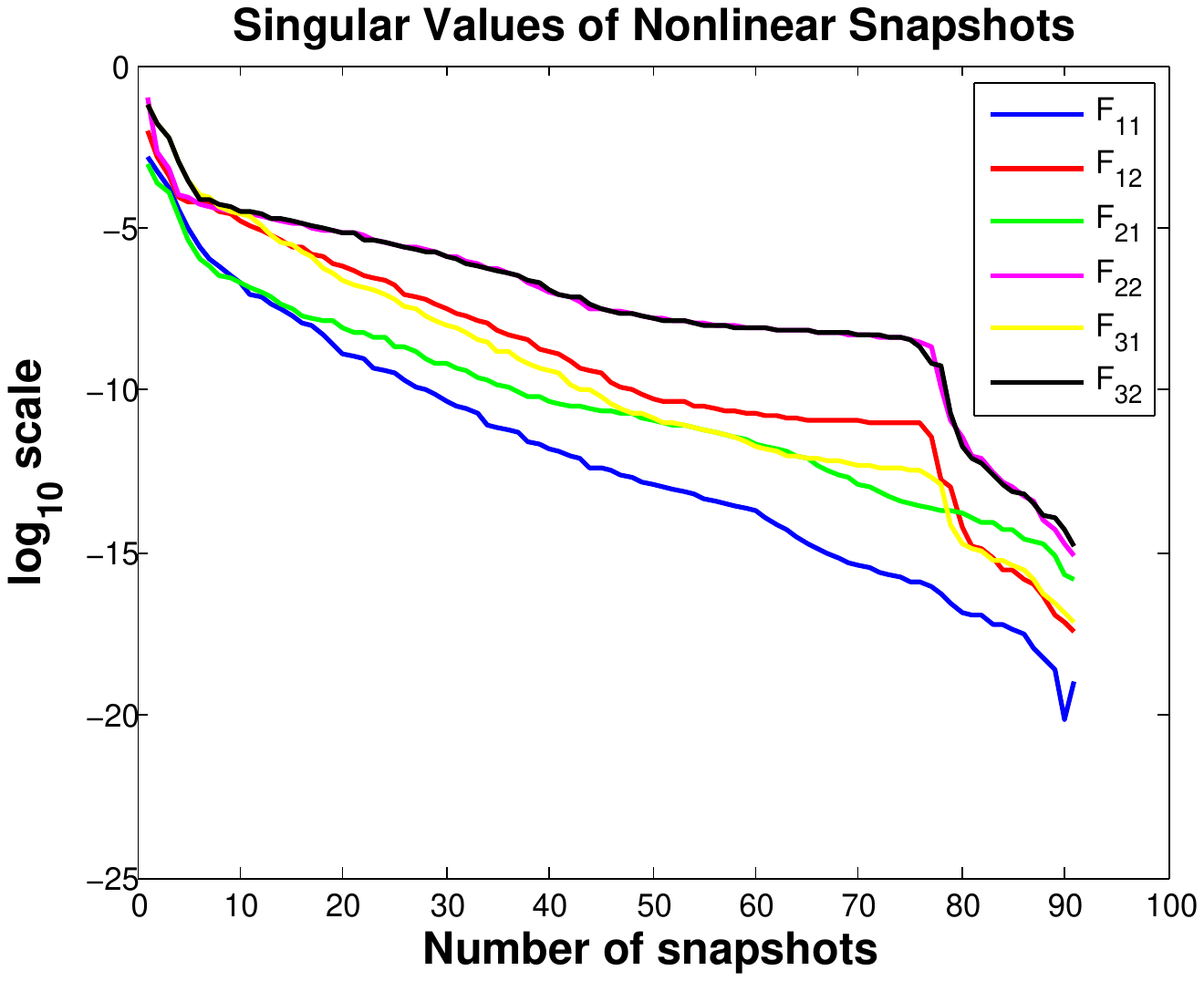}
\caption{\label{fig:DEIM7}The decay around the singular values of the snapshots solutions for $u,v,~\phi$ and nonlinear functions for $\Delta t = 960 s$.}
\end{figure}

The POD basis functions were constructed using $91$ snapshots obtained from the numerical solution of the full - order ADI FD SWE model at equally spaced time steps in the interval $[0,24h]$. Figure \ref{fig:DEIM7} shows the decay around the eigenvalues of the snapshot solutions for $u,~v,~\phi$ and the nonlinear snapshots $F_{11},~F_{12},$ $F_{21},~F_{22}$, $F_{31},~F_{32}$.

\begin{figure}[h]
\centering
\includegraphics[trim = 15mm 60mm 12mm 65mm, clip, width=7.4cm]{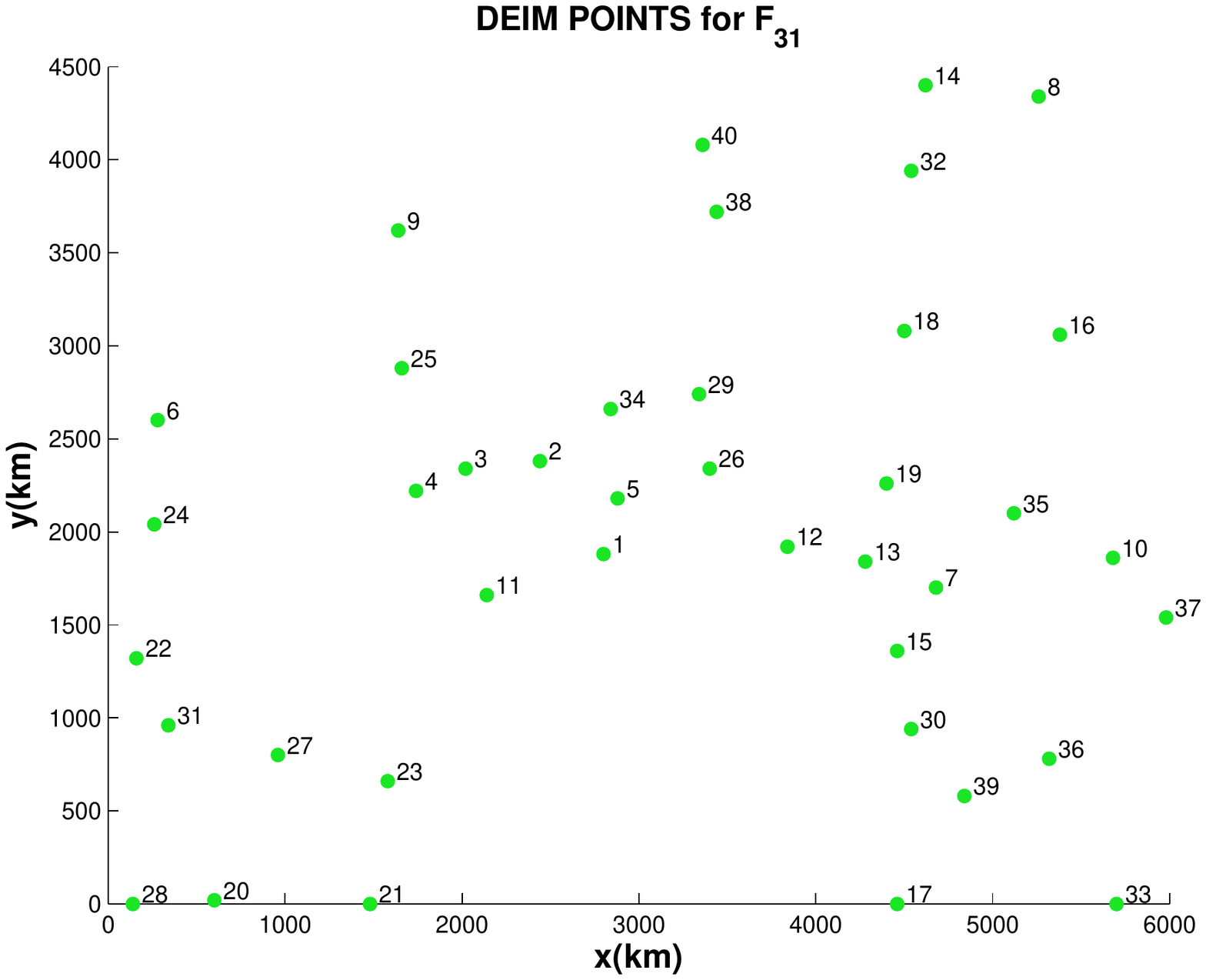}
\includegraphics[trim = 15mm 60mm 12mm 65mm, clip, width=7.4cm]{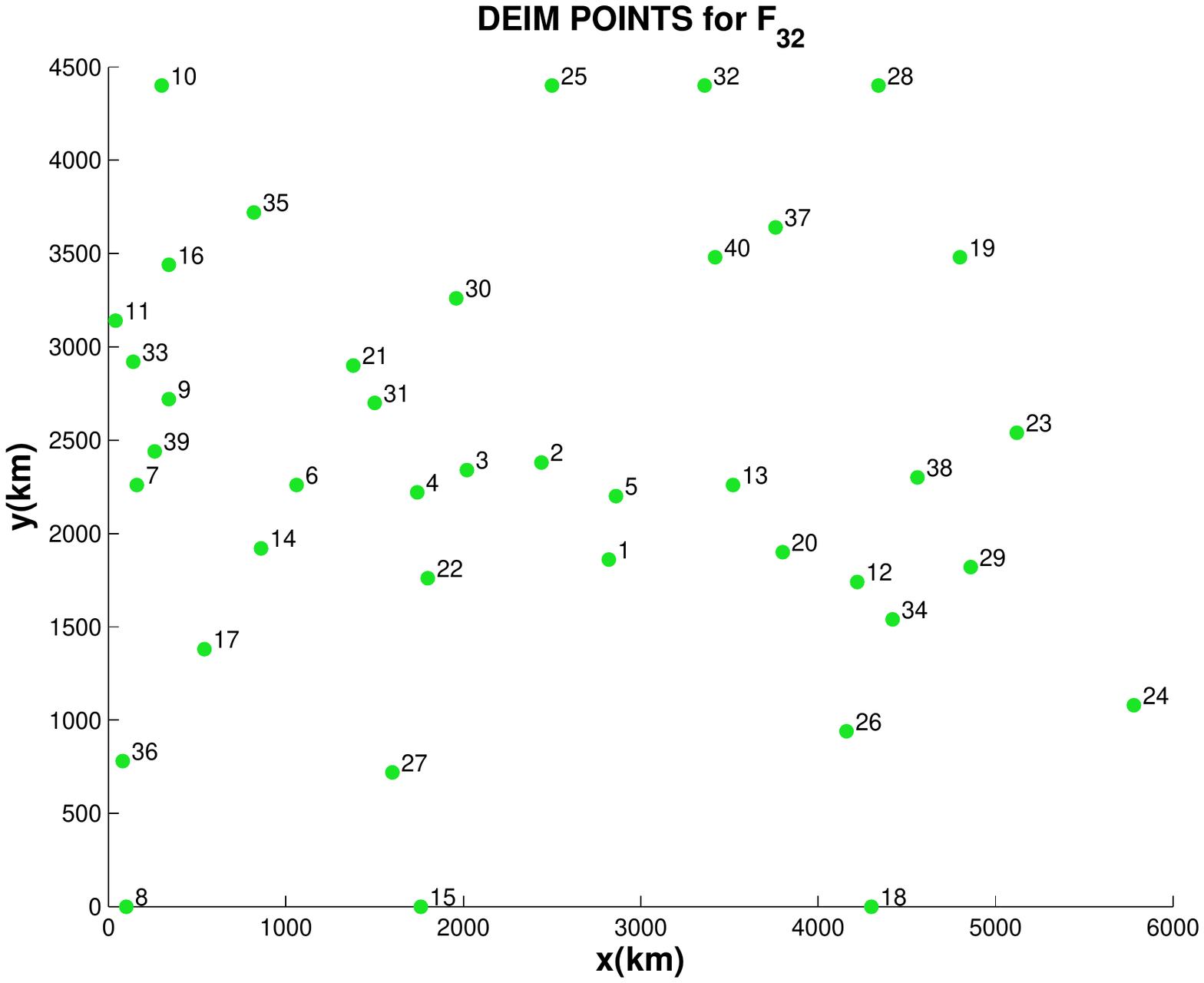}
\caption{\label{fig:DEIM8}First $40$ points selected by DEIM for the nonlinear functions $F_{31}$ (left) and $F_{32}$ (right)}
\end{figure}

The dimension of the POD bases for each variable was taken to be $35$, capturing more than $99.9\%$ of the system energy. We applied the DEIM algorithm for interpolation indices to improve the efficiency of the POD approximation and to achieve a complexity reduction of the nonlinear terms with a complexity proportional to the number of reduced variables. Figure \ref{fig:DEIM8} illustrates the distribution of the first $40$ spatial points selected from the DEIM algorithm using the POD bases of $F_{31}$ and $F_{32}$ as inputs.

\begin{figure}[h]
\centering
\includegraphics[trim = 15mm 60mm 12mm 65mm, clip, width=5.1cm]{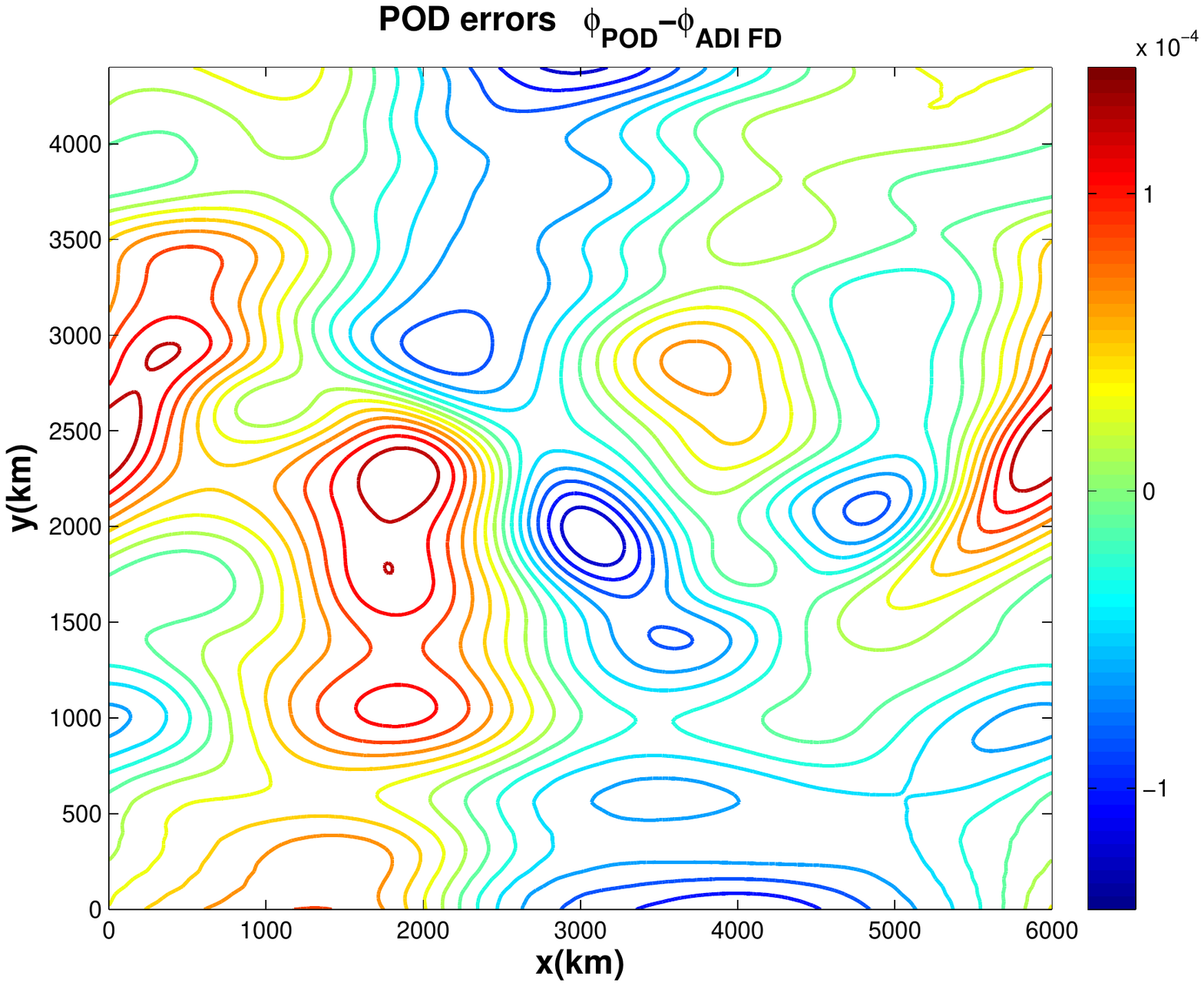}
\includegraphics[trim = 15mm 60mm 12mm 65mm, clip, width=5.1cm]{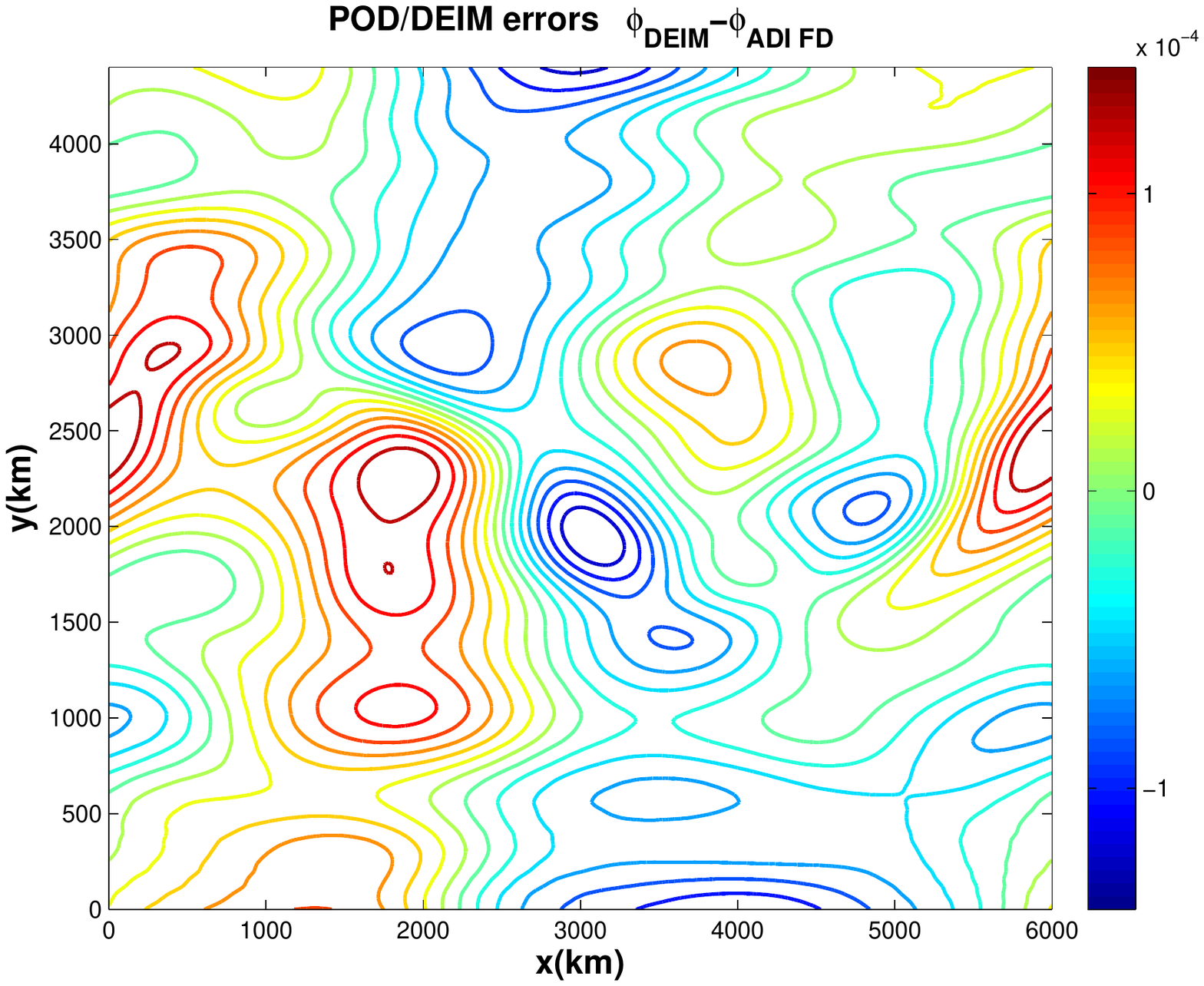}
\includegraphics[trim = 15mm 60mm 12mm 65mm, clip, width=5.1cm]{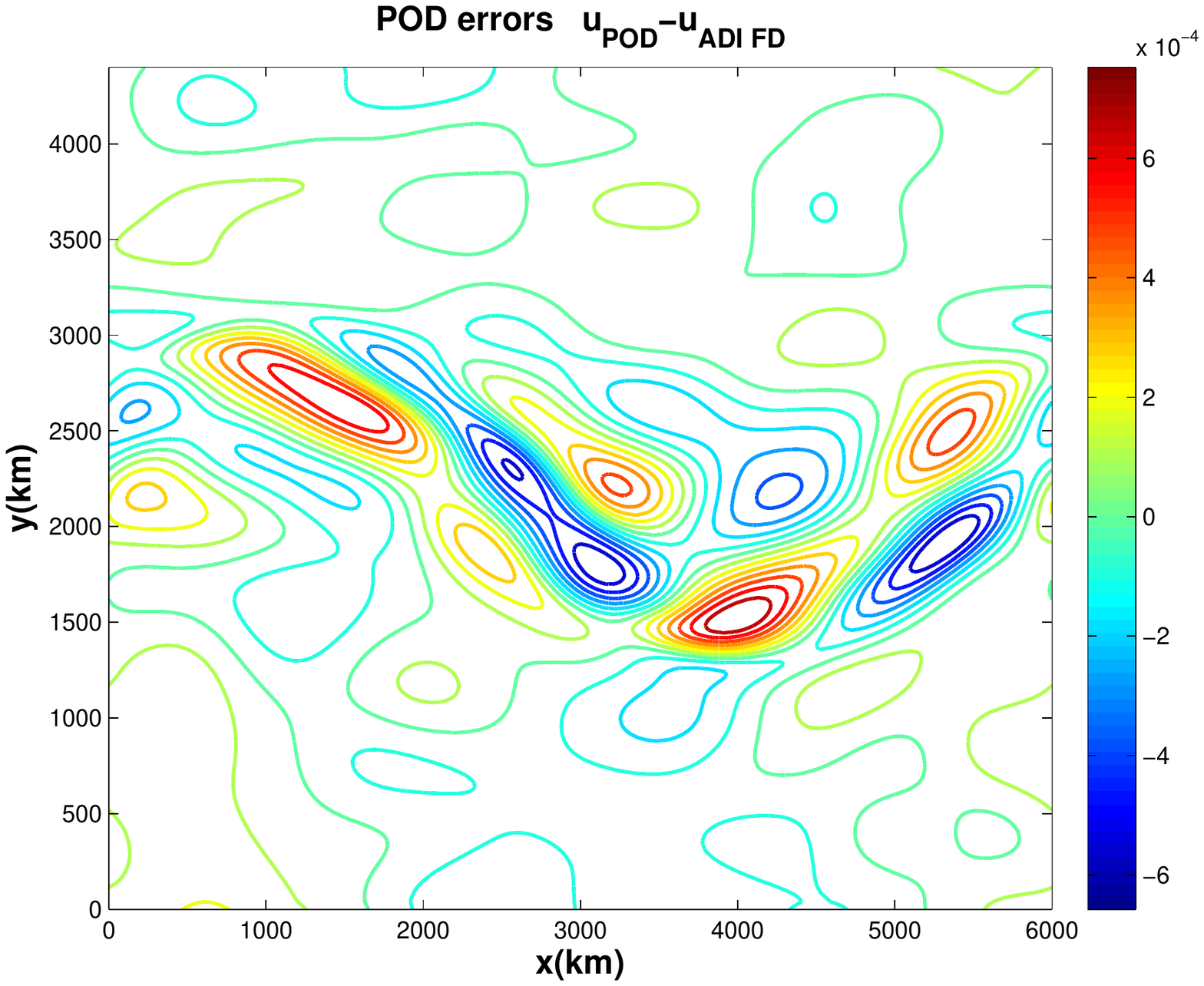}
\includegraphics[trim = 15mm 60mm 12mm 65mm, clip, width=5.1cm]{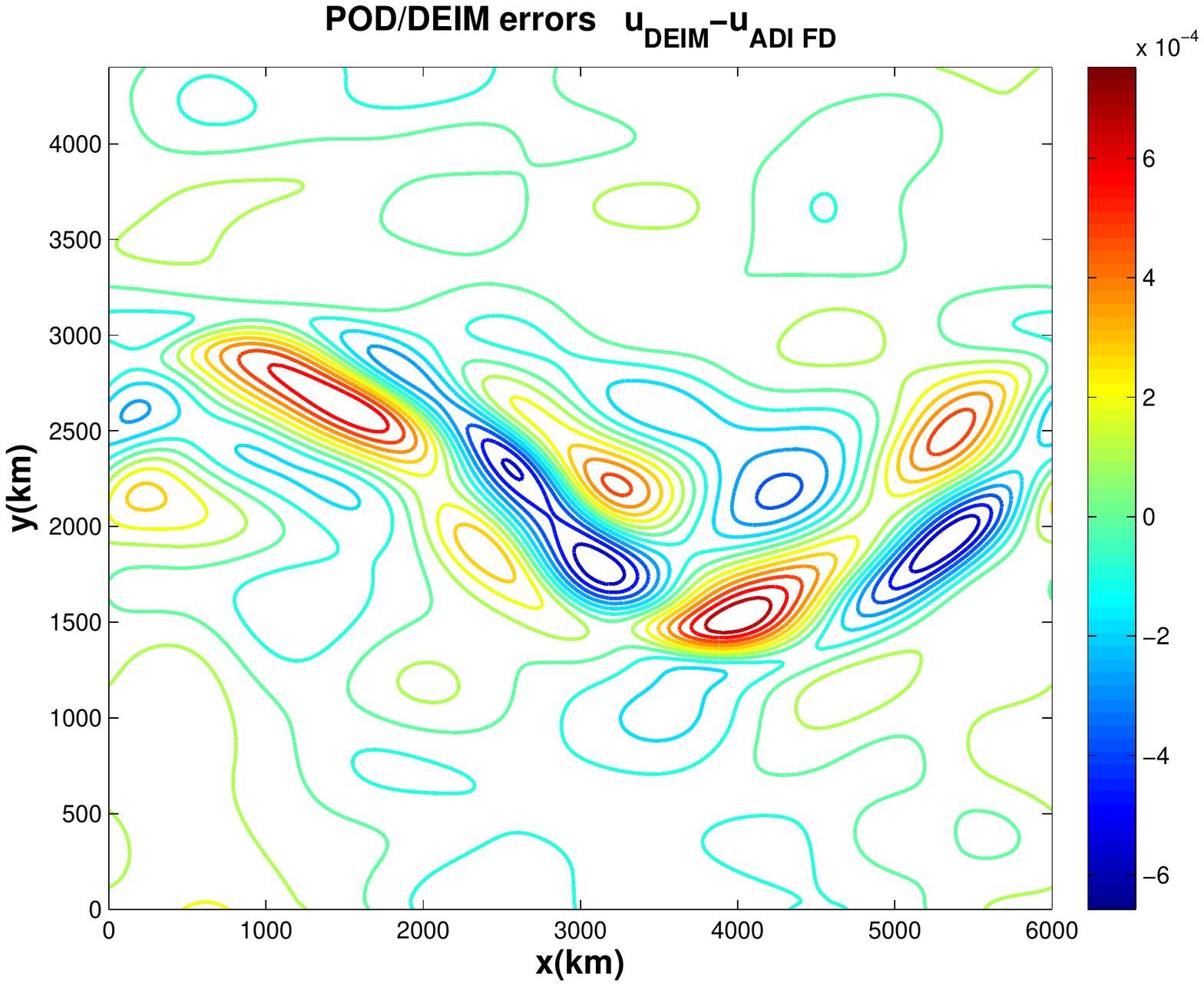}
\includegraphics[trim = 15mm 60mm 12mm 65mm, clip, width=5.1cm]{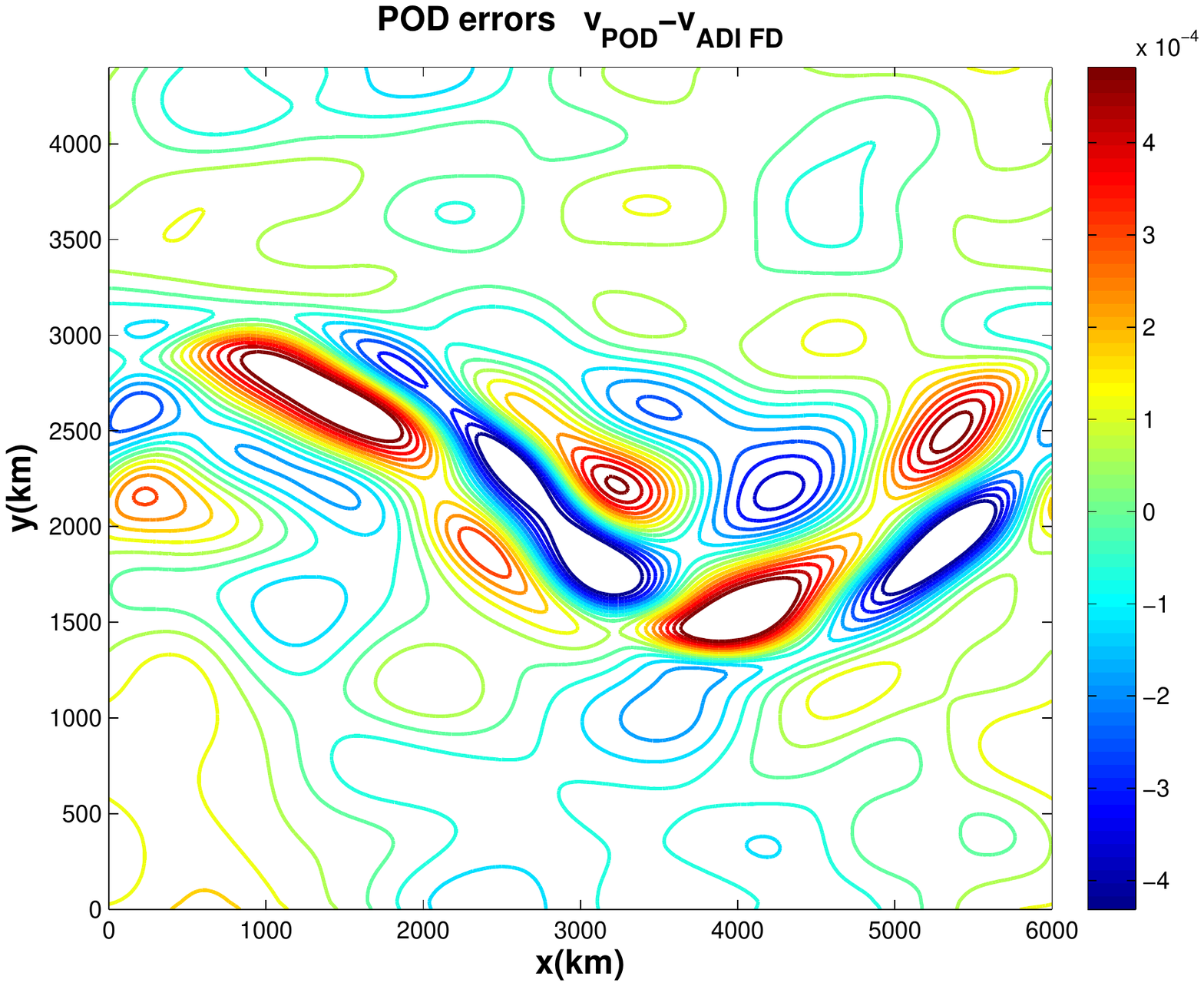}
\includegraphics[trim = 15mm 60mm 12mm 65mm, clip, width=5.1cm]{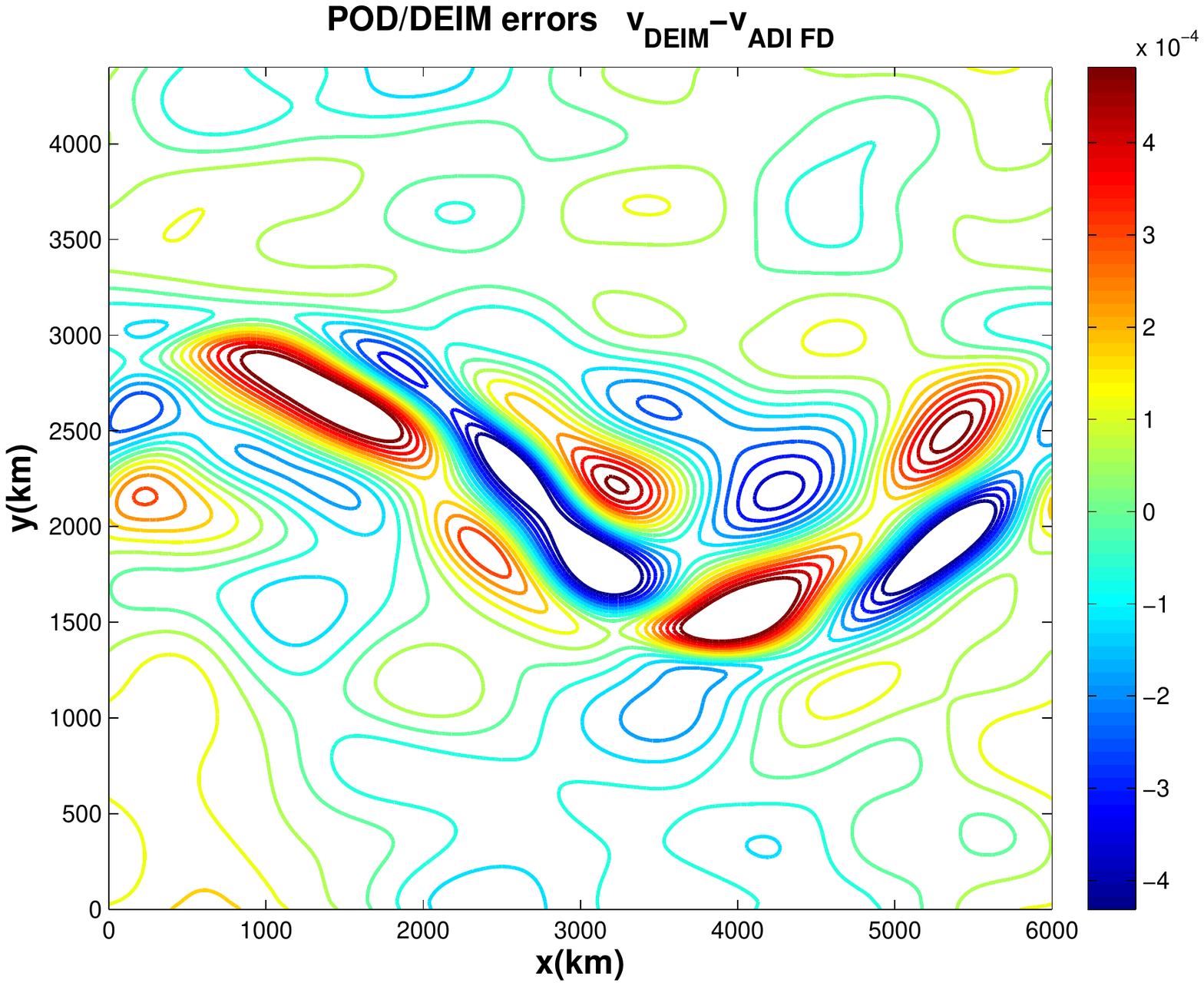}
\caption{\label{fig:DEIM9}Local errors between  POD, POD/DEIM ADI SWE solutions and the ADI FD SWE solutions at $t = 24h$ ($\Delta t = 960 s$). The number of DEIM points was taken $90$.}

\end{figure}

We emphasize the performances of POD/DEIM method in comparison with the POD approach using the numerical solution of the ADI FD SWE model. Figure \ref{fig:DEIM9} depicts the grid point local error behaviors between POD, POD/DEIM ADI SWE solutions and ADI FD SWE solutions, where we used $90$ DEIM points.

%

Using the following norms
$$ \frac{1}{NT}\sum_{i=1}^{t_f}\frac{||w^{ADI~FD}(:,i)-w^{POD~ADI}(:,i)||_2}{||w^{ADI~FD}(:,i)||_2},~~
\frac{1}{NT}\sum_{i=1}^{t_f}\frac{||w^{ADI~FD}(:,i)-w^{POD/DEIM~ADI}(:,i)||_2}{||w^{ADI~FD}(:,i)||_2},$$
$i=1,2,..,t_f$ we calculated the average relative errors in Euclidian norm for all three variables of SWE model $w=u,v,\phi $. The results are presented in Table \ref{tab:table1}.
\begin{table}[H]
\centerline{
\scalebox{0.7}{
\begin{tabular}{|c|c|c|}\hline \hline
 &  POD ADI SWE & POD/DEIM ADI SWE\\ \hline
$E_{\phi}$ & 7.127e-005 & 1.106e-004\\ \hline
$E_u$ & 4.905e-003 & 6.189e-003\\ \hline
$E_v$ & 6.356e-003 & 9.183e-003\\ \hline
\end{tabular}}}
\caption{\label{tab:table1}Average relative errors for each of the model variables. The POD bases dimensions were taken $35$ capturing more than $99.9\%$ of the system energy. $90$ DEIM points were chosen.}
\end{table}

In addition to the ADI FD SWE scheme we propose an Euler explicit FD SWE scheme as the starting point for a POD, POD/DEIM reduced model. The POD bases were constructed using the same $91$ snapshots as in the POD ADI SWE case, only this time the Galerkin projection was applied to the Euler FD SWE model. The DEIM algorithm was used again and the numerical results are provided in Table \ref{tab:table2}. This time we employed the root mean square error calculation in order to compare the POD and POD/DEIM techniques at time $t=24h$.

\begin{table}[h]
\centerline{
\scalebox{0.7}{
\begin{tabular}{|c|c|c|c|c|c|}\hline \hline
 & ADI SWE & POD ADI SWE & POD/DEIM ADI SWE & POD EE SWE & POD/DEIM EE SWE\\ \hline
CPU time seconds & 73.081 & 43.021 & 0.582& 43.921& 0.639\\ \hline
$RMSE_{\phi}$ & - & 5.416e-005 & 9.668e-005& 1.545e-004& 1.792e-004\\ \hline
$RMSE_u$ & - & 1.650e-004 & 2.579e-004& 1.918e-004& 3.126e-004\\ \hline
$RMSE_v$ & - & 8.795e-005 & 1.604e-004& 1.667e-004& 2.237e-004\\ \hline\hline
\end{tabular}}}
\caption{\label{tab:table2}CPU time gains and the root mean square errors for each of the model variables at $t=t_f$. The POD bases dimensions were taken as $35$ capturing more than $99.9\%$ of the system energy. $90$
DEIM points were chosen.}
\end{table}

Applying DEIM method to POD ADI SWE model we reduced the computational time by a factor of $73.91$. In the case of the explicit scheme the DEIM algorithm decreased the CPU time by a factor of $68.733$. The POD/DEIM EE SWE model was solved using the Runge-Kutta-Fehlberg method (RKF45). Due to the large number of spatial discretization points, small number of time steps and only one Newton iteration threshold imposed when solving the nonlinear algebraic systems of POD and POD/DEIM ADI SWE schemes made these two implicit schemes faster than than the POD and POD/DEIM EE SWE explicit schemes. The numerical results obtained showed also that the implicit schemes are slightly more accurate than the explicit ones.

\begin{figure}[h]
\centering
\includegraphics[trim = 15mm 60mm 12mm 65mm, clip, width=10.4cm]{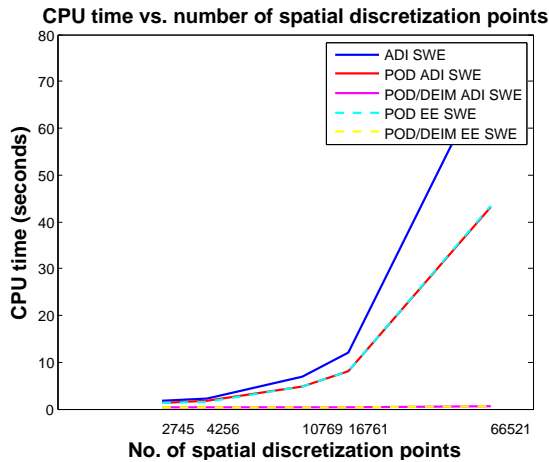}
\caption{\label{fig:DEIM10}Cpu time vs. Spatial discretization points; POD DIM = 35, No. DEIM points = 90.}
\end{figure}

Figure \ref{fig:DEIM10} illustrates the efficiency of POD/DEIM methods as a function of spatial discretization points. Once the number of spatial discrete points is larger than $10000$ the POD/DEIM schemes are faster than the POD schemes by a factor of $10$, for $90$ points selected by DEIM algorithm.

Next we carried out a second experiment to test the performances of POD/DEIM methods. We increased the number of time steps as well as the number of snapshots used to generate the POD bases. Thus, we took $NT=181$ and the number of snapshots $n_s=181$. Due to the rather demanding memory requirement we had to decrease the number of spatial discretization points. As a  consequence we choose $N_x=151$ and $N_y=111,$ with $\Delta x = \Delta y =40km$ and $\Delta t = 480s.$

We solved again the SWE model using the ADI FD SWE scheme in order to generate the $181$ snapshots required for POD and POD/DEIM reduced systems. This time the Courant-Friedrichs-Levy (CFL) condition was
 $$\sqrt{gh}(\frac{\Delta t}{\Delta x}) < 1.797.$$

\begin{figure}[h]
\centering
\includegraphics[trim = 15mm 60mm 12mm 65mm, clip, width=7.4cm]{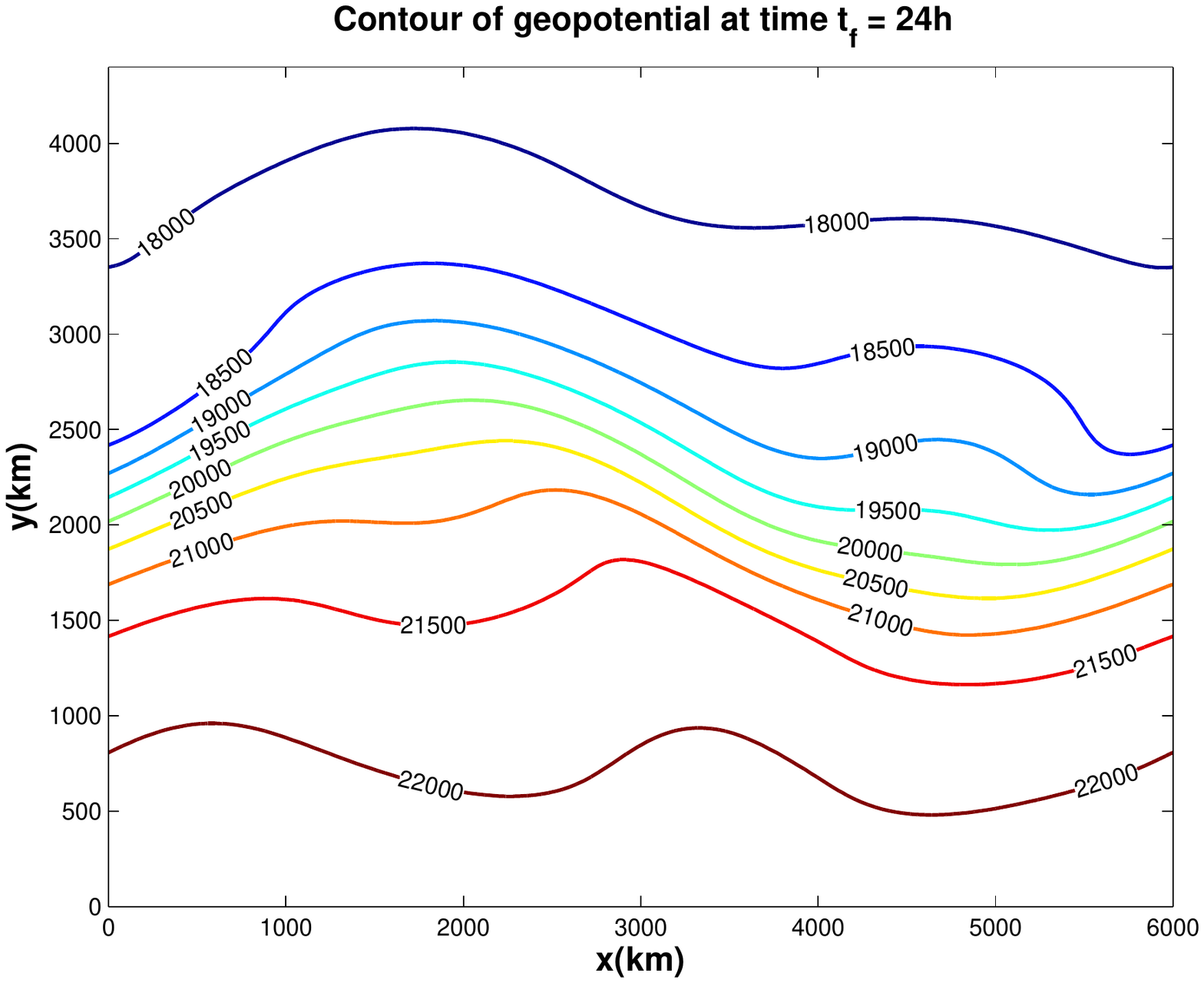}
\includegraphics[trim = 15mm 60mm 12mm 65mm, clip, width=7.4cm]{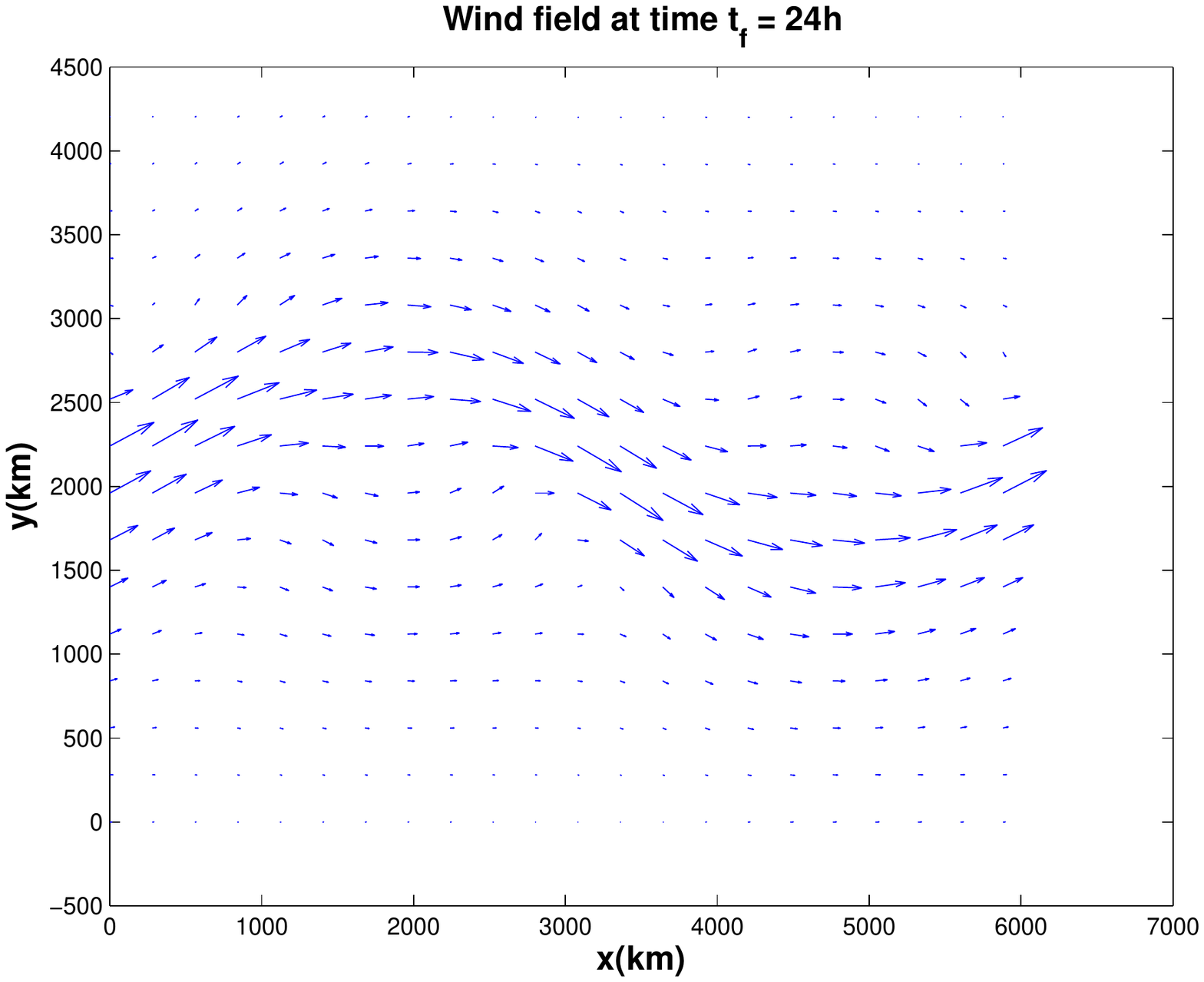}
\caption{\label{fig:DEIM11}The geopotential field (left) and the wind field (the velocity unit is 1km/s) at $t = t_f =24h$ obtained using the ADI FD SWE scheme for $\Delta t = 480 s$. }
\end{figure}

 The results obtained are similar with the ones obtained for a CFL condition $\sqrt{gh}(\frac{\Delta t}{\Delta x}) < 7.188$ underlying the performance of fully implicit Gustafsson scheme. The geopotential and wind field at final time $t_f=24h$ are depicted in Figure \ref{fig:DEIM11}.

\begin{figure}[h]
\centering
\includegraphics[trim = 15mm 60mm 12mm 65mm, clip, width=8.7cm]{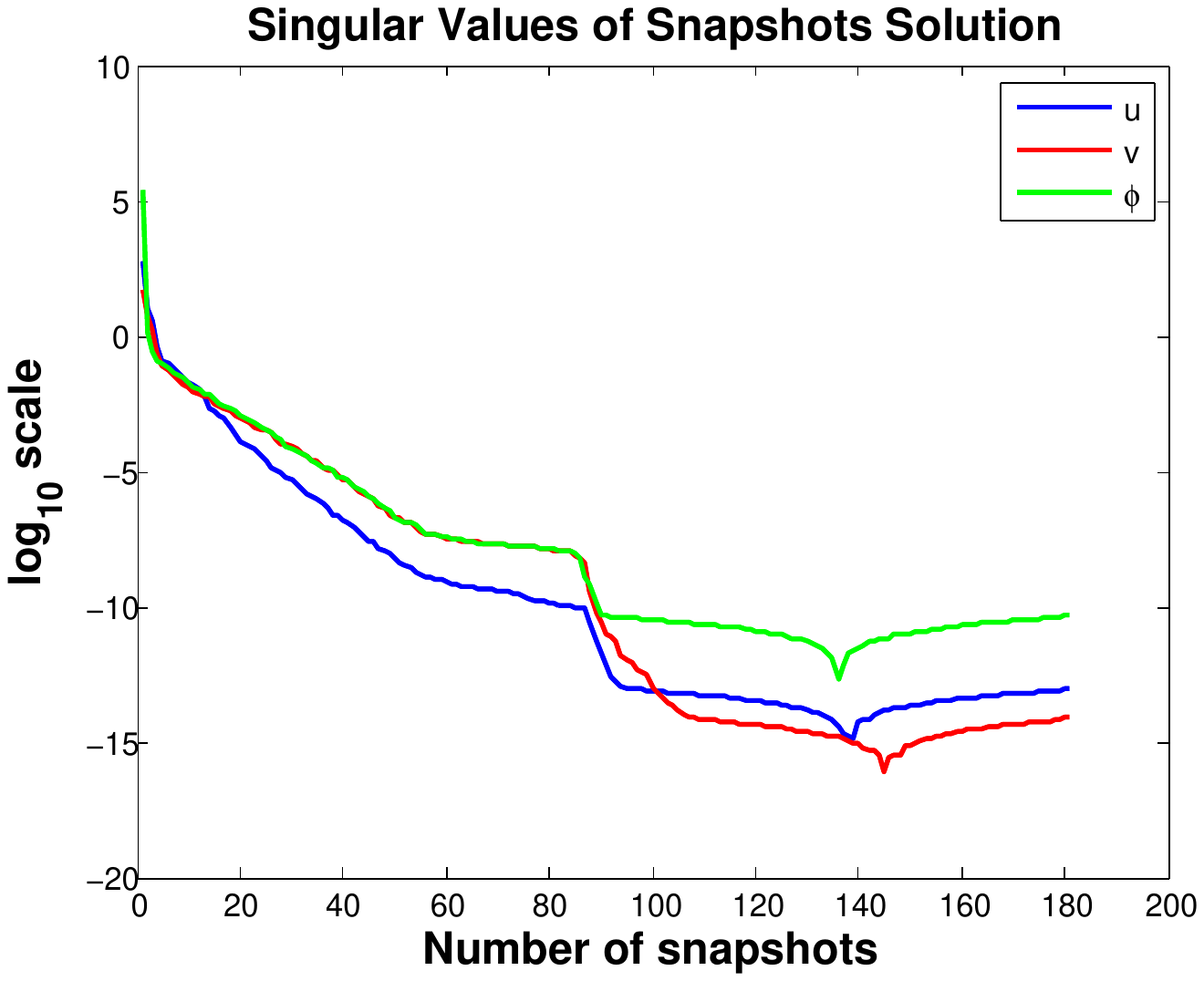}
\includegraphics[trim = 15mm 30mm 12mm 65mm, clip, width=6.2cm]{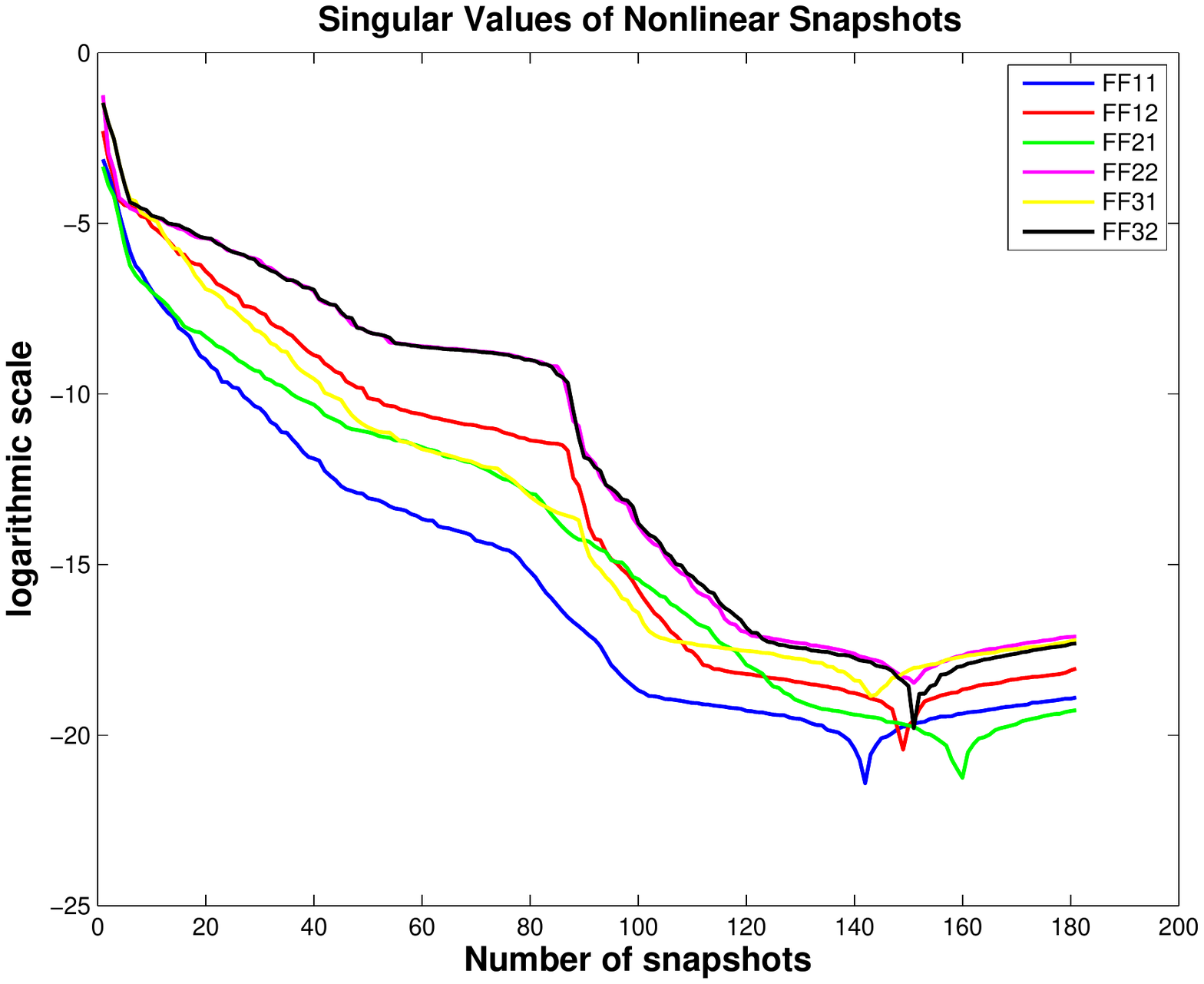}
\caption{\label{fig:DEIM12}Singular values of the snapshots solutions for $u,v,~\phi$ and nonlinear functions for $\Delta t = 480 s$.}
\end{figure}

Figure \ref{fig:DEIM12} shows the decay of the singular values of the snapshot solutions for $u,~v,~\phi$ and the nonlinear snapshots $F_{11},~F_{12},~F_{21},$ $F_{22},~F_{31},~F_{32}$. We noticed that the singular values of $F_{31}$ and $F_{32}$ are decreasing slowly when compared with the singular values of the other nonlinear functions.

\begin{figure}[h]
\centering
\includegraphics[trim = 15mm 60mm 12mm 65mm, clip, width=7.4cm]{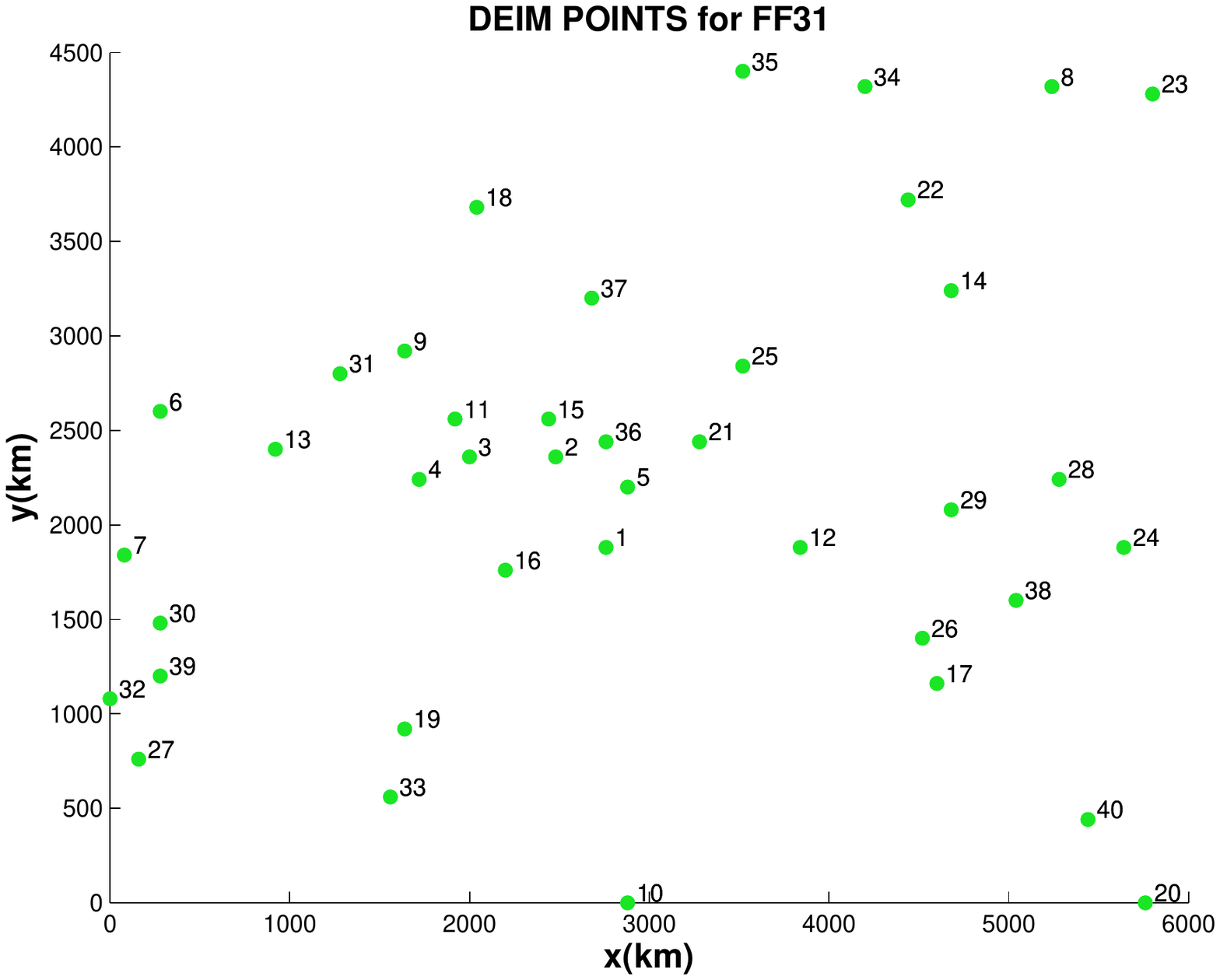}
\includegraphics[trim = 15mm 60mm 12mm 65mm, clip, width=7.4cm]{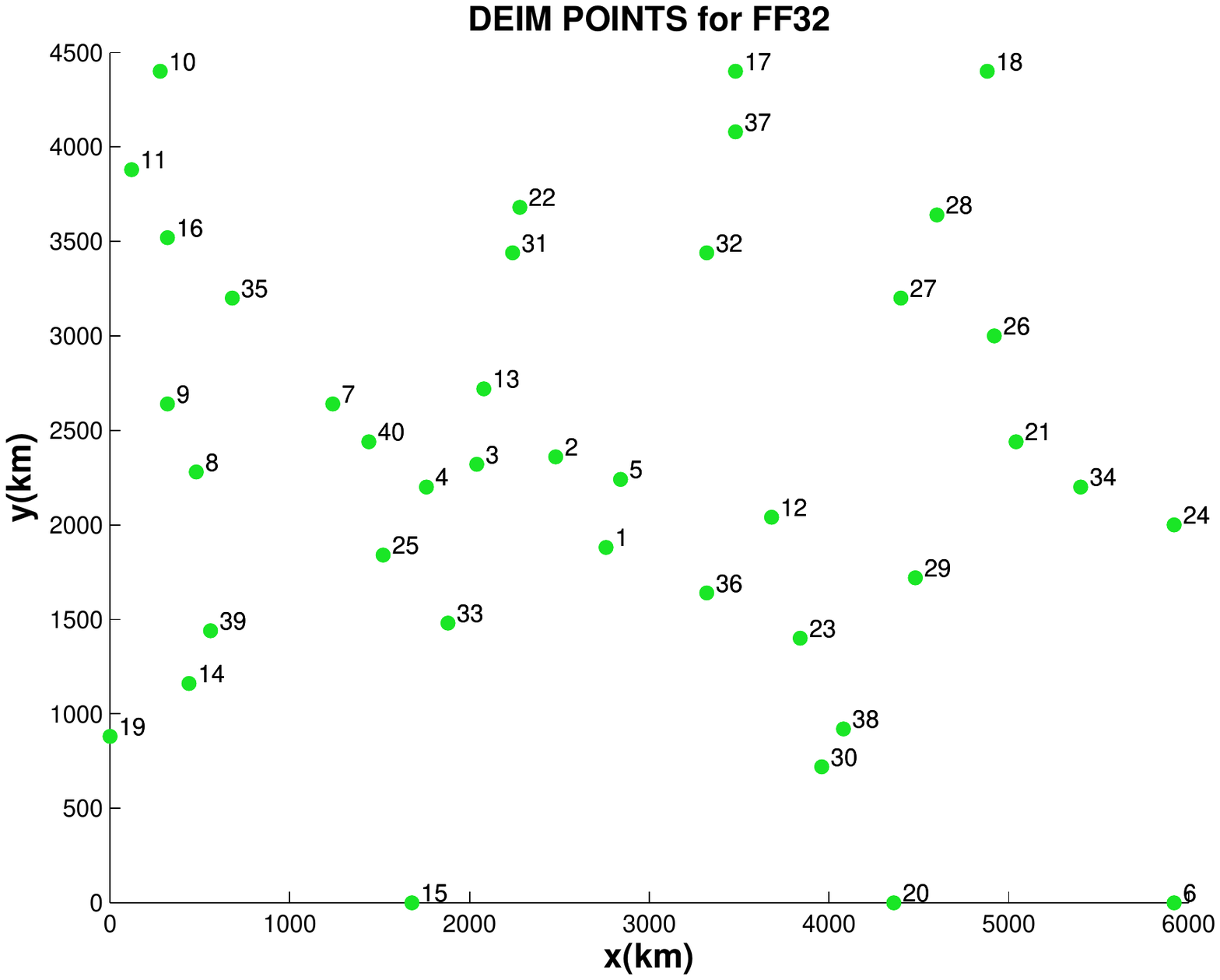}
\caption{\label{fig:DEIM13}First $40$ points selected by DEIM for the nonlinear functions FF31 (left) and FF32 (right)}
\end{figure}

The dimension of the POD bases for each variable was taken $35$. Next we apply the DEIM algorithm using as input the POD bases corresponding to the nonlinear functions. The first $40$ points selected by the discrete empirical interpolation method for $F_{31}$ and $F_{32}$ are illustrated in Figure \ref{fig:DEIM13}.

\begin{figure}[h]
\centering
\includegraphics[trim = 15mm 60mm 12mm 65mm, clip, width=5.1cm]{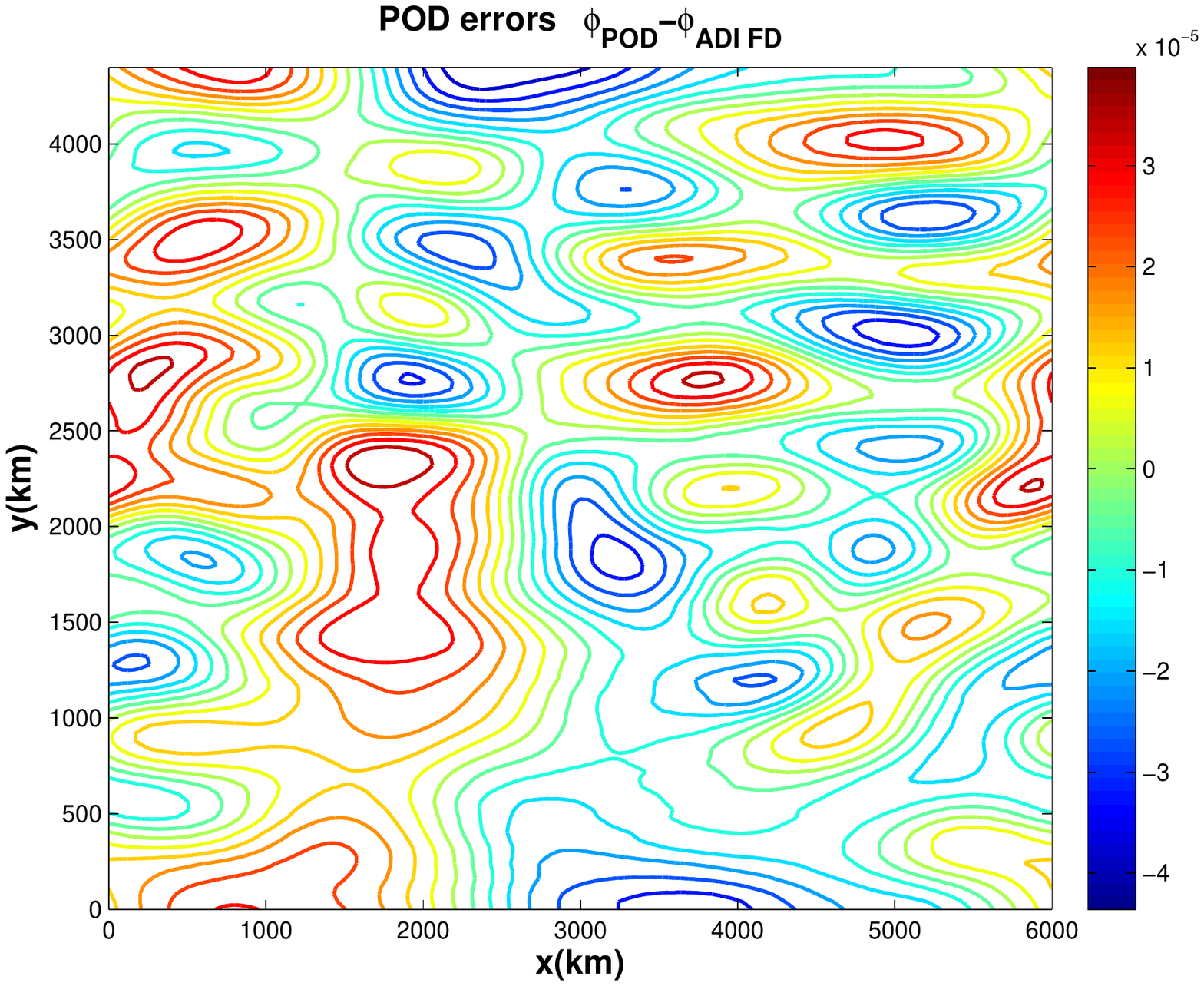}
\includegraphics[trim = 15mm 60mm 12mm 65mm, clip, width=5.1cm]{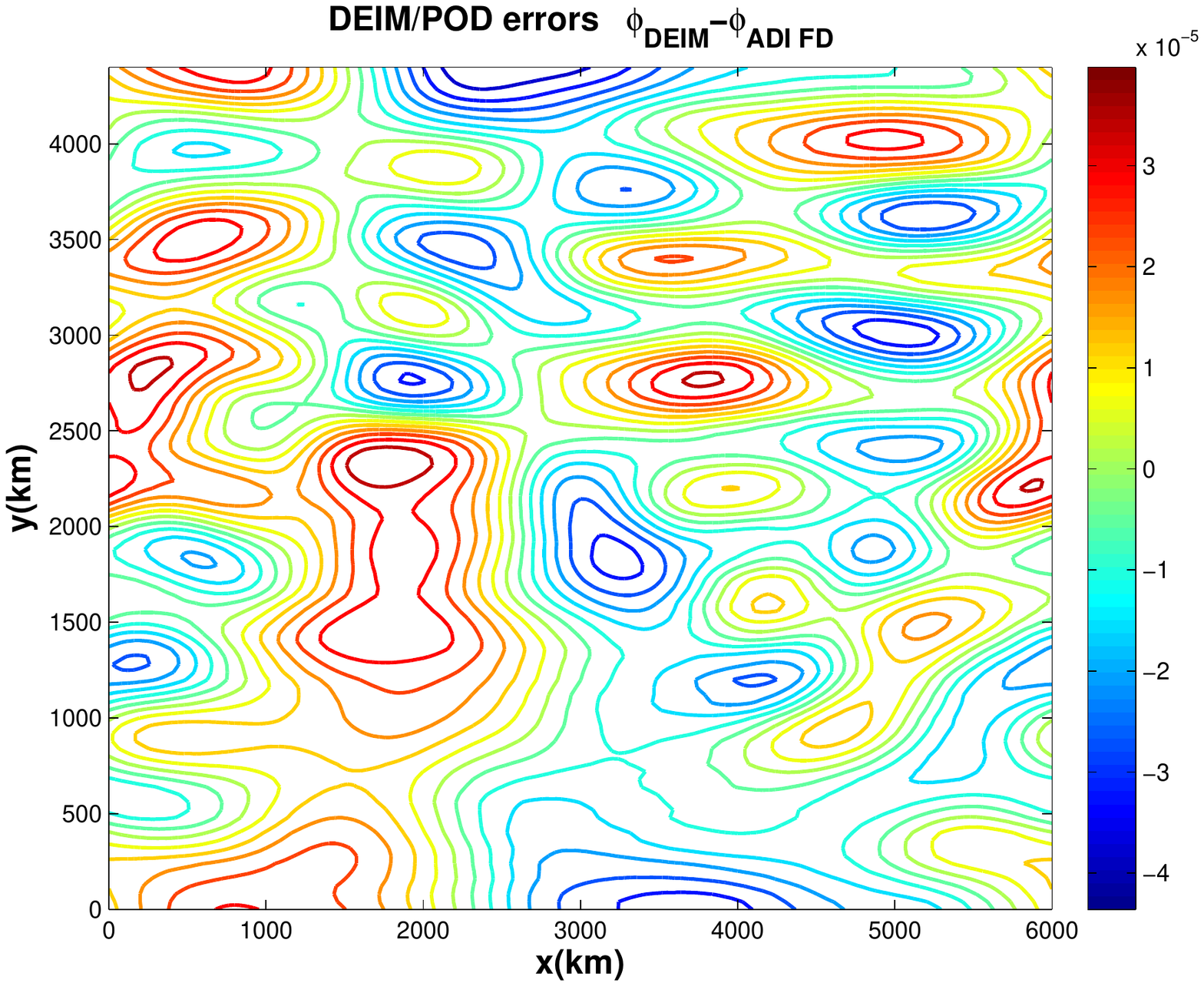}
\includegraphics[trim = 15mm 60mm 12mm 65mm, clip, width=5.1cm]{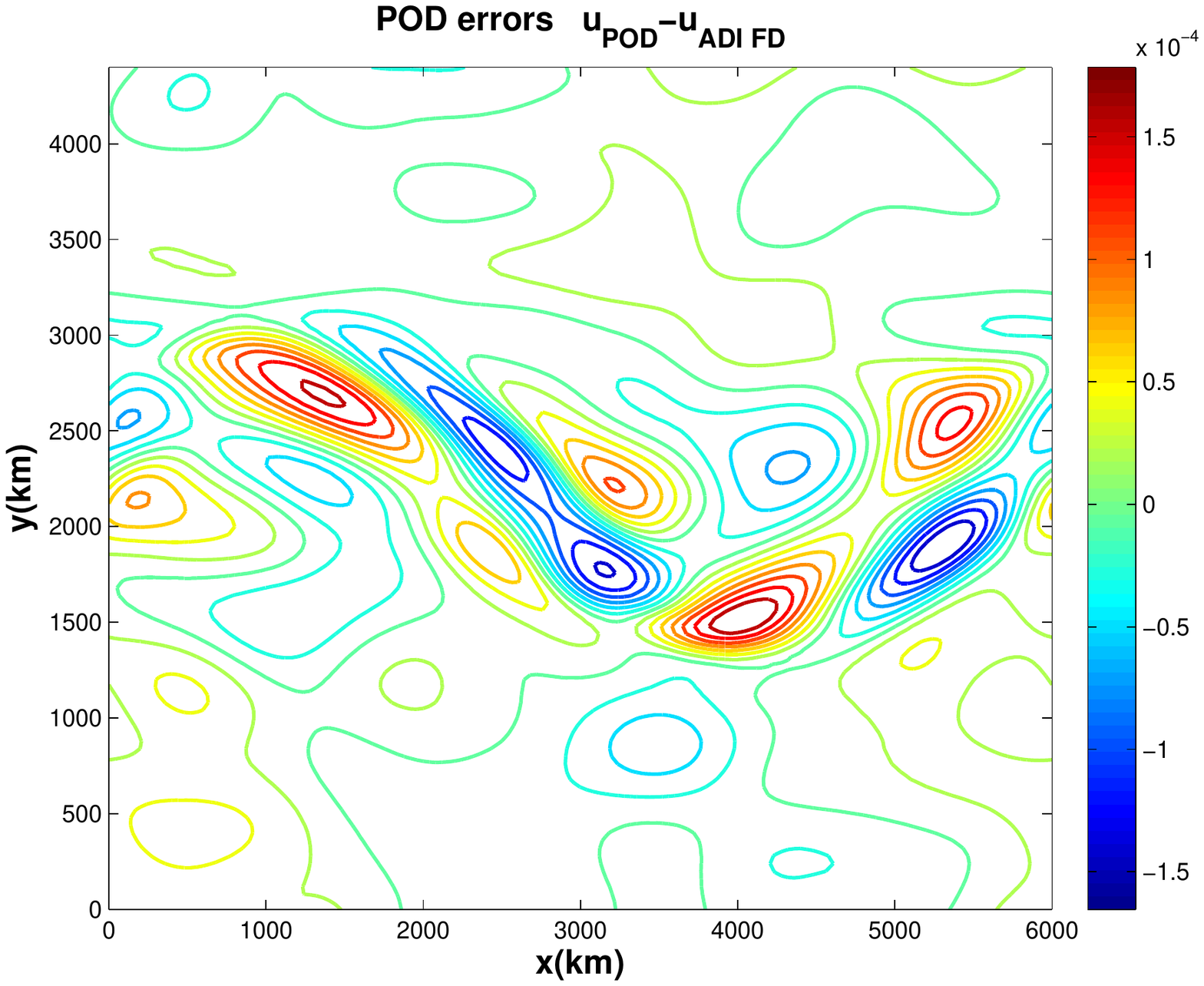}
\includegraphics[trim = 15mm 60mm 12mm 65mm, clip, width=5.1cm]{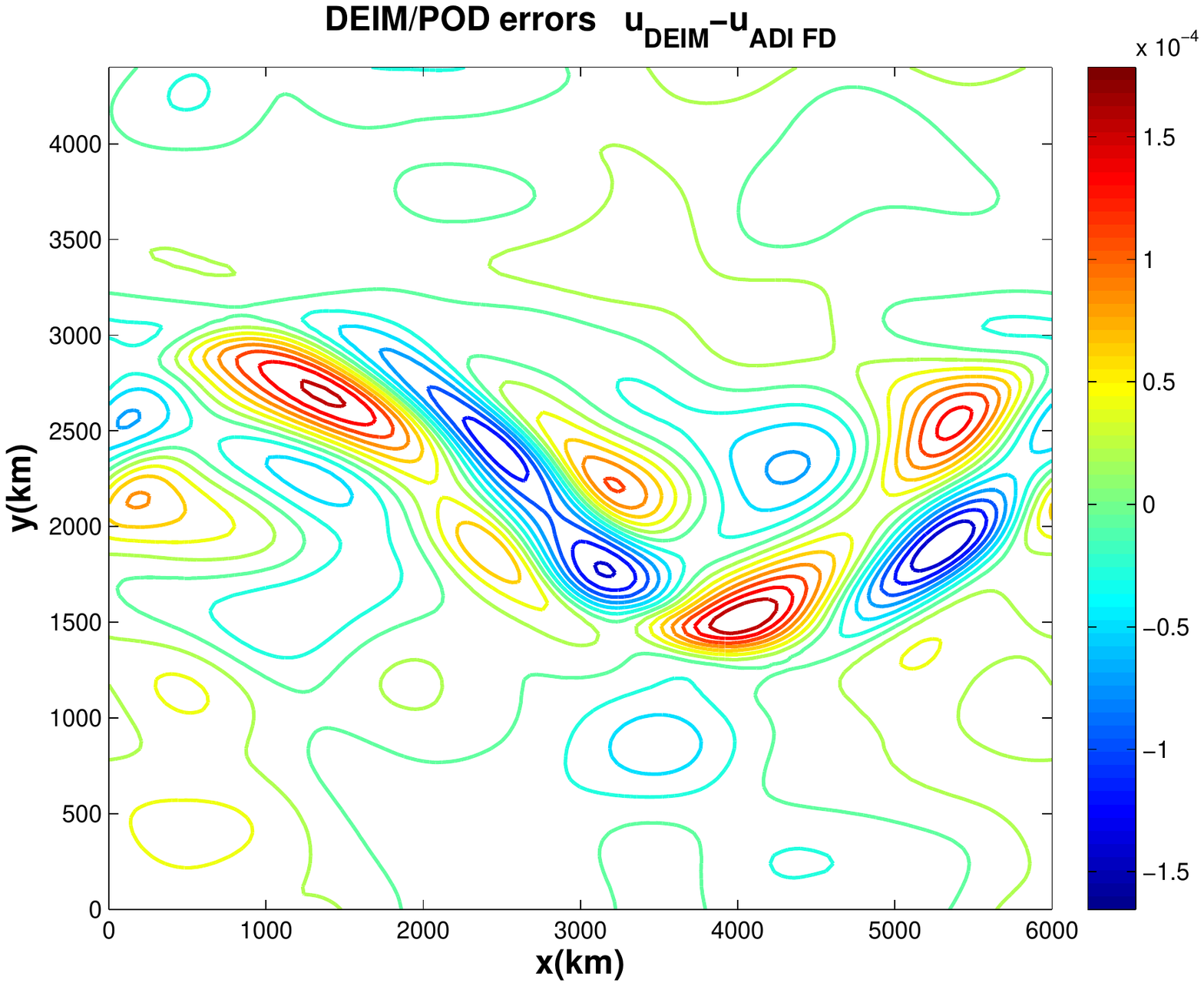}
\includegraphics[trim = 15mm 60mm 12mm 65mm, clip, width=5.1cm]{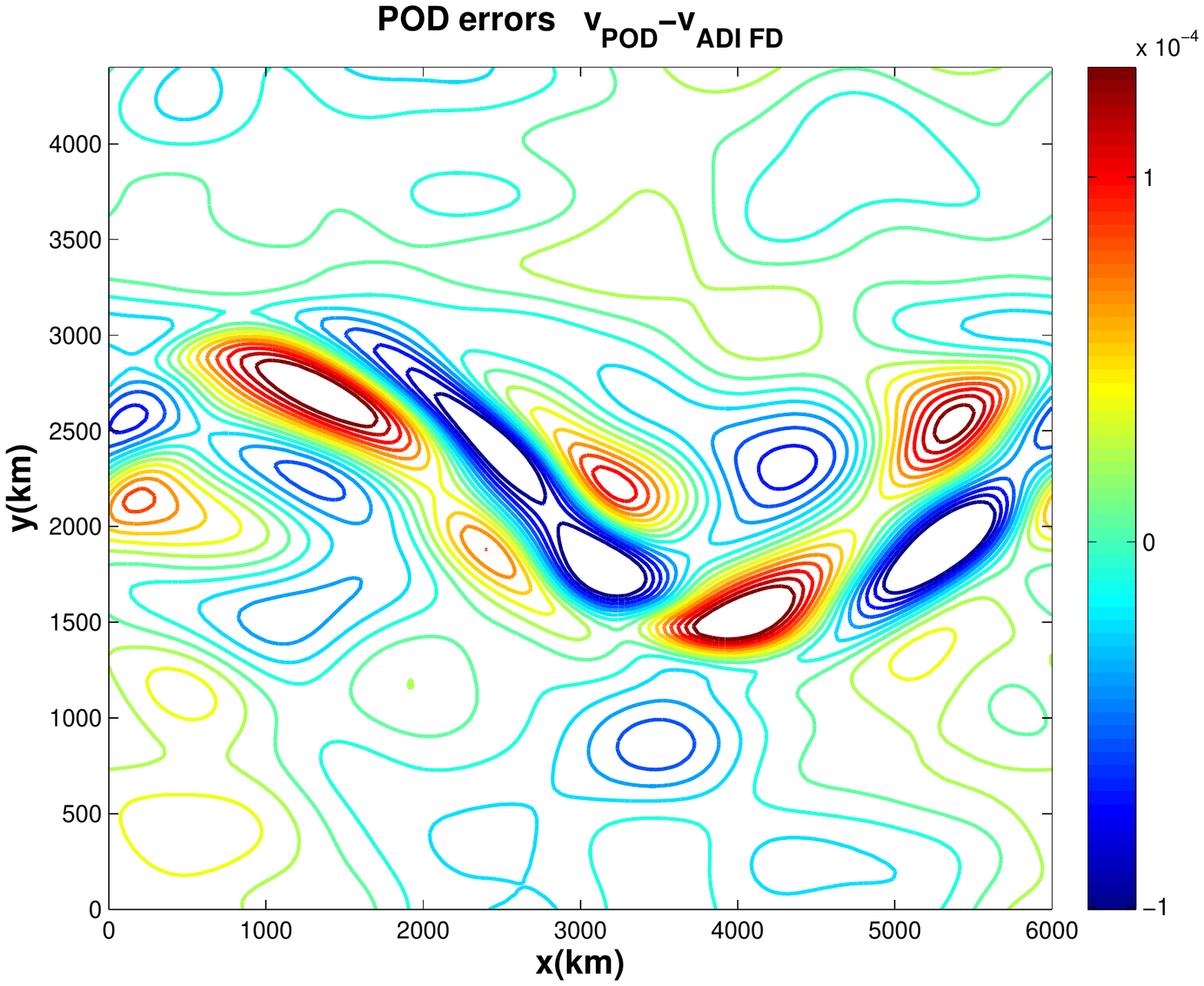}
\includegraphics[trim = 15mm 60mm 12mm 65mm, clip, width=5.1cm]{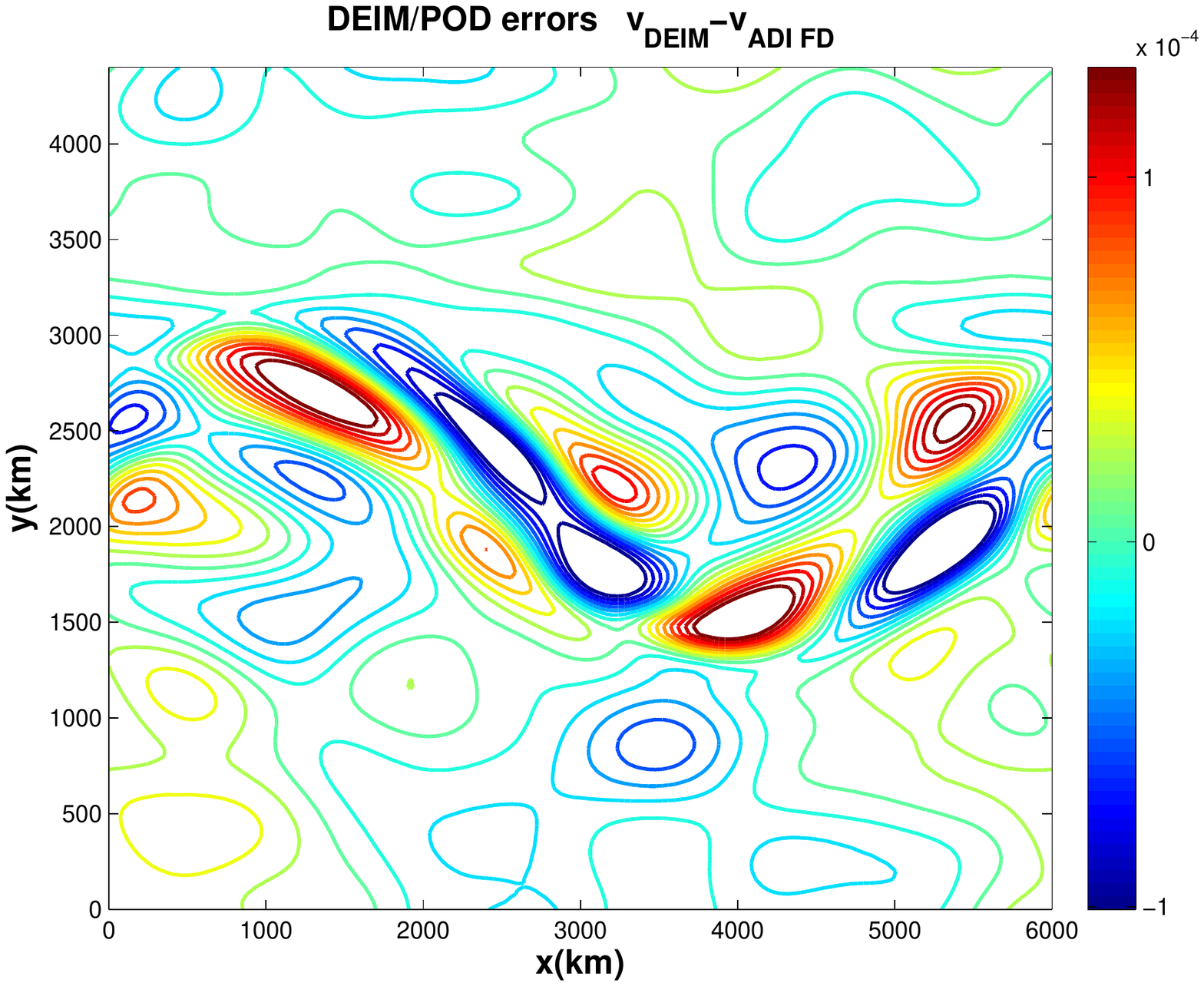}

\caption{\label{fig:DEIM14}Local errors between  POD, POD/DEIM ADI SWE solutions and the ADI FD SWE solutions at $t = 24h$ ($\Delta t = 480 s$). The number of DEIM points was taken $80$.}
\end{figure}

Next we determineed solutions of the POD ADI SWE model and the POD/DEIM ADI SWE model using $80$ DEIM points. The solutions of POD/DEIM implicit scheme are very accurate, local errors depicted in Figure \ref{fig:DEIM14}, average relative errors in Table \ref{tab:table3} and RMSE results in Table \ref{tab:table4} confirm it and showing that POD/DEIM ADI SWE scheme is much faster and almost as accurate as POD ADI SWE scheme.

\begin{figure}[h]
\centering
\includegraphics[trim = 15mm 60mm 12mm 65mm, clip, width=5.1cm]{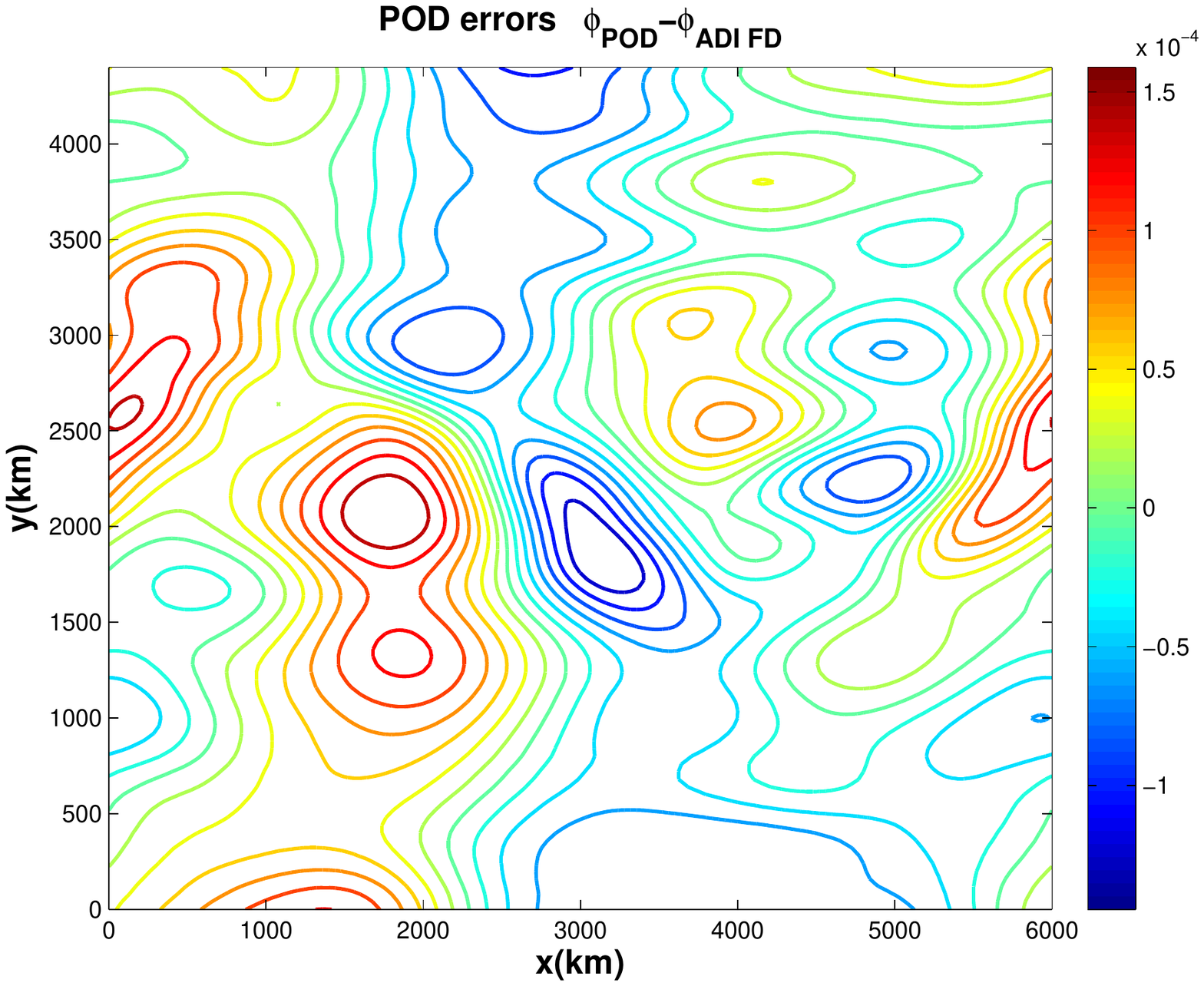}
\includegraphics[trim = 15mm 60mm 12mm 65mm, clip, width=5.1cm]{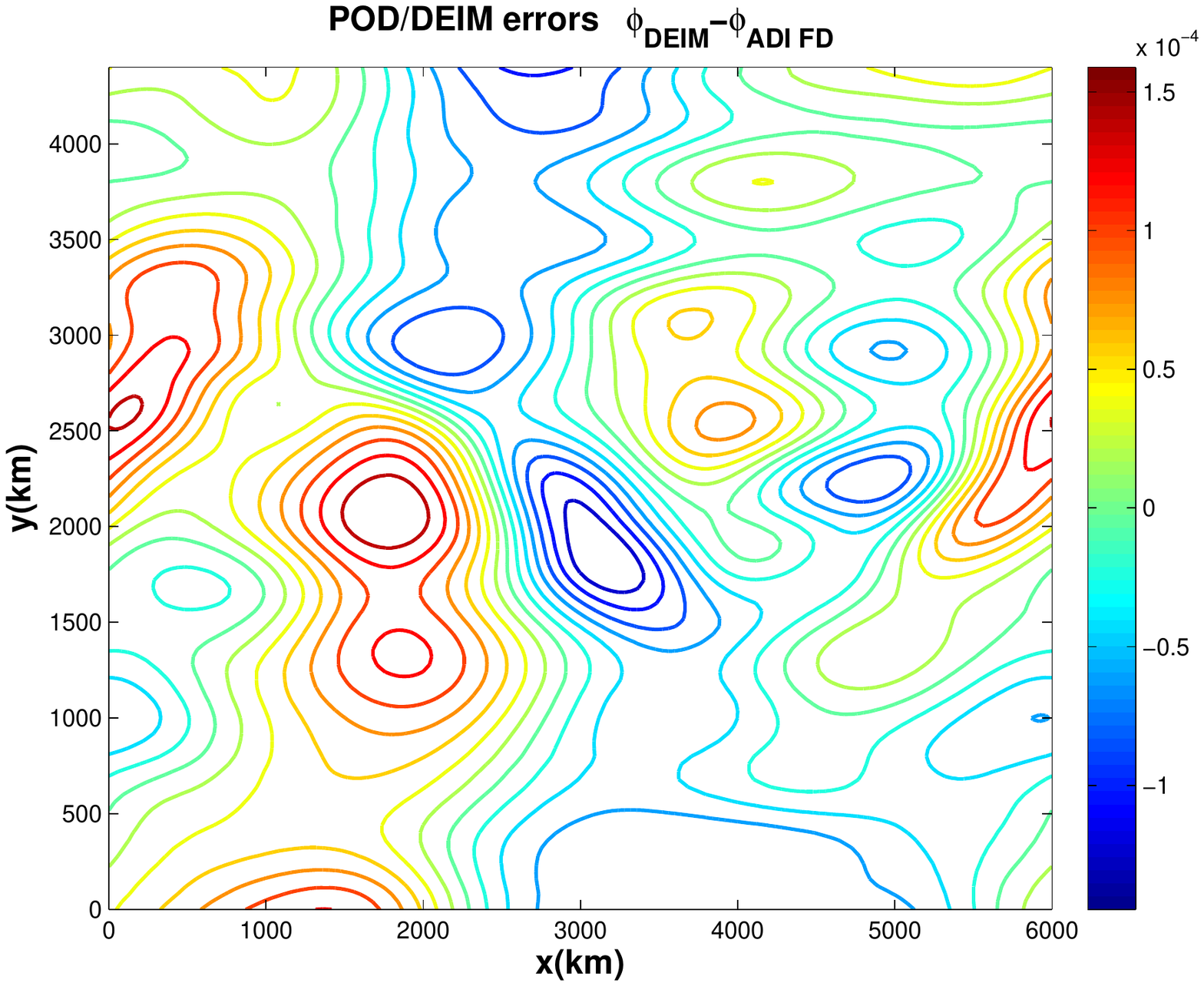}
\includegraphics[trim = 15mm 60mm 12mm 65mm, clip, width=5.1cm]{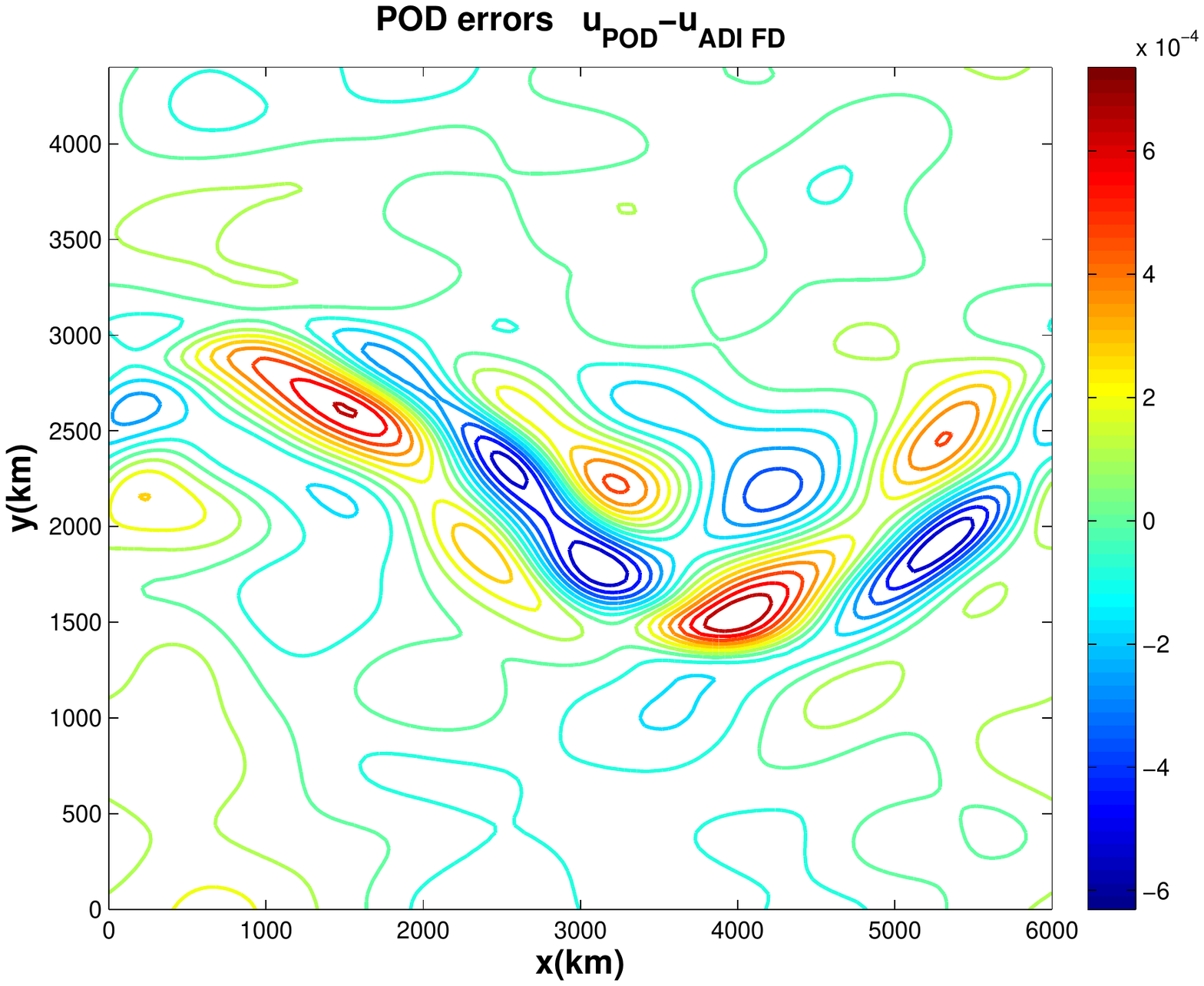}
\includegraphics[trim = 15mm 60mm 12mm 65mm, clip, width=5.1cm]{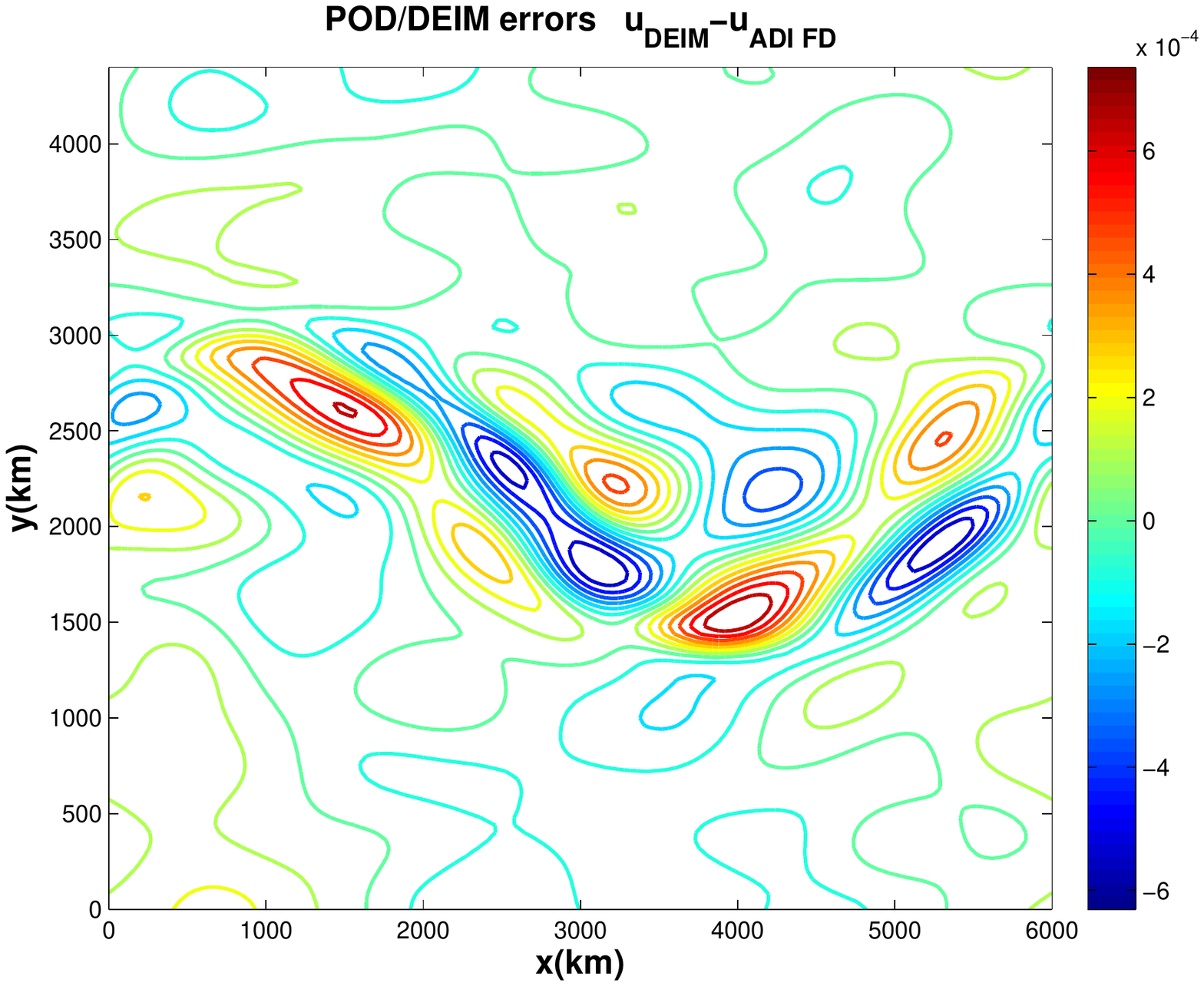}
\includegraphics[trim = 15mm 60mm 12mm 65mm, clip, width=5.1cm]{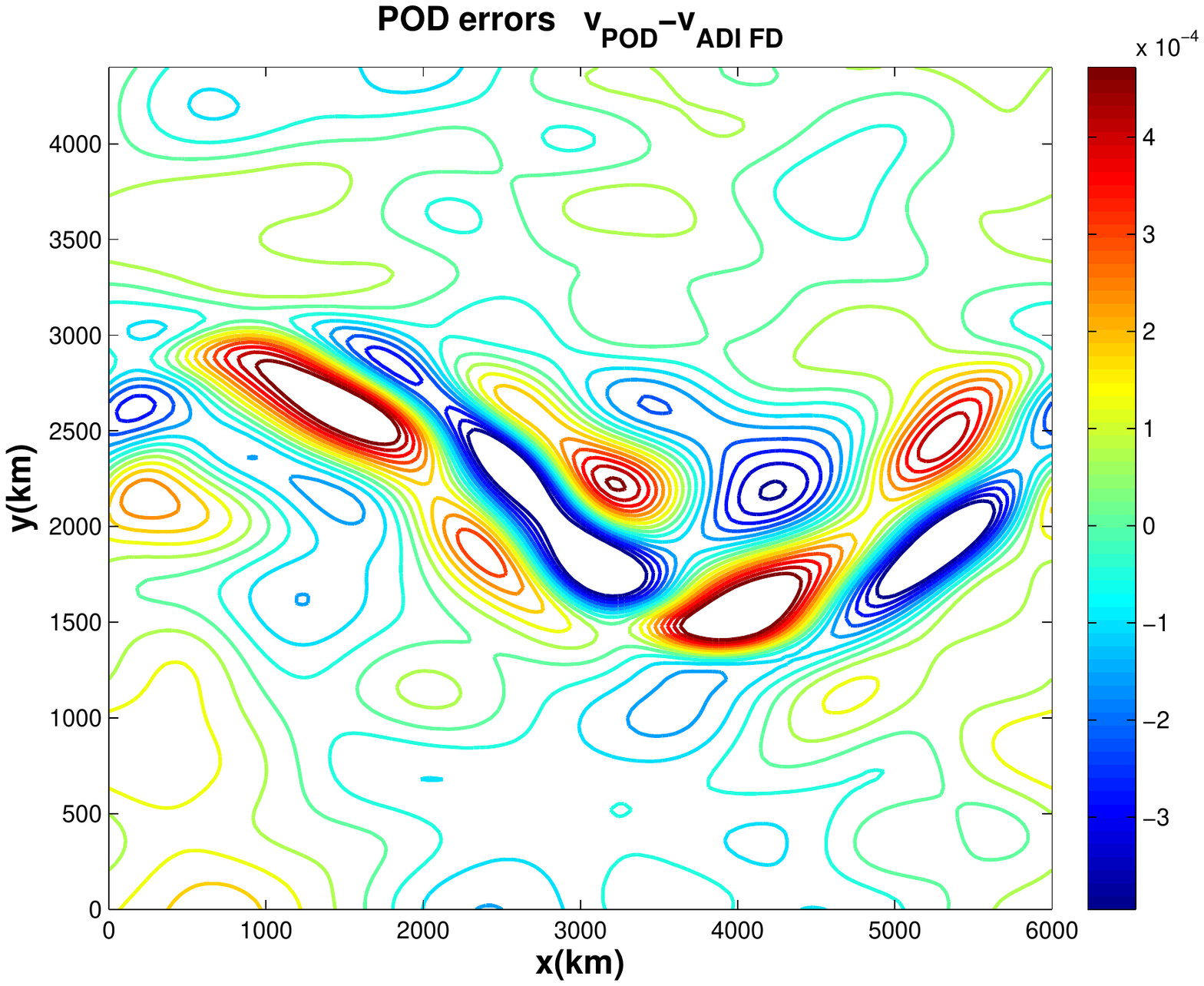}
\includegraphics[trim = 15mm 60mm 12mm 65mm, clip, width=5.1cm]{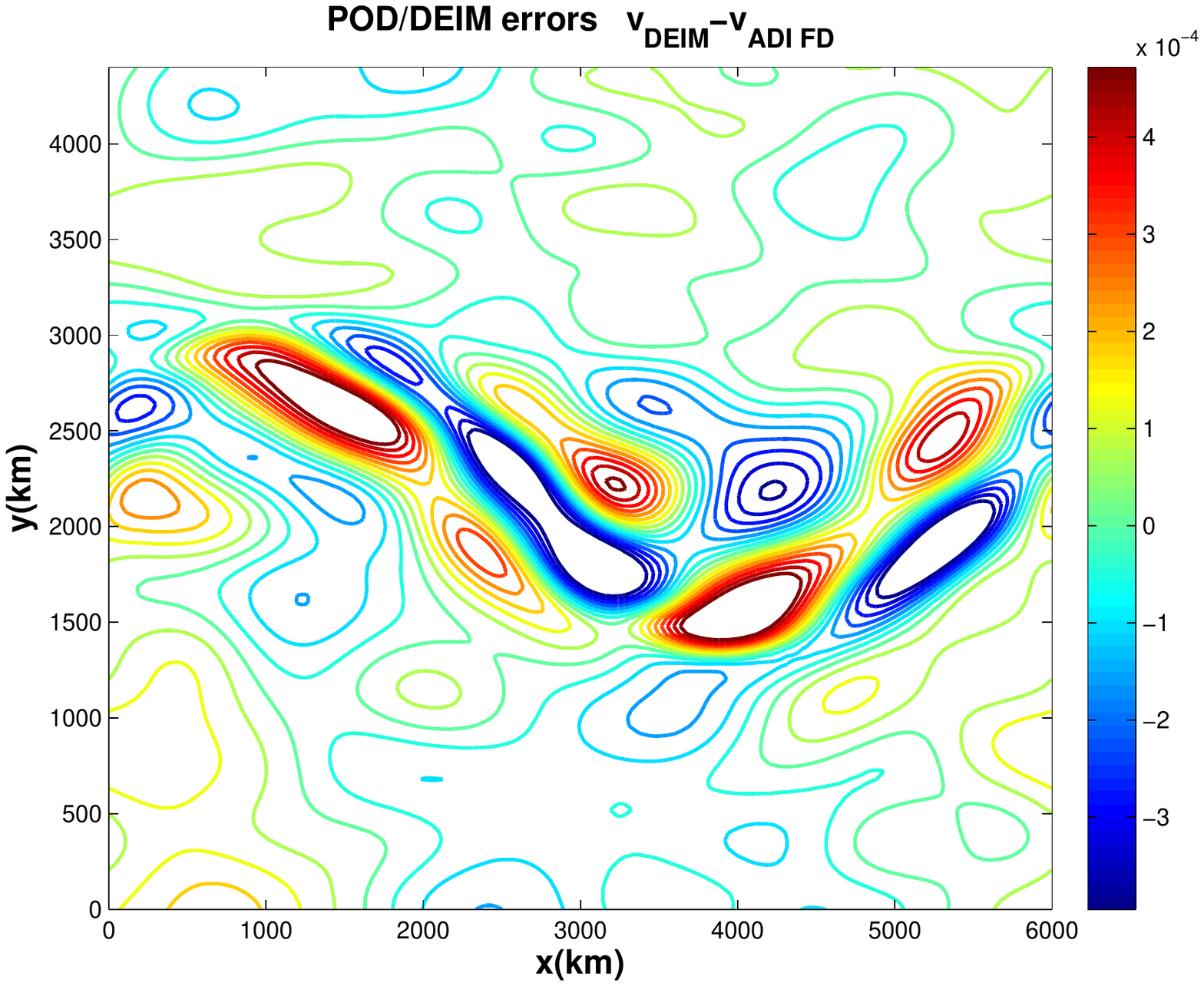}
\caption{\label{fig:DEIM15}Local errors between  POD, POD/DEIM ADI SWE solutions and the ADI FD SWE solutions at $t = 24h$ ($\Delta t = 960 s$). The number of snapshots is $91$ and the number of DEIM points was taken $80$.}
\end{figure}

Compared with the first experiment we reduced the number of spatial discretization points by a factor of $4$. This does not affect the magnitude of the local errors even if they were decreased for both POD and POD/DEIM ADI methods with factors between $3$ and $4$, when compared with the results obtained in the first case. The increased number of time steps and snapshots is responsible for improving the accuracy of the solutions. This can be observed in Figure \ref{fig:DEIM15} where we illustrate the local errors for the POD and POD/DEIM ADI SWE solutions using the same configuration as in the second experiment but having decreased the number of time steps and snapshots at $91$.

%

Table \ref{tab:table3} compares the accuracy of POD and POD/DEIM ADI SWE schemes measuring the average relative errors of the solutions with respect to the ADI FD SWE solutions.

\begin{table}[h]
\centerline{
\scalebox{0.7}{
\begin{tabular}{|c|c|c|}\hline \hline
 &  POD ADI SWE & DEIM/POD ADI SWE\\ \hline
$E_{\phi}$ & 2.648e-005 & 3.073e-005\\ \hline
$E_u$ & 1.279e-003 & 1.292e-003\\ \hline
$E_v$ & 2.207e-003 & 2.471e-003\\ \hline
\end{tabular}}}
\caption{\label{tab:table3}Average relative errors for each of the model variables at $t=t_f$,$\Delta t = 480s$. The POD bases dimensions were taken $35$ capturing more than $99.9\%$ of the system energy. $80$
DEIM points were chosen.}
\end{table}

Once again we calculate the solution of SWE model using the POD and POD/DEIM EE SWE schemes. By employing the DEIM method on the POD ADI SWE model we reduced the CPU time with a factor of $13$. In the case of explicit scheme the DEIM algorithm decreased the computational time by a factor of $21$. The adaptative Runge-Kutta-Fehlberg method involved in the explicit reduced order models (ROMs) was faster than the Newton method used to solve the nonlinear algebraic systems in implicit ROMs mostly because we doubled the number of time steps and thus the RKF45 didn't need to generate a large amount of intermediary time steps as it did in the first experiment in order to generate an accurate solution. From Table \ref{tab:table4} we notice that the RMSEs for both implicit and explicit POD/DEIM schemes are almost similar with the ones generated by the implicit and explicit POD systems with respect to the ADI FD SWE numerical solutions.

Collecting the results obtained from experiments $1$ and $2$ we conclude that the POD and POD/DEIM ADI SWE schemes are more accurate than the POD and POD/DEIM EE SWE schemes.

\begin{table}[h]
\centerline{
\scalebox{0.7}{
\begin{tabular}{|c|c|c|c|c|c|}\hline \hline
 & ADI SWE & POD ADI SWE & DEIM/POD ADI SWE & POD EE SWE & DEIM/POD EE SWE\\ \hline
CPU time & 24.353 & 8.848 & 0.686& 8.019& 0.386\\ \hline
$RMSE_{\phi}$ & - & 1.607e-005 & 1.743e-005& 3.999e-005& 4.074e-005\\ \hline
$RMSE_u$ & - & 4.076e-005 & 4.233e-005& 5.089e-005& 5.780e-005\\ \hline
$RMSE_v$ & - & 2.397e-005 & 2.755e-005& 4.613e-005& 4.919e-005\\ \hline\hline
\end{tabular}}}

\caption{\label{tab:table4}CPU time gains and the root mean square errors for each of the model variables at $t=t_f$, $\Delta t =480s$. The POD bases dimensions were taken $35$ capturing more than $99.9\%$ of the system energy. $80$
DEIM points were chosen.}
\end{table}

\begin{figure}[h]
\centering
\includegraphics[trim = 35mm 80mm 22mm 85mm, clip, width=7.7cm]{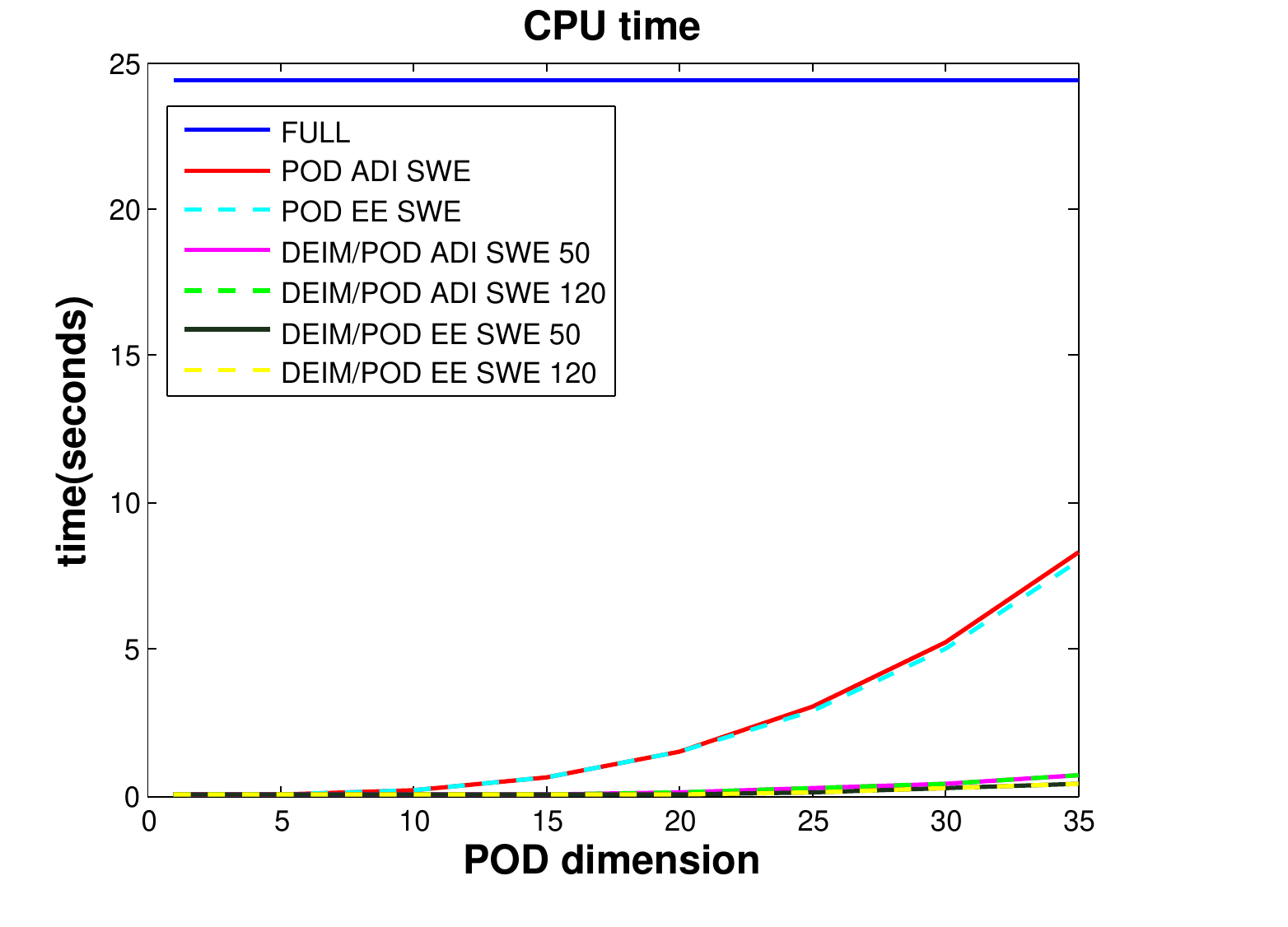}
\includegraphics[trim = 35mm 80mm 22mm 85mm, clip, width=7.7cm]{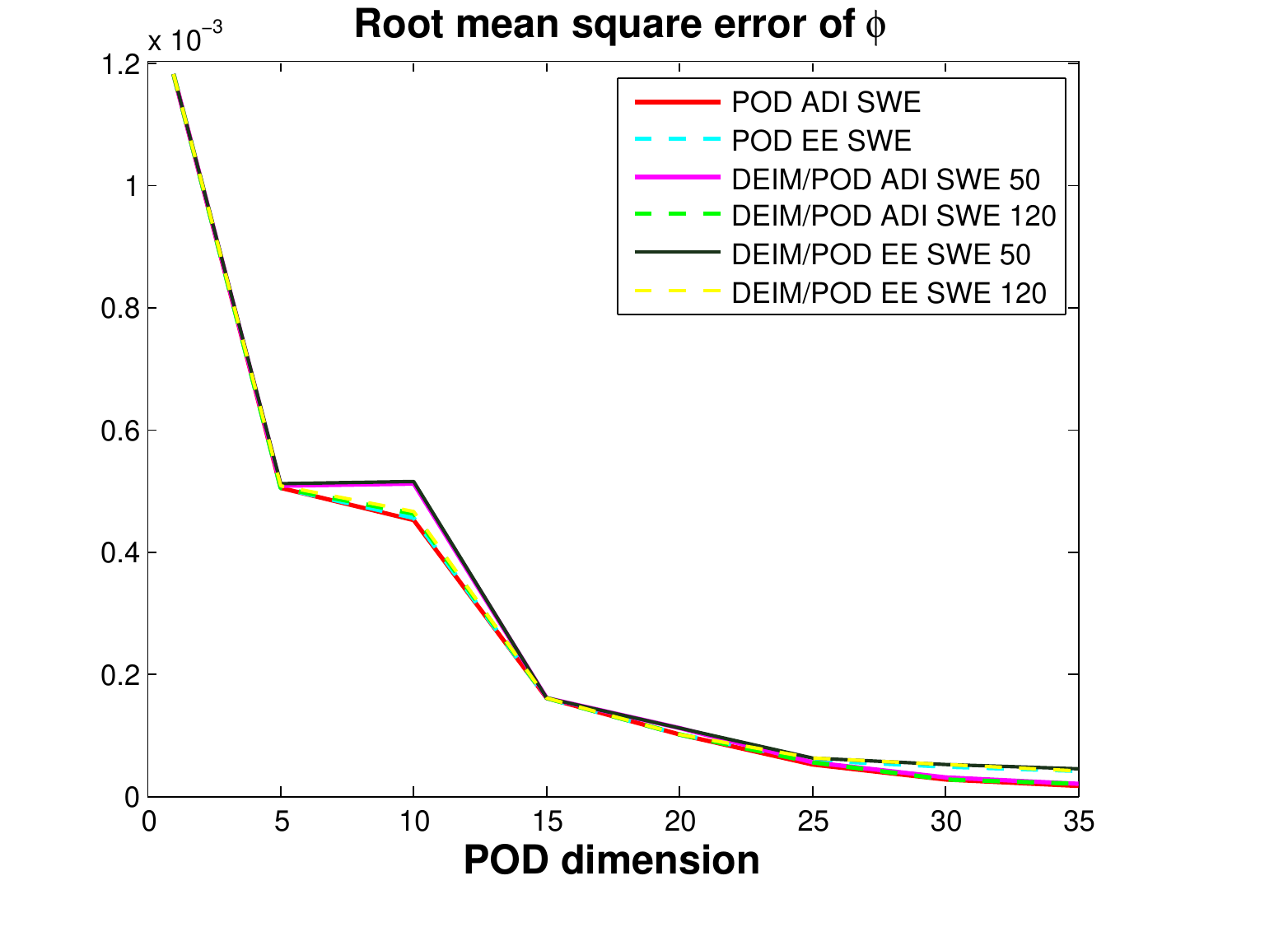}
\caption{\label{fig:DEIM16}CPU time of the full system, POD reduced systems and POD - DEIM reduced systems (left); Root mean square error of $\phi$ calculated for POD/DEIM and POD reduced systems with respect to ADI FD SWE solutions}
\end{figure}

Next we evaluate the efficiency of POD/DEIM method as a function of POD dimension. Figure \ref{fig:DEIM16} gives the root mean square errors of $\phi$ and the corresponding average CPU times for different dimensions of POD and DEIM approximations.

The results in Tables \ref{tab:table5} and \ref{tab:table6} show the performances of POD/DEIM method. Once the POD dimension exceeds $15(5)$ in case of DEIM/POD ADI (EE) SWE scheme the CPU time is decreased at least by a factor of $10$. When DEIM dimension reached $50$, we notice that RMSE results between POD and POD/DEIM are almost identical. All the methods performed well when POD dimension exceeded $25$ leading to RMSE results of order O$(10^{-5})$.

\begin{table}[h]
\centerline{
\scalebox{0.5}{
\begin{tabular}{|c|c|c|c|c|c|c|}\hline \hline
 PODDIM & POD ADI SWE & DEIM/POD ADI SWE50&DEIM/POD ADI SWE120 & POD EE SWE & DEIM/POD EE SWE50&DEIM/POD EE SWE120\\ \hline
$1$ &0.0009 &0.0003 &0.0002 &0.0004 &0.0001 &0.0002  \\ \hline
$5$ &0.0264 &0.0109 &0.0111 &0.0235 &0.0014 &0.0014  \\ \hline
$10$ &0.1896 &0.0244 &0.0249 &0.1838 &0.0066 &0.0071  \\ \hline
$15$ &0.6478 &0.0599 &0.0615 &0.6211 &0.0238 &0.0251  \\ \hline
$20$ &1.5336 &0.1245 &0.1283 &1.4795 &0.0586 &0.0615  \\ \hline
$25$ &2.9991 &0.2326 &0.2403 &2.8901 &0.1165 &0.1237  \\ \hline
$30$ &5.1956 &0.3840 &0.4013 &5.0250 &0.2202 &0.2292  \\ \hline
$35$ &8.8480 &0.6718 &0.6971 &8.0190 &0.3773 &0.3992  \\ \hline\hline
\end{tabular}
}}
\caption{\label{tab:table5}Comparison between CPU times of POD and DEIM/POD implicit and explicit schemes. The computational time of the full model (ADI FD SWE) was $24.3530$. }
\end{table}
\begin{table}[h]
\centerline{
\scalebox{0.5}{
\begin{tabular}{|c|c|c|c|c|c|c|}\hline \hline
 PODDIM & POD ADI SWE & DEIM/POD ADI SWE50&DEIM/POD ADI SWE120 & POD EE SWE & DEIM/POD EE SWE50&DEIM/POD EE SWE120\\ \hline
$1$ &1.182e-003 &1.182e-003 &1.182e-003 &1.182e-003 &1.182e-003 &1.182e-003  \\ \hline
$5$ &5.018e-004 &5.087e-004 &5.047e-004 &5.037e-004 &5.110e-004 &5.070e-004  \\ \hline
$10$ &4.526e-004 &5.122e-004 &4.609e-004 &4.553e-004 &5.155e-004 &4.635e-004  \\ \hline
$15$ &1.587e-004 &1.682e-004 &1.589e-004 &1.584e-004 &1.669e-004 &1.595e-004  \\ \hline
$20$ &1.009e-004 &1.101e-004 &1.012e-004 &9.876e-005 &1.097e-004 &9.876e-005  \\ \hline
$25$ &5.075e-005 &5.526e-005 &5.438e-005 &5.909e-005 &6.319e-005 &6.281e-005  \\ \hline
$30$ &2.765e-005 &2.898e-005 &2.800e-005 &4.788e-005 &5.149e-005 &4.923e-005  \\ \hline
$35$ &1.607e-005 &1.846e-005 &1.835e-005 &3.999e-005 &4.439e-005 &4.175e-005  \\ \hline \hline
\end{tabular}
}}
\caption{\label{tab:table6}Comparison between RMSE of POD and DEIM/POD implicit and explicit schemes. }
\end{table}

\begin{figure}[h]
\centering
\includegraphics[trim = 35mm 80mm 22mm 85mm, clip, width=5.1cm]{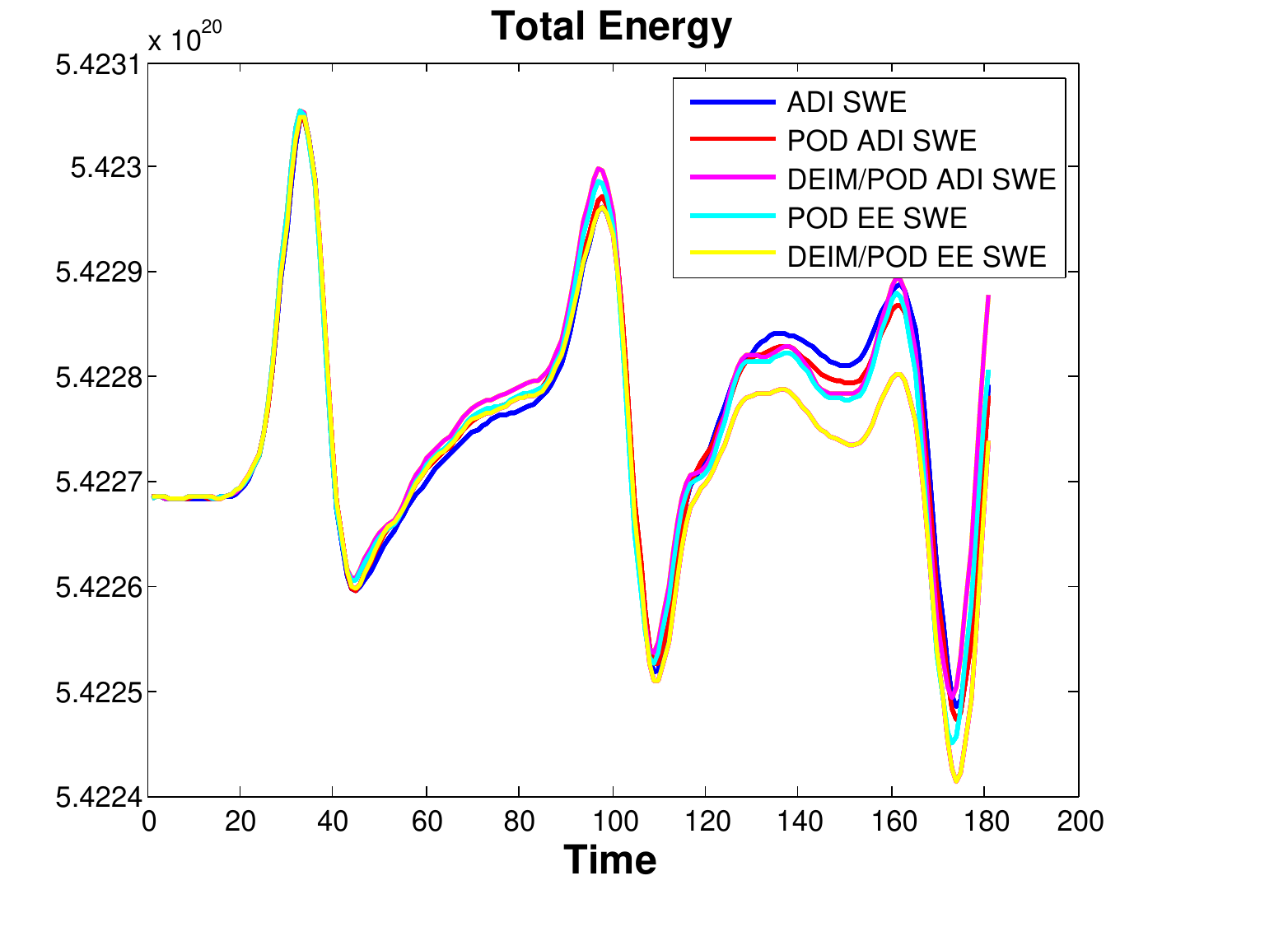}
\includegraphics[trim = 35mm 80mm 22mm 85mm, clip, width=5.1cm]{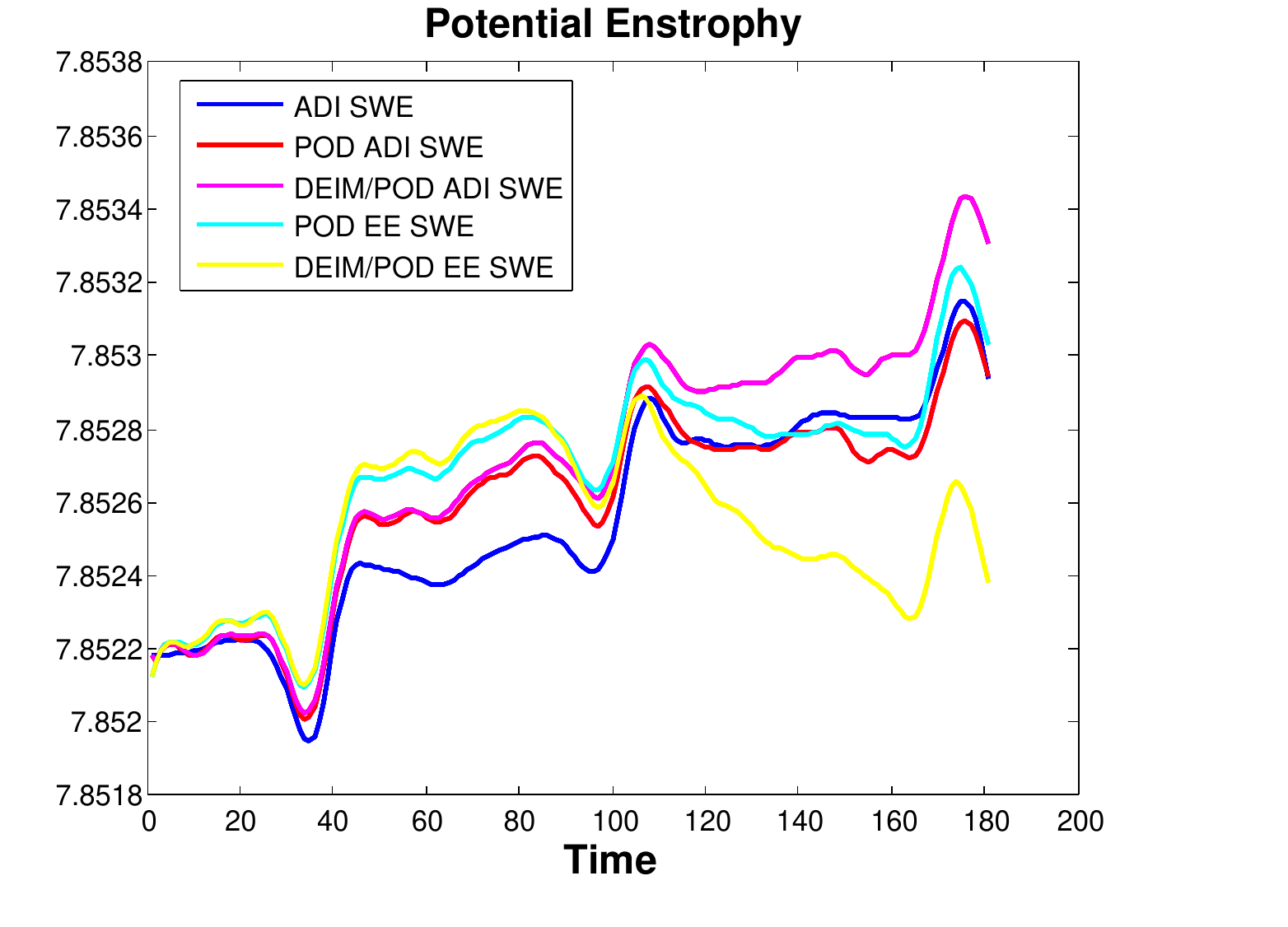}
\includegraphics[trim = 35mm 80mm 22mm 85mm, clip, width=5.1cm]{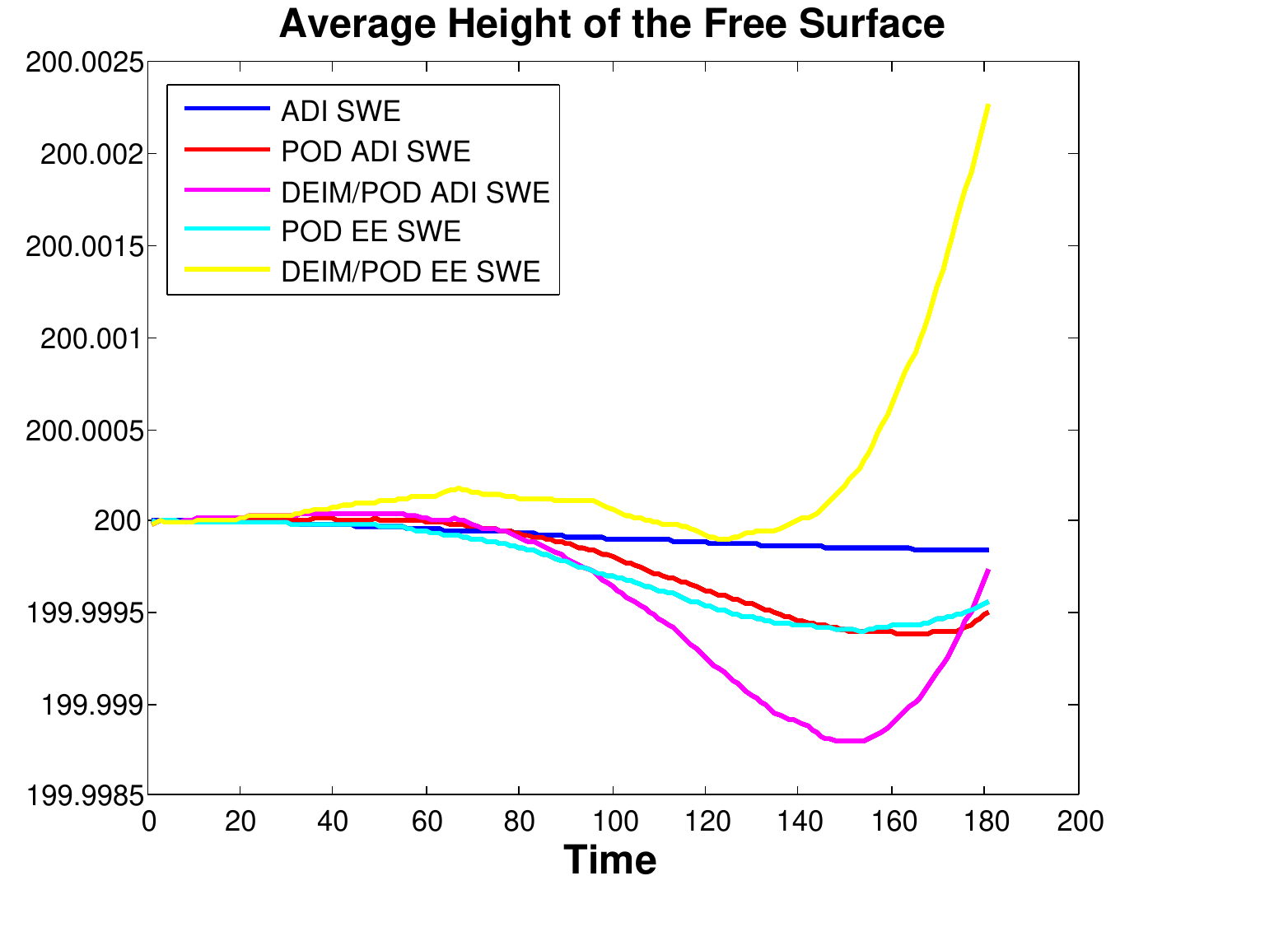}
\caption{\label{fig:DEIM17}Shallow Water Equations invariants, POD dimension =$35$, No of DEIM points = $120$}
\end{figure}

Numerical experiments carried out for a 1-day integration showed that the POD/DEIM ADI SWE scheme as well as the POD/DEIM EE SWE discrete model conserve the average height of the free surface and the potential enstrophy while another integral invariant of the SWE model, the total energy, is not preserved largely due to the absence of a staggered C-grid in the numerical discretization. Arakawa in \cite{Araka1997} showed that when the finite difference Jacobian expression for the advection term is restricted to a form which properly represents the interaction between grid points the computational instability is prevented thereby  preserving all the integral invariants. Thus the POD and POD/DEIM systems behave in a similar way  as the ADI FD SWE full scheme in the matter of integral invariants conservations and Figure \ref{fig:DEIM17} depicts their evolution in time.

Tables \ref{tab:table7} - \ref{tab:table9} present integral invariants measures for all the schemes involved in this study using $max - min$ evaluation and Euclidian norm with respect to SWE invariants calculated with ADI FD SWE full scheme.
\begin{table}[h]
\centerline{
\scalebox{0.7}{
\begin{tabular}{|c|c|c|c|c|c|}\hline \hline
 & ADI SWE & POD ADI SWE & DEIM/POD ADI SWE & POD EE SWE & DEIM/POD EE SWE\\ \hline
Max-Min & 0.0017 & 0.0063 & 0.0125& 0.0060& 0.0237\\ \hline
Norm & 0.0000 & 0.0331 & 0.0693& 0.0348& 0.0795\\ \hline\hline
\end{tabular}}}
\caption{\label{tab:table7}Average Height of the conservation of the mass}
\end{table}
\begin{table}[h]
\centerline{
\scalebox{0.7}{
\begin{tabular}{|c|c|c|c|c|c|}\hline \hline
 & ADI SWE & POD ADI SWE & DEIM/POD ADI SWE & POD EE SWE & DEIM/POD EE SWE\\ \hline
Max-Min & 0.0012 & 0.0011 & 0.0014& 0.0011& 0.0008\\ \hline
Norm & 0.0000 & 0.0014 & 0.0023& 0.0024& 0.0040\\ \hline\hline
\end{tabular}}}
\caption{\label{tab:table8}Potential Enstrophy}
\end{table}
\begin{table}[h]
\centerline{
\scalebox{0.7}{
\begin{tabular}{|c|c|c|c|c|c|}\hline \hline
 & ADI SWE & POD ADI SWE & DEIM/POD ADI SWE & POD EE SWE & DEIM/POD EE SWE\\ \hline
Max-Min & 5.6440e+016 & 5.7392e+016 & 5.6140e+016& 6.0441e+016& 6.3439e+016\\ \hline
Norm & 0.0000e+000 & 1.4289e+016 & 3.3597e+016& 2.9898e+016& 5.3457e+016\\ \hline\hline
\end{tabular}}}
\caption{\label{tab:table9}Total Energy}
\end{table}

\begin{figure}[H]
\centering
\includegraphics[trim = 35mm 80mm 22mm 85mm, clip, width=5.1cm]{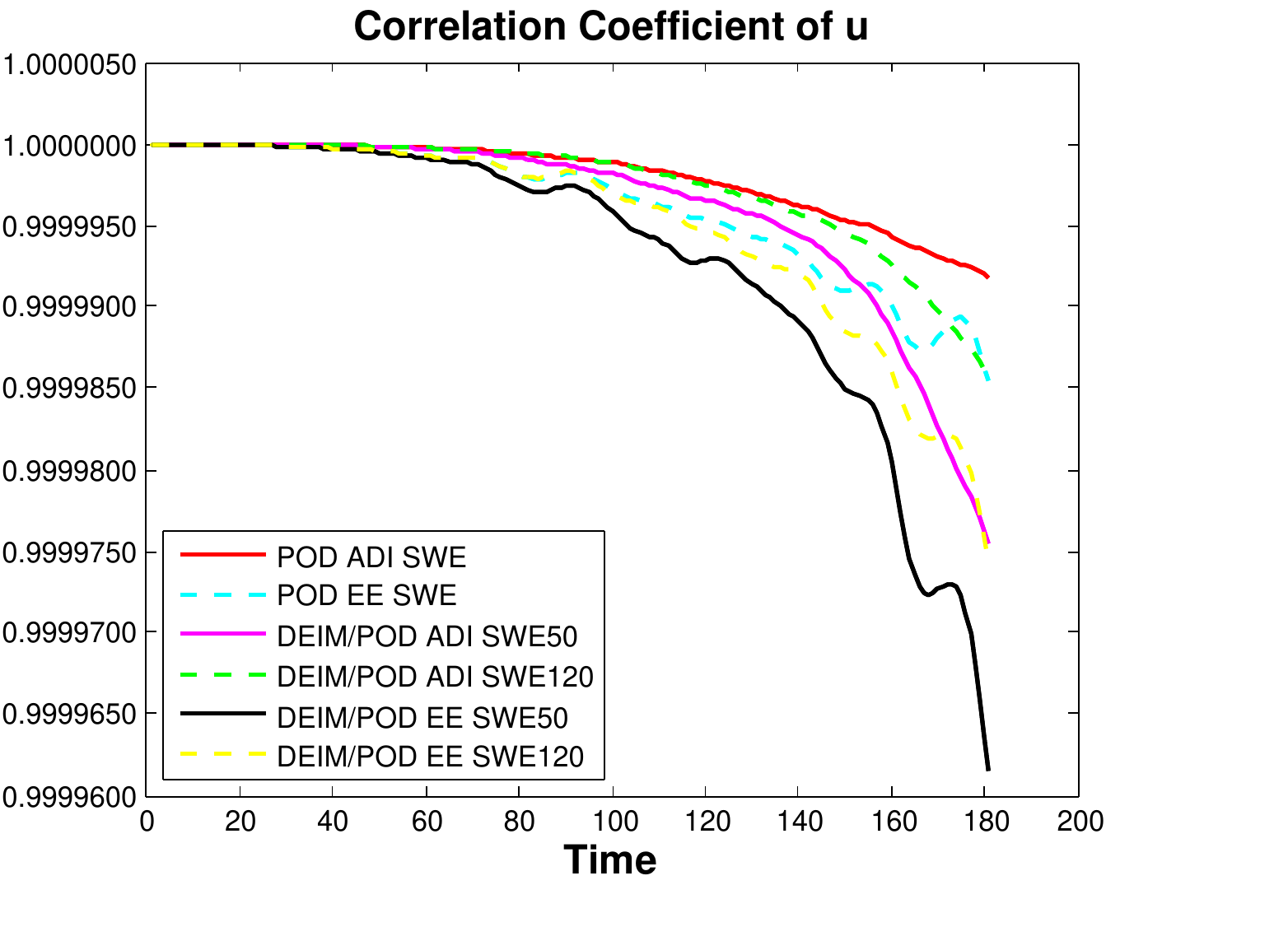}
\includegraphics[trim = 35mm 80mm 22mm 85mm, clip, width=5.1cm]{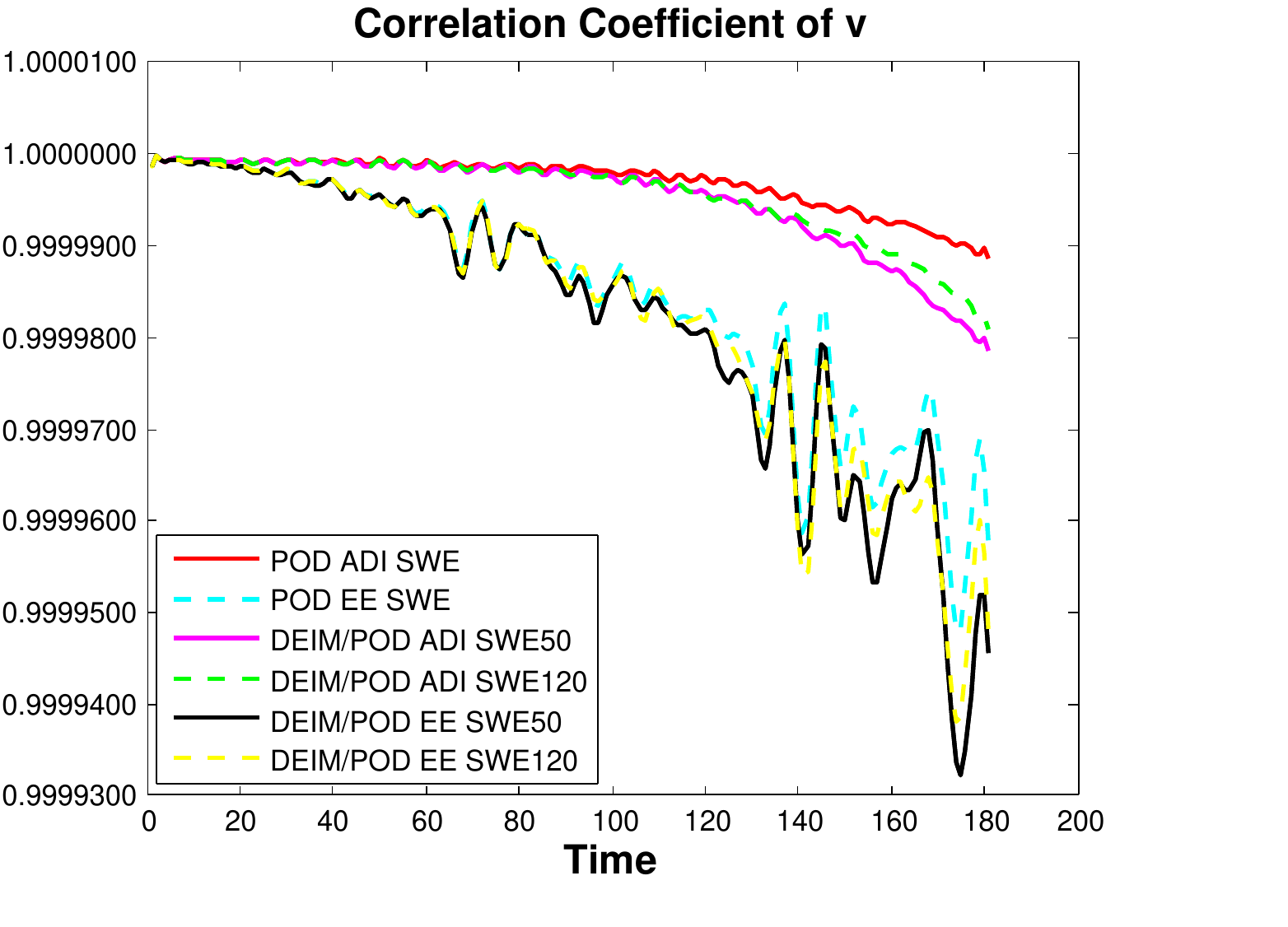}
\includegraphics[trim = 35mm 80mm 22mm 85mm, clip, width=5.1cm]{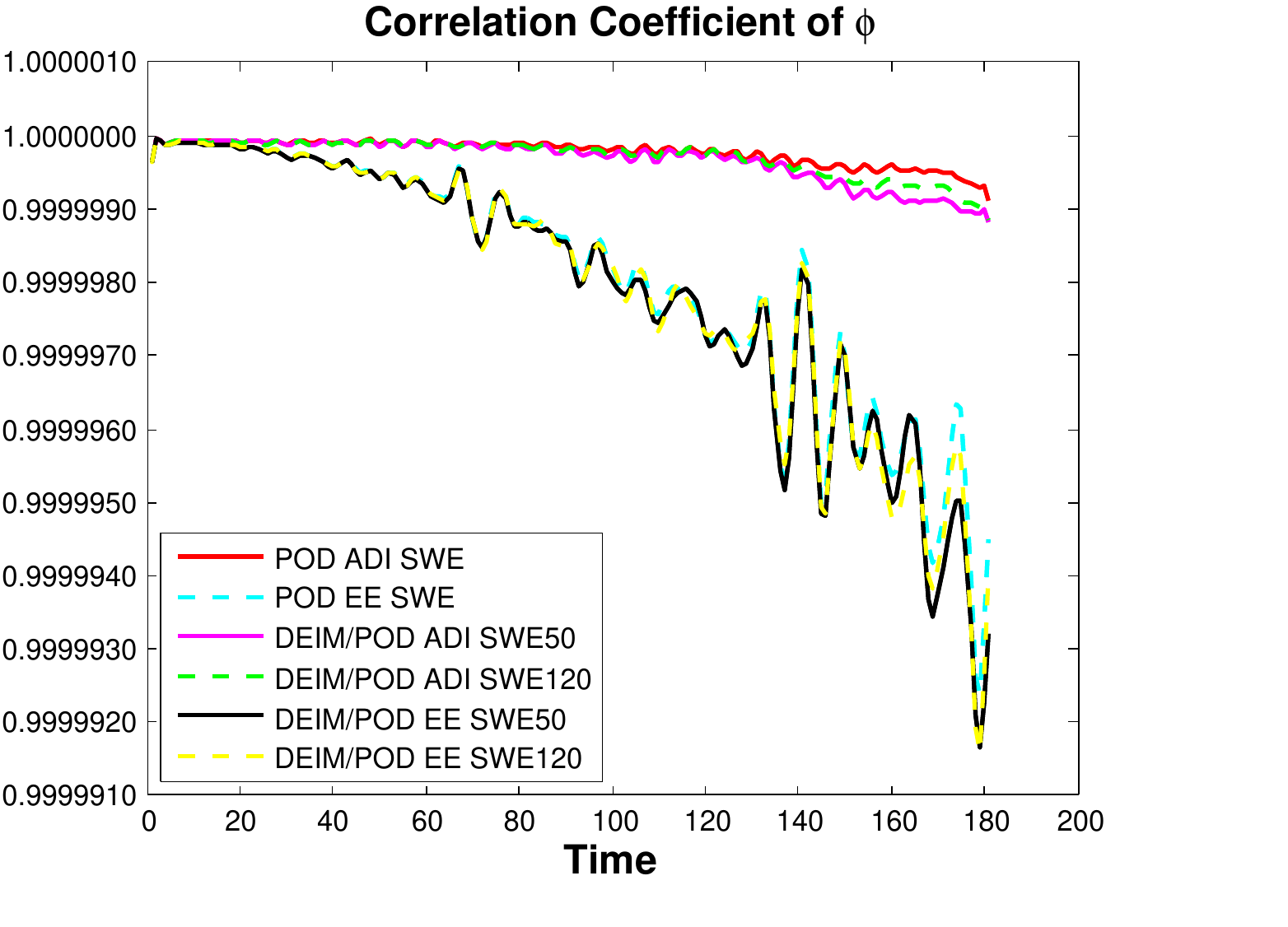}
\caption{\label{fig:DEIM18}Correlation coefficients for the SWE variables, POD dimension =$35$.}
\end{figure}

In Figure \ref{fig:DEIM18}, the Pearson correlation coefficient defined below is used as an additional metric to evaluate the quality of POD/DEIM schemes

\begin{equation*}
r_{i}=\frac{cov_{12}^{i}}{\sigma_{1}^{i}\sigma_{2}^{i}},~i=1,..,NT,\end{equation*}
where
\begin{equation*}
\sigma_{1}^{i}=\sum_{j=1}^{j=n_{xy}}\left(W_{i,j}-\overline{W}_{j}\right)^{2},\,\,\,\,\sigma_{2}
=\sum_{j=1}^{j=n_{xy}}\left(W_{i,j}^{scheme}-\overline{W^{scheme}}_{j}\right)^{2},\,\,\, i=1,\ldots,NT,\end{equation*}

\begin{equation*}
cov_{12}=\sum_{j=1}^{j=n_{xy}}\left(W_{i,j}-\overline{W}_{j}\right)\left(W_{i,j}^{scheme}-
\overline{W^{scheme}}_{j}\right),\,\,\, i=1,\ldots,NT,\end{equation*}
where $W=u,v,~\phi$ represents the ADI FD SWE solution and $W^{scheme}=u^{scheme},v^{scheme},~\phi^{scheme}$ the solution calculated with one of the following schemes: POD ADI SWE, POD/DEIM ADI SWE, POD EE SWE and POD/DEIM EE SWE using $50$ and $120$ DEIM points. $\overline{W}_{j}$ and $\overline{W^{scheme}}_{j}$ are corresponding means over the simulation period $\left[0,t_f\right]$ at spatial node $j$.

\begin{figure}[h]
\centering
\includegraphics[trim = 44mm 80mm 45mm 75mm, clip, width=5.2cm]{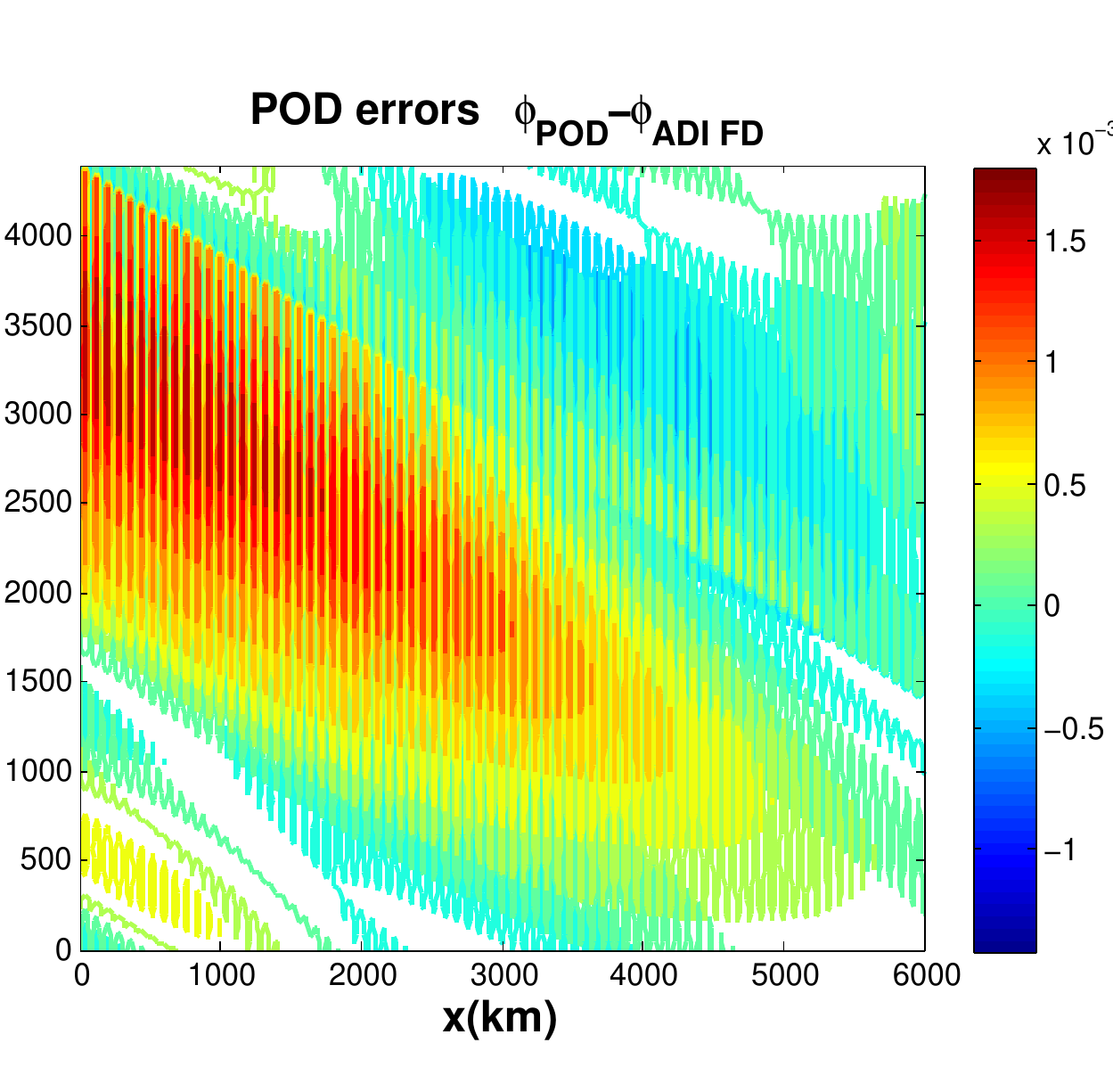}
\includegraphics[trim = 44mm 80mm 45mm 75mm, clip, width=5.2cm]{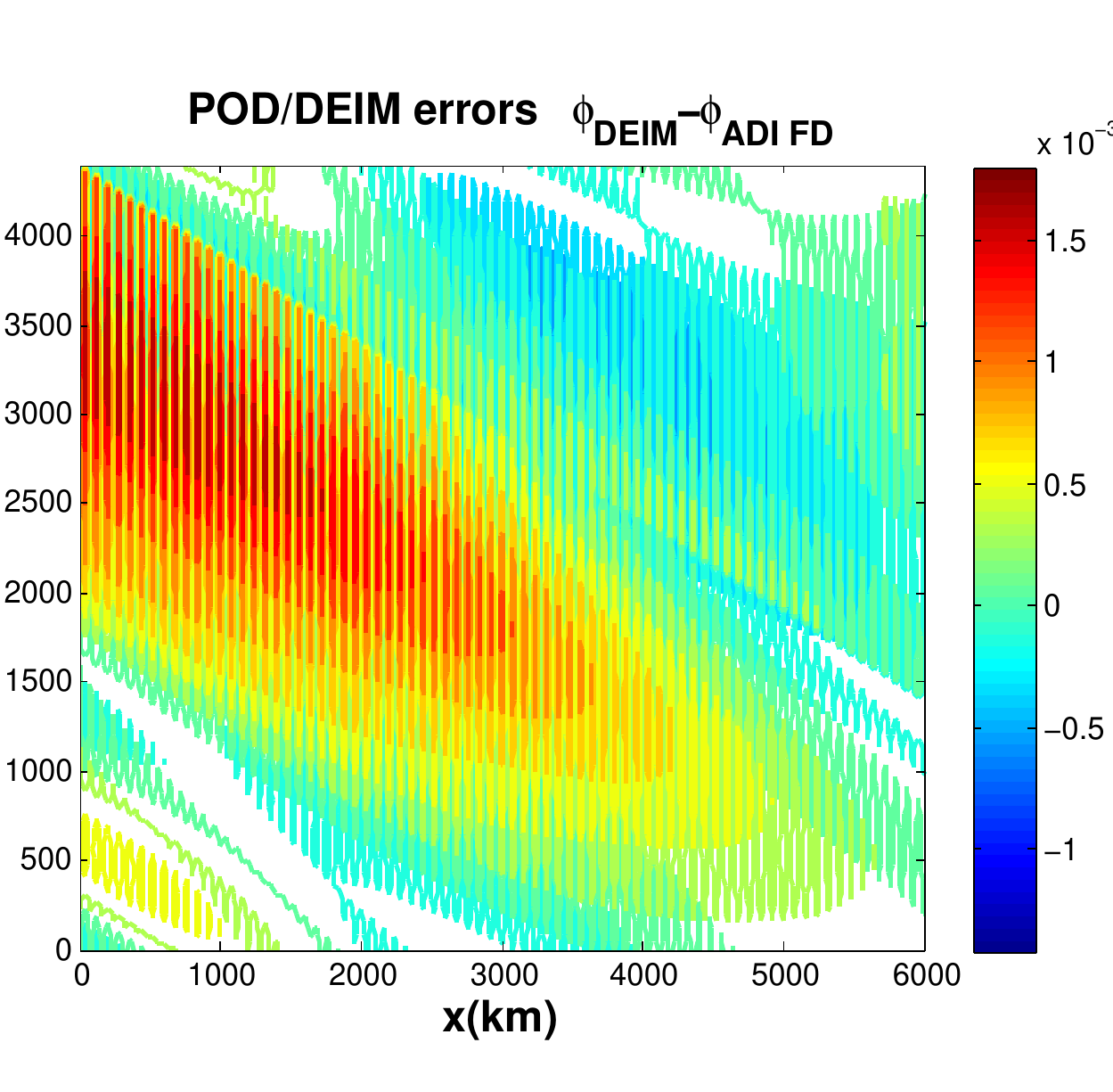}
\includegraphics[trim = 44mm 80mm 45mm 75mm, clip, width=5.2cm]{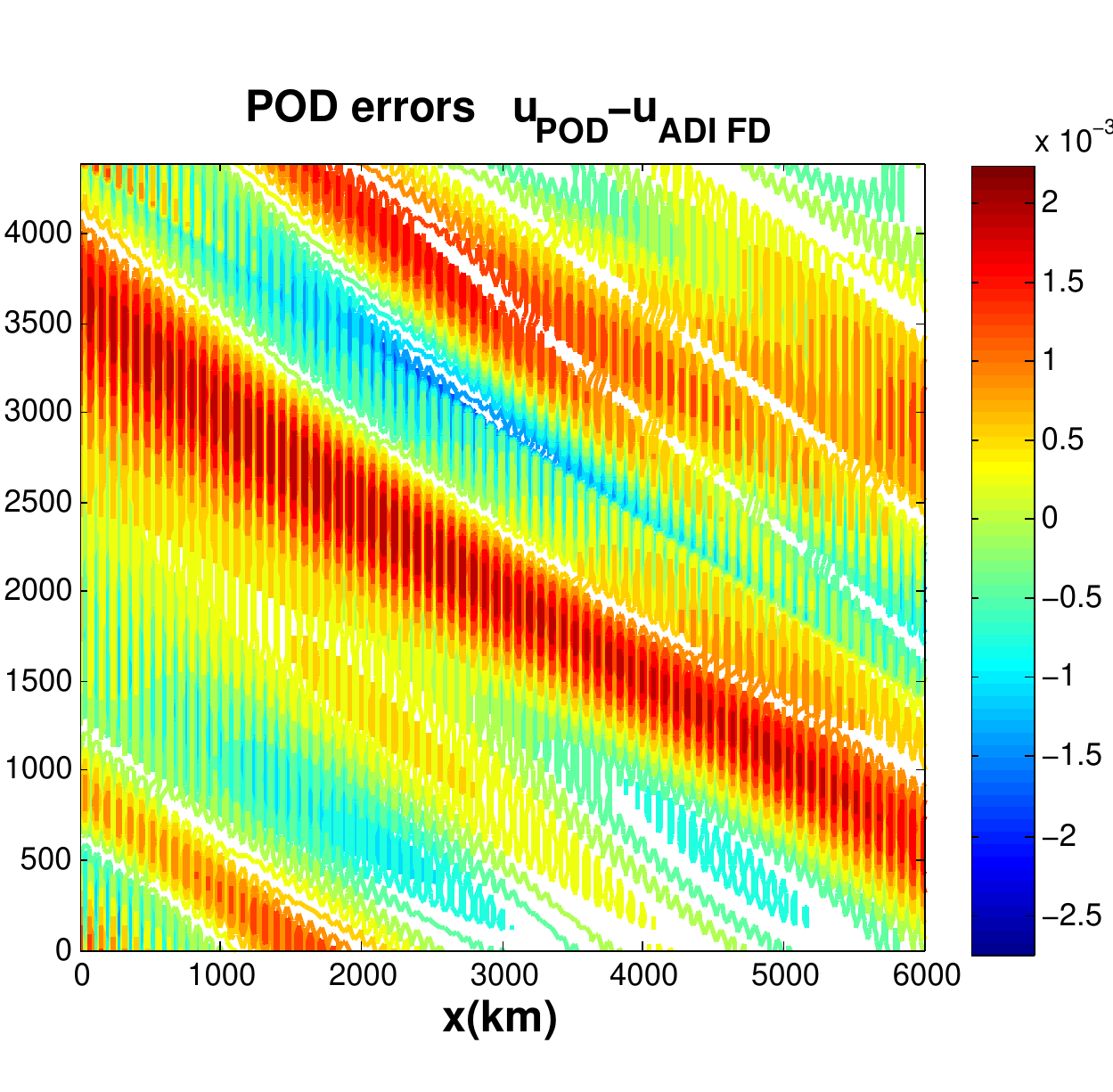}
\includegraphics[trim = 44mm 80mm 45mm 75mm, clip, width=5.2cm]{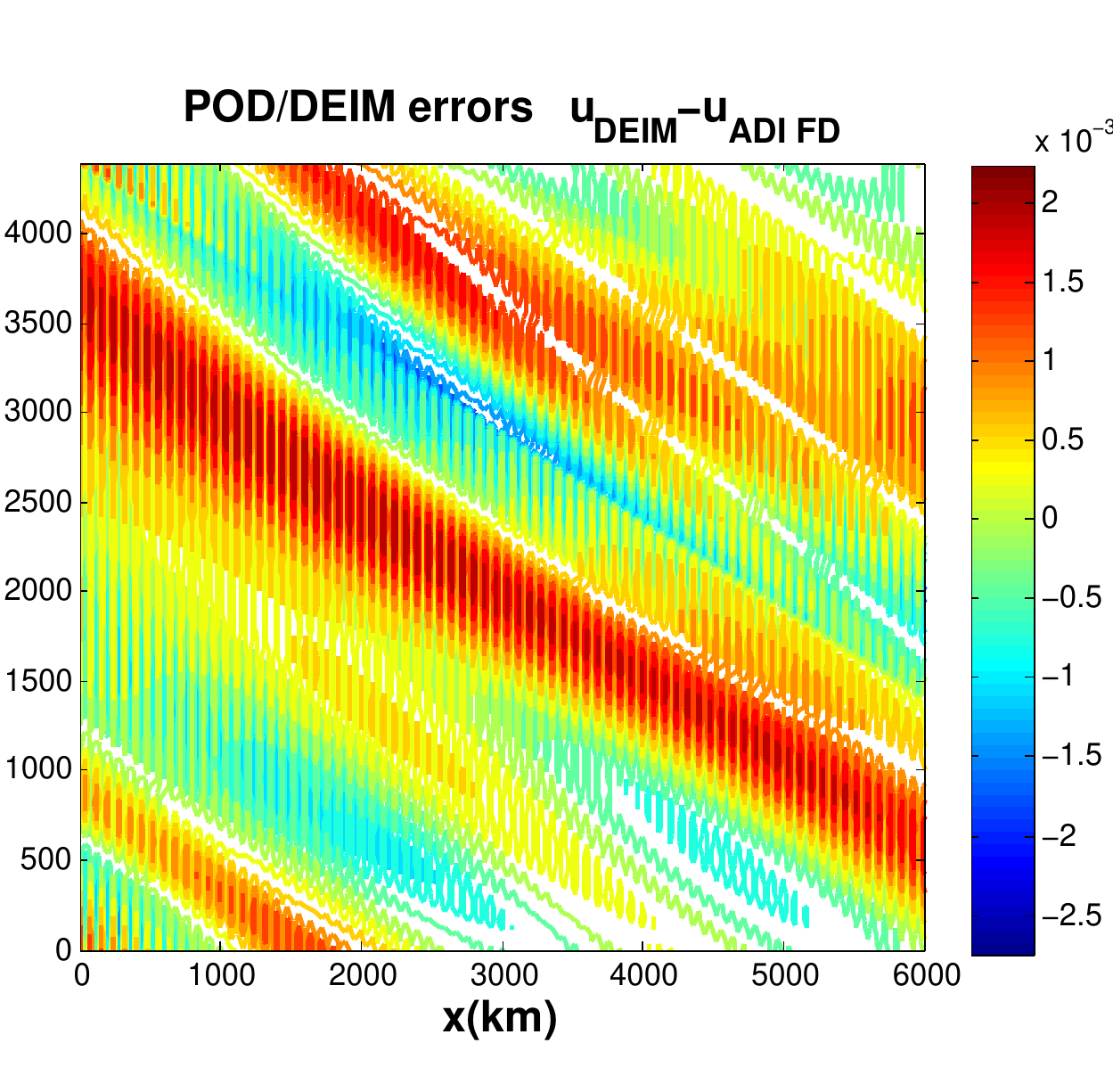}
\includegraphics[trim = 44mm 80mm 45mm 75mm, clip, width=5.2cm]{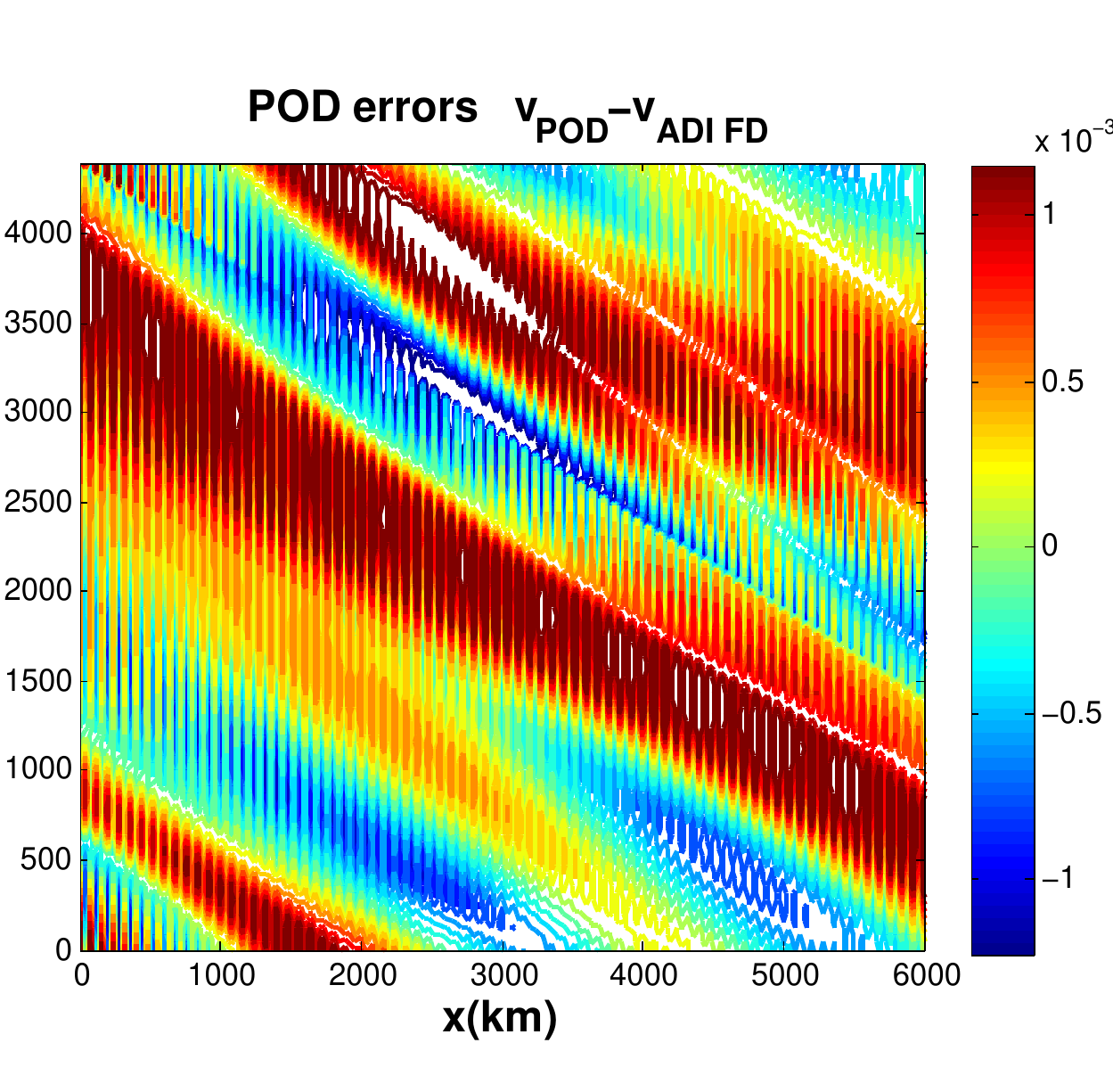}
\includegraphics[trim = 44mm 80mm 45mm 75mm, clip, width=5.2cm]{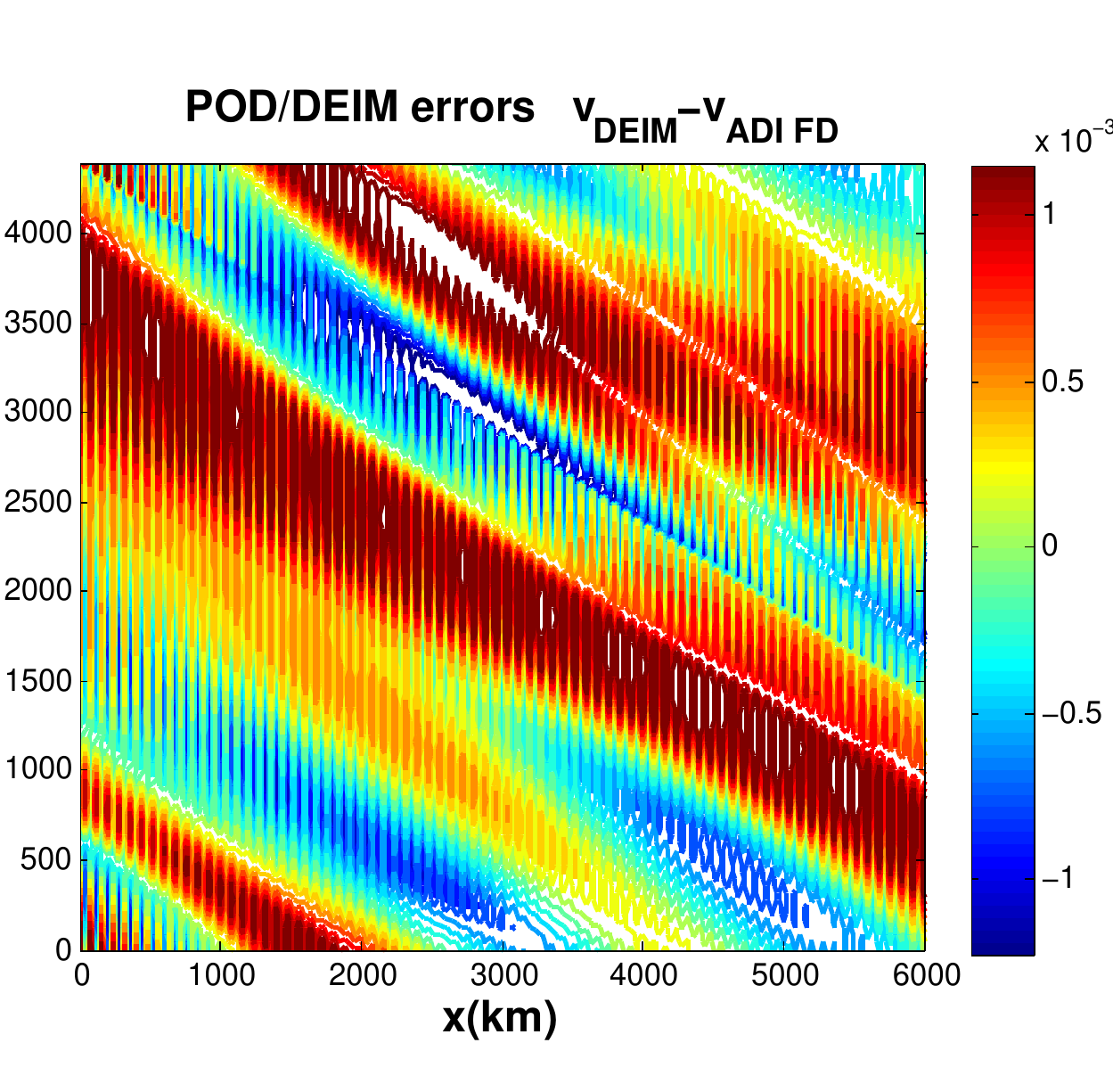}
\caption{\label{fig:DEIM19}Local errors between  POD, POD/DEIM ADI SWE solutions and the ADI FD SWE solutions at $t = 24h$ ($\Delta t = 960 s$) in context of parameter settings. The number of DEIM points was taken $90$.}

\end{figure}

We also tested the SWE discrete schemes in context of parametric variation of the Coriolis parameter. For these tests we used a mesh of $301\times 221$ points and we run the ADI FD scheme to generate $91$ snapshots using $\hat f=10^{-4}$. Next we derived the POD bases and than we solved the POD ADI SWE and POD/DEIM SWE discrete problems for $\hat f =3\cdot 10^{-4}$. Figure \ref{fig:DEIM19} depicts the grid point local error behaviours between POD, POD/DEIM ADI SWE solutions and ADI FD SWE solutions obtained by assigning the Coriolis parameter the value $3\cdot 10^{-4}$. The DEIM scheme used $90$ interpolations points.

\begin{figure}[h]
\centering
\includegraphics[trim = 35mm 80mm 22mm 85mm, clip, width=9.1cm]{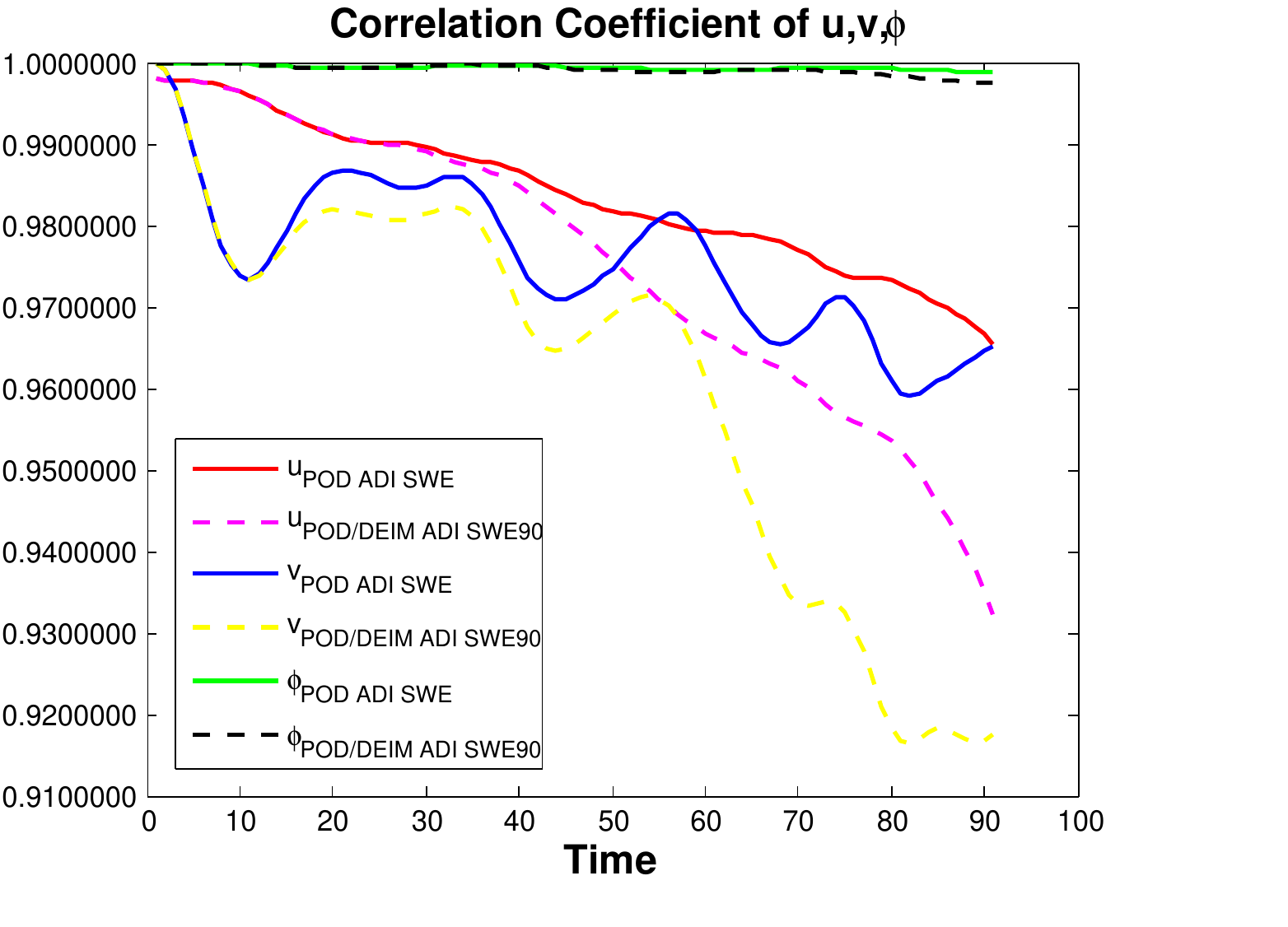}

\caption{\label{fig:DEIM20}Correlation coefficients for the SWE variables in parameter settings, POD dimension =$35$.}

\end{figure}

 Tables \ref{tab:table10} and \ref{tab:table11} contain the average relative errors and RMSEs for $u,v$ and $\phi$. Correlation coefficients are shown in Figure \ref{fig:DEIM20}.
 \begin{table}[h]
\centerline{
\scalebox{0.9}{
\begin{tabular}{|c|c|c|}\hline \hline
 &  POD ADI SWE & POD/DEIM ADI SWE\\ \hline
$E_{\phi}$ & 0.124508 & 0.14588\\ \hline
$E_u$ & 0.22403 & 0.27408\\ \hline
$E_v$ & 1.4169e-003 & 1.6294e-003\\ \hline
\end{tabular}}}
\caption{\label{tab:table10}Average relative errors for each of the model variables. The POD bases dimensions were taken to be $35$ capturing more than $99.9\%$ of the system energy. $90$ DEIM points were chosen.}
\end{table}
 \begin{table}[h]
\centerline{
\scalebox{0.9}{
\begin{tabular}{|c|c|c|}\hline \hline
 &  POD ADI SWE & POD/DEIM ADI SWE\\ \hline
$RMSE_{\phi}$ & 5.6992e-004 & 8.3716e-004\\ \hline
$RMSE_u$ & 8.8999e-004 & 1.2342e-003\\ \hline
$RMSE_v$ & 3.56849e-004 & 5.2638e-003\\ \hline
\end{tabular}}}
\caption{\label{tab:table11}Root mean square errors for each of the model variables at time $t = 24h$. The POD bases dimensions were taken $35$ and $90$ DEIM points were chosen.}
\end{table}

We can improve the accuracy of POD, POD/DEIM solutions and reduce the computational time by taking into account different POD bases dimensions according to the eigenvalues decay.  The final numerical test presented here is dedicated to the effect of different dimensions of POD bases in POD and POD/DEIM Euler explicit SWE schemes. As we expected, after a short analysis of eigenvalues spectrum we chose the dimensions of POD bases to be $31,32$ and 34. The results are presented in Table \ref{tab:table12} and we notice that the errors are smaller then the ones in Table \ref{tab:table2} where we used $35$ modes for each POD basis.

\begin{table}[h]
\centerline{
\scalebox{0.6}{
\begin{tabular}{|c|c|c|c|c|}\hline \hline
 & POD EE SWE (35 modes) & POD/DEIM EE SWE (35 modes) & POD EE SWE (31,32,34 modes) & POD/DEIM EE SWE (31,32,34 modes)\\ \hline
$RMSE_{\phi}$ & 1.545e-004&1.792e-004 & 1.1605e-004& 1.4246e-004\\ \hline
$RMSE_u$& 1.918e-004&3.126e-004&9.3842e-005 & 1.712e-004\\ \hline
$RMSE_v$& 1.667e-004 & 2.2374e-004& 9.551e-005&1.0791e-004\\ \hline\hline
\end{tabular}}}
\caption{\label{tab:table12}The root mean square errors for each of the model variables at $t=24h$. Comparison between the errors calculated with the same number of modes (first and second columns) and different number of modes (third and fourth columns).  We used a mesh of $301\times 221$ points and $90$ DEIM points were chosen.}
\end{table}

\section{Conclusions}
To obtain the approximate solution in case of both POD and POD/DEIM reduced systems, one must store POD or POD/DEIM solutions of order $O(kNT)$, $k$ being the POD bases dimension and $NT$ the number of time steps in the integration window. The coefficient matrices that must be retained while solving the POD reduced system are of order of $O(k^2)$  for projected linear terms and $O(n_{xy}k)$ for the nonlinear term, where $n_{xy}$ is the space dimension.

In the case of solving POD/DEIM reduced system the coefficient matrices that need to be stored are of order of $O(k^2)$ for projected linear terms and $O(mk)$ for the nonlinear terms, where $m$ is the number of DEIM points determined by the DEIM indexes algorithm, $m\ll n_{xy}.$ Therefore DEIM improves the efficiency of the POD approximation and achieves a complexity reduction of the nonlinear term with a complexity proportional to the number of reduced variables.

We proved the efficiency of DEIM using two different schemes, the ADI FD SWE fully implicit model and the Euler explicit FD SWE scheme. We noticed, as we expected, that POD/DEIM CPU time is most sensitive to the number of spatial discretization points. The largest reduction of the CPU time was obtained in first experiment where it was reduced by a factor of $73.91$ when using the POD/DEIM ADI SWE scheme while in the case of POD/DEIM EE SWE model we decreased the CPU time by a factor of $68.733$. Also we noticed that the approximation errors of POD/DEIM and POD reduced systems are almost identical once the dimension of DEIM attained the value of $50$, for any of the methods used, either explicit or implicit. In the second experiment, we increased the number of time steps and snapshots and consequently the solutions accuracy was higher in comparison with the results obtained in the first experiment.

In future research we plan to apply the DEIM technique to different inverse problems such as POD 4-D VAR of the limited area finite element shallow water equations and adaptive POD 4-D VAR applied to a finite volume SWE model on the sphere.

We are also interested to compare the Discrete Empirical Interpolation Method with the 'Best Points' Interpolation (BPIM) one in proper orthogonal decomposition framework applied to SWE equation. BPIM was proposed by Nguyen et al \cite{NPP2008} where the interpolation points are defined as a solution of a least-squares minimization problem. Thus, BPIM replaces the greedy algorithm used in Empirical Interpolation Method(EIM) by an optimization problem which provides higher accuracy at the cost of greater computational complexity. For instance Galbally et al. \cite{Gal2010} made a comparison between gappy POD, EIM and BPIM techniques for a nonlinear combustion problem governed by an advection diffusion PDE.


%
%
{\centering
\section*{Acknowledgments}

Prof. I.M. Navon acknowledges the support of NSF grant ATM-0931198. The authors acknowledge the generous help of Professor Danny C. Sorensen Noah Harding Professor of Computational and Applied Mathematics at Rice University, that provided us with a 1-D Matlab code of POD DEIM of the Burgers equation. The contribution of two anonymous reviewers to improving and clarifying the presentation of our manuscript is acknowledged.}

\section*{Appendix}

This appendix contains a definition for the componentwise multiplication operation and a more symbolic representation of the Gustafsson's nonlinear ADI finite difference shallow water equations scheme defined in (2-3), Section 2. The componentwise multiplication is also known as the Hadamard product.
\vskip 0.3cm

{\bf Definition} {\it  The componentwise multiplication is a binary operation that takes two matrices of the same dimensions, $A,~B\in \mathbb{R}^{m \times n}$, $m,n \in \mathbb{N}^{*}$, and produces another matrix of the same dimensions $A*B\in \mathbb{R}^{m \times n}$ with elements satisfying the following relationship}
$$(A*B)_{i,j}=(A)_{i,j}\cdot(B)_{i,j}.$$
\vskip 0.3cm
Now, recall that the number of mesh points is $n_{xy}=N_x \cdot N_y$. We shall denote by $w_{i,j}^n(i=1,..,N_x;~j=1,..,N_y;~n=1,..,NT$) an approximation to $w(i\Delta x,j\Delta y,n\Delta t)=(u,v,\phi)(i\Delta x,j\Delta y,n\Delta t)$. The basic difference operators are
$$D_{0x}w_{i,j}^n=(w_{i+1,j}^n-w_{i-1,j}^n)/(2\Delta x),$$
$$D_{+x}w_{i,j}^n=(w_{i+1,j}^n-w_{i,j}^n)\Delta x, $$
$$D_{-x}w_{i,j}^n=(w_{i,j}^n-w_{i-1,j}^n)\Delta x, $$
respectively, with similar definitions for $D_{0y},D_{+y},D_{-y}$. We also define the operators $P_{i,j}^n$ and $Q_{i,j}^n$ by
$$ P_{i,j}^n=\Delta t/2(A(w_{i,j}^n)D_{0x}+C_j^{(1)}),$$
$$ Q_{i,j}^n=\Delta t/2(B(w_{i,j}^n)D_{j}+C_j^{(2)}),$$
with $A,~B$ defined at the beginning of Section (2),
$$D_j=\begin{cases} D_{0y}, & ~j=2,..,N_y-1, \\ D_{+y}, & ~j=1, \\ D_{-y}, & ~j=N_y,\end{cases},$$
(owing to the boundary conditions in the $y-$ direction)
$$ C_{j}^{(1)}=-\left(\begin{array}{ccc}
           0&0&0\\
           -f_j&0&0\\
           0&0&0 \end{array}\right),
~C_{j}^{(2)}=-\left(\begin{array}{ccc}
           0&f_j&0\\
           0&0&0\\
           0&0&0 \end{array}\right)
$$
and $f_j=f(j\Delta y),~j=1,..,N_y.$

Thus we rewrite the Gustafsson's  nonlinear  ADI  difference  scheme given in (\ref{eq2}-\ref{eq3})

\begin{equation}\label{eq12} (I-P_{i,j}^{n+1/2})w_{i,j}^{n+1/2}=(I+Q_{i,j}^n)w_{i,j}^n,\end{equation}
\begin{equation}\label{eq13} (I-Q_{i,j}^{n+1})w_{i,j}^{n+1}=(I+Q_{i,j}^{n+1/2})w_{i,j}^{n+1/2},\end{equation}
$$i=1,..,N_x-1;~j=1,..,N_y;~n=1,2,..,NT-1.$$
The periodic boundary conditions in the $x-$direction allowed us to use only central differences to approximate the derivatives with respect to $x$ and eliminated the need of calculating the SWE solutions for $i=N_x$.

The nonlinear algebraic equations do not apply to the $v$ component for $j=1,~j=N_y$, but we used the condition $v_{i,1}^n=v_{i,N_y}^n=0,~i=1,..,N_x-1,~n=1,2,..,NT$.

\end{document}